\documentclass[12pt]{article}

\usepackage{amsbsy}
\usepackage{amsfonts}
\usepackage{amsmath}
\usepackage{graphicx}
\usepackage{array}
\usepackage{booktabs}
\usepackage{float}
\usepackage[usenames, dvipsnames]{color}
\usepackage{latexsym,amssymb,theorem,epsfig}

\newcommand{\mathsym}[1]{{}}
\def\id{\protect{{1 \kern-.28em {\rm l}}}}

\def\be{\begin{equation}}
\def\ee{\end{equation}}

\def\bea{\begin{eqnarray}}
\def\eea{\end{eqnarray}}

\def\p{{\partial}}

\makeatletter
\renewcommand\section{\@startsection {section}{1}{\z@}%
                                   {-3.5ex \@plus -1ex \@minus -.2ex}%
                                   {2.3ex \@plus.2ex}%
                                   {\normalfont\large\bfseries}}
\renewcommand\subsection{\@startsection{subsection}{2}{\z@}%
                                   {-3.25ex\@plus -1ex \@minus -.2ex}%
                                   {1.5ex \@plus .2ex}%
                                   {\normalfont\normalsize\bfseries}}

\makeatother

\def\eps{{\epsilon}}

\def\Tr{{\rm Tr}}

\def \foot {\footnote}

\def \tr {{\rm tr}}
\def \ha {{1 \over 2}}

\def \N {{\mathcal N}}

\def \L {\Lambda}

\def\a{\alpha}
\def\b{\beta}

\def\p{\phi}

\def \del{\partial}
\def \a {\alpha}
\def \aa {{\a'}}
\def\g{\gamma}
\def\s{\sigma}

\def\ov{\over}

\def\b{\beta}
\def\l{\lambda}
\def\eps{\epsilon}

\def \k {\kappa}
\def\foot{\footnote}

\def \Tr {{\rm Tr}}
\def \P {\Phi}
\def \l  {\lambda}

\def \N {{\mathcal N}}

\def \N {{\mathcal N}}

\def \m {\mu}

\def \la {\label}

\def \l {\lambda}
\def\foot{\footnote}

\newcommand{\rf}[1]{(\ref{#1})}
\def \ov {\over}

\def\N{{\cal N}}

\def \ha{{1\ov 2}}
\def \r {\rho}

\def \no {\nonumber}

\def \del {\partial}

\def \la {\label}

\def \l {\lambda}
\def\foot{\footnote}

 \def \p {\phi}
 \def \r {\rho}

\def \ov {\over}

\def \varpi {{\rm w}}

\def \ep {\epsilon}

\def \DD {{\rm D}}

\def\Tr{{\rm Tr}}

\def \s {\sigma}
\def\eps{{\epsilon}}

 \def \n {\nu}

\def \vp {\varphi}

\def \cC {{\cal C}}

\def \ed {\end{document}}

\def \ha {{{\textstyle{1 \ov2}}}}
\def \fo {{\textstyle{1 \ov4}}}

\def \CC {{\rm C}}

\def\Tr{{\rm Tr}}

\def\NeqFour{{{\cal N}=4}}

\def\NeqEight{{{\cal N}=8}}
\def\NeqTwo{{{\cal N}=2}}

\def\NeqOne{{{\cal N}=1}}

\def\rep#1{{\bf #1}}
\def\rk{{\rm k}}

\def\spa#1.#2{\left\langle#1\,#2\right\rangle}
\def\spb#1.#2{\left[#1\,#2\right]}
\def\f{\tilde f}

 \def \eqref  {\rf}

\def \ha {{\textstyle {1 \ov 2}}}
\def \trd {{\textstyle {3 \ov 2}}}

\def\tlambda{{\tilde\lambda}}

\def \nf   {$\N=4$\ } 
\def \cC {B} \def \bC {C}
\def \vp {\varphi}
\def \Im {{\rm Im\,}}
\def \nv { n_{\rm v}}

\def \AC {A^{(c)}}  \def \AS  {A^{(s)}} \def \G {\Gamma}
\def\taua{{{\rm t}}}
\def\bartaua{{{\bar {\rm t}}}}
\def \qq {{\rm q}} 

\begin{document}

%%%%%%%%%%%%%%%%%%%%%%%%%%%%%%%%%%%%%%

\overfullrule=0pt
\parskip=2pt
\parindent=12pt
\headheight=0in \headsep=0in \topmargin=0in \oddsidemargin=0in

\vspace{ -3cm}
\thispagestyle{empty}
\vspace{-1cm}

\rightline{SU-ITP-13/04}

\rightline{Imperial-TP-AT-2013-02}

\def \ka {{\varkappa}}

\begin{center}
\vspace{1cm}
{\Large\bf  
 On %(the relation between)
  the $U(1)$ duality anomaly  and 
 the S-matrix\\
 \vspace {.2cm}
  of  ${\cal N}=4$  supergravity  
}

\end{center}

\vspace{.2cm}

\begin{center}
 
 J.J.M. Carrasco$^{a}$, R. Kallosh$^{b}$, R. Roiban$^{c}$  and   A.A. Tseytlin$^{d,}$\footnote{Also at Lebedev  Institute, Moscow. }

\end{center}

\begin{center}
{
\em 
\vskip 0.08cm
\vskip 0.08cm 
$^{a,b}$Stanford Institute for Theoretical Physics and Department of Physics,\\
Stanford University, Stanford, CA 94305-4060, USA\\
\vskip 0.08cm
\vskip 0.08cm 
$^{c}$Department of Physics, The Pennsylvania  State University,\\
University Park, PA 16802 , USA\\
\vskip 0.08cm
\vskip 0.08cm 
$^{d}$Blackett Laboratory, Imperial College,
London SW7 2AZ, U.K.
 }
 \end{center}

\vspace{1.5cm}

\vspace{.2cm}

\begin{abstract}

\noindent
%%%%%%%%%%%%%%%%%%%%%%%%%%%%%%%%%
${\cal N}=4$  Poincar\'e supergravity has a global $SU(1,1)$  duality  
symmetry that acts manifestly only on shell  as it  involves  duality 
rotations of vector fields.  A $U(1)$ subgroup
of  this   symmetry is known to be anomalous at the quantum level 
in the presence of a non-trivial  gravitational background.  We first  
derive  this anomaly from a novel perspective, by relating it to  a similar 
anomaly in conformal supergravity where $SU(1,1)$
acts off shell, using the fact  that  ${\cal N}=4$ Poincar\'e supergravity
has a superconformal formulation.
 We explicitly  construct  the corresponding 
local and nonlocal anomalous terms  in the one-loop  effective action. 
We then  study how this anomaly  is reflected in the  supergravity  S-matrix. 
 Calculating one-loop  ${\cal N}=4$  supergravity scattering amplitudes  (with and
without additional matter multiplets) using color/kinematics duality and 
the  double-copy construction  we  find that 
a particular  $U(1)$ symmetry which was present in the tree-level amplitudes 
 is broken at the quantum  level.  
This breaking manifests  itself  in the appearance  of new one-loop ${\cal N}=4$  
supergravity amplitudes  that  have  non-vanishing soft-scalar limits
(these amplitudes   are absent  in  ${\cal N}>4$ supergravities).   
We   discuss the relation between  these  symmetry-violating amplitudes and the  
corresponding  $U(1)$ anomalous  term in the  one-loop   supergravity effective action. 
%%%%%%%%%%%%%%%%%%%%%%%%%%%%%%%%%

\end{abstract}

\newpage

\tableofcontents

\setcounter{equation}{0} 
\setcounter{footnote}{0}
\setcounter{section}{0}

\renewcommand{\theequation}{1.\arabic{equation}}
 \setcounter{equation}{0}

\newpage

%%%%%%%%%%%%%%%%%%%%%%%%%
\section{Introduction}
\label{intro}
%%%%%%%%%%%%%%%%%%%%%%%%%

Extended Poincar\'e supergravity theories and their ultraviolet
behavior have been a subject of  renewed interest
in recent years. Although, in general, supersymmetry softens
ultraviolet divergences, it cannot overcome the effects
of the two-derivative stress tensor coupling characteristic to
gravitational theories. It has been suggested that, apart
from supersymmetry, 
%U-duality symmetries -- 
non-compact global  %(``U-duality'')  
symmetries \cite{jul0, jul} of the equations of motion of
${\cal N}\ge 4$ classical supergravity  theories  also play an
important role in constraining the ultraviolet behavior
of theories exhibiting them.
These duality symmetries, involving electric/magnetic duality transformations
of abelian vector fields, are specific to four-dimensional extended supergravity theories;
% while the details of their appearance remains to be fully
%clarified, 
their origin  is related to  dimensional reduction from 
higher-dimensional supergravities. 
%For ${\cal N}\ge 4$ 
%supergravities the U-duality group is non-abelian.

The largest  duality group is that of $\NeqEight$ supergravity,
$E_{7(7)}$. Other 
%four-dimensional pure
supergravities can be obtained as 
%through factorized or non-factorized 
consistent truncations from  $\NeqEight$ supergravity and their
duality groups may be thought of as subgroups of $E_{7(7)}$.  While
the  duality groups of ${\cal N}>4$ supergravities
are expected to be preserved at  the quantum level due to the absence of
anomalies, an abelian subgroup of the
classical $SU(1,1)$ duality group of $\NeqFour$ supergravity was
argued in \cite{mar} to be anomalous. %have a chiral anomaly.

Quite generally, one may say that an anomaly is present whenever a
classical (global or local) symmetry is broken by quantum
corrections. It may either be a symmetry of the Lagrangian or of the
equations of motion or a symmetry of scattering
amplitudes (i.e. a symmetry realized on asymptotic states). 
It would undoubtedly be interesting to identify the consequences of an
anomaly such as that of $U(1)\subset SU(1,1)$
at the level of the scattering matrix of the theory and to 
address  an  open question of whether its presence has any
consequences  for  the ultraviolet properties  
of higher-order S-matrix elements
%AT
 (for a recent discussion see \cite{Bossard:2012xs}  and references therein). 
 %RR
 Anticipating our results, we will find a close relation between the $U(1)$ anomaly 
 and special classes of amplitudes of $\NeqFour$ supergravity which vanish
 at tree-level but not at one-loop; though the double-copy construction of 
 supergravity amplitudes they are related to non-supersymmetric gauge 
 theory amplitudes that have similar properties. 

As it is well-known, $\NeqFour$ supergravity can be formulated in  different  
classically-equivalent ways. In  the  ``covariant''  formulation,
the global symmetry $SU(1,1)$ acts linearly  on two complex scalar
fields $\P_\a$  (satisfying $\P_1^*\P_1 - \P_2^*\P_2=1$)
while an auxiliary local $U(1)$ symmetry  (with field-dependent composite  gauge field)
acts  also on other fields. The physical complex scalar
field  of  the $\NeqFour$  supergravity  multiplet
parameterizes the coset $SU(1,1)/U(1)$.
An alternative  ``unitary-gauge'' formulation  corresponds  to a particular 
gauge choice for the auxiliary $U(1)$  gauge symmetry. 
 The action
of the global $U(1)$ symmetry on vector fields has the interpretation
of an electric/magnetic duality rotation.
The
$U(1)$ anomaly has different -- but related -- interpretations in
these two formulations  \cite{dfg,gdw}: in the former it is the chiral anomaly
of the auxiliary $U(1)$  gauge symmetry,
%\footnote{Such an anomaly does not lead to an inconsistent theory because the
%auxiliary $U(1)$ local symmetry does not have an independent gauge field.}
while in the latter it is the chiral anomaly of a particular  global $U(1)$  duality 
subgroup whose precise embedding into   $SU(1,1)$   depends on the gauge choice for the auxiliary
gauge symmetry. %After gauge fixing,

%%%%%%%%%%%%%%%%%%%%%%%%%%%%%%%%%%%%%%%
%The essential point in Marcus' calculation of the anomaly  \cite{mar}
The crucial point  in the   computation of the  duality symmetry anomaly   
 in the  ``unitary-gauge'' formulation  in~\cite{mar}  was  the 
 inclusion of the  vector field   contribution  to the anomaly. 
 As the vectors  transform under the duality in a nonlocal way,  via a $\delta F = \eps F^*$  
 rotation of their on-shell  field  strength, this was done   indirectly, via  a topological   
 count of anomaly for self-dual  tensors. 
 The corresponding  anomaly 
% this was done is the treatment of vector fields as self-dual
%tensor fields. In the absence of matter ($\NeqFour$ vector) multiplets
%one 
%AT
may indeed  be interpreted  as the vector field   contribution to the chiral gravitational   anomaly, i.e.  
as   the non-vanishing of the expectation value of the divergence of the  corresponding  current  
%corresponding to this  field strength duality transformation 
%duality transformation on vector fields.
in a non-trivial gravitational  background as was  understood  in unrelated  work in \cite{dolg,end,reu,erd}. 
This anomaly was also  rederived using a Lorentz non-covariant doubled-vector formulation  in \cite{bhn}. 

Here we shall present a novel way   of understanding 
this anomaly, by relating it to
a similar anomaly \cite{rvn}  in the  conformal supergravity  (CSG)  \cite{bdw}, 
which admits a formulation where  $SU(1,1)$ is a global off-shell symmetry.  The key  
observation   is  that  the   classical 
${\cal N}=4$ Poincar\'e supergravity~(PSG) has  a superconformal formulation
\cite{roo} as the $\NeqFour$ conformal supergravity coupled to six  $\NeqFour$   vector 
multiplets  (with the  higher-derivative action of pure  $\NeqFour$    CSG  not   added). 
%%%%%%%%
%Since the superconformal formulation of the theory is
%classically-equivalent  ?!?!     why then same anomalies ?????to the more standard ones, one may
%expect that their anomalies are also the same.
%This     relation  suggests that one  may count
%%%%%%%%
In this superconformal framework the six vector fields, which 
(upon gauge-fixing of the conformal symmetry and S-supersymmetry and
elimination of the auxiliary fields) become the six vector fields of the PSG
multiplet, couple  via their field strengths  $F$   to the six (anti)self-dual rank-2 tensors  $T$ 
appearing in the CSG multiplet   and that provides a  possible   link  between   the  $F$  and $T $   
contributions to the corresponding anomalies. 

%Since the self-dual tensors $T$  transform under
%the $U(1)$ symmetry, so must the (anti)self-dual components of the
%field-strengths of the six vector fields.   No, they couple via scalars 
%hints at an explanation for interpreting  the vector fields as
%self-dual tensors in the counting of the chiral $U(1)$
%anomaly in Poincar\'e supergravity.

For $\NeqEight$ Poincar\'e  supergravity, it was argued in \cite{Elvang:2010kc,
ArkaniHamed:2008gz, Kallosh:2008rr}
that, at the tree-level, the $E_{7(7)}$ duality  symmetry group implies that all
scattering amplitudes vanish in the single soft scalar limit (i.e. the limit in 
which the momentum of a scalar field goes to zero).
This may  be viewed  as a consequence of the on-shell $SU(8)$
R-symmetry group of the theory. 
It has also been argued in \cite{ArkaniHamed:2008gz}
and further detailed in \cite{Brodel:2009hu} that the
double soft scalar limit (i.e. the limit in which the momenta of two
scalar fields vanish simultaneously) probes the
commutation relations of the duality group. 
%%%
%To this end it is
%necessary that the soft scalars share three of their four $SU(8)$
%indices~\cite{Brodel:2009hu}.
%%%

A similar analysis has not yet been carried out  for $\NeqFour$  Poincar\'e   supergravity.
As  was  shown in \cite{cgr}, the $\NeqFour$ supergravity
can be interpreted as an orbifold truncation of $\NeqEight$
supergravity. From this perspective, the one physical  complex scalar field
of the former theory is a linear combination of the scalar fields of
the latter. As such, the tree-level scattering amplitudes of the 
$\NeqFour$ supergravity are a subset of those of the $\NeqEight$
supergravity and thus they should vanish in the soft scalar limit.
%
% R JHEP
%
Probing the commutation relations of the ${\cal N} = 4$ duality group SU(1,1) through a
double-soft limit would require a careful construction and analysis of non-MHV one-loop 
amplitudes with at least six external particles.
%%
%% However, it  is difficult to probe the commutation relations of the
%% $\NeqFour$ duality group $SU(1,1)$ (through, e.g.,  a double-soft limit of a 
%% one-loop six-point amplitude) as the required non-MHV amplitudes are not
%% currently known. 
%
% Curiously, it appears that the double-soft limit of an anomalous 5-point amplitude
% (hh t tbar tbar) which should give a 3-point anomalous amplitude (hh tbar) is in fact zero.
% 
%However, it  appears difficult to probe the commutation relations of the
%$\NeqFour$ duality group $SU(1,1)$ as 
%only one scalar field survives the orbifold projection which,
%in   the  $\NeqEight$ context,  is insufficient~\cite{Brodel:2009hu} for probing 
%any of the commutation relations of $E_{7(7)}$.
%%

The next step would be 
to study directly the soft scalar limits of the 
{\it one-loop}  amplitudes in $\NeqFour$ supergravity. As we shall
see below, the $U(1)\subset SU(1,1)$ symmetry that requires the vanishing of the
tree-level soft scalar limits is {\it broken}  at one-loop
level and there exist amplitudes   with nontrivial $U(1)$ charge which are non-vanishing; 
we shall refer to them as ``anomalous amplitudes". The mechanism through which 
these amplitudes are non-vanishing is quite similar to the one   leading to the chiral 
anomaly --  a divergence in a loop integral is compensated by a zero 
in its coefficient (in the amplitude  case from the momenta of states running in 
%AAT
the loop).\foot{One may,  in  a sense,   interpret these  anomalous amplitudes 
as representing  an  anomaly  in an  on-shell    bosonic symmetry 
 (which is a remnant of the extended {supersymmetry} of the supersymmetric  Yang-Mills theory)
of  tree-level  scattering amplitudes of  pure  Yang-Mills theory. Indeed,  at  tree level the S-matrices  of 
the pure Yang-Mills  and  the super Yang-Mills   theories   are the same (in the bosonic vector sector), 
 but this is no longer so at  the one-loop level.
 }
We will  construct examples of such amplitudes and  find  their soft scalar limit.
We will then use the resulting soft scalar functions  to construct the
one-loop all-multiplicity $U(1)$-violating amplitudes with
all external legs belonging to a single on-shell chiral multiplet.

We shall also consider the   scattering amplitudes   in 
$\NeqFour$ Poincar\'e supergravity coupled to
 an arbitrary number $n_\text{v}$ of   abelian  $\NeqFour$  matter (vector) multiplets. 
While in the presence of matter multiplets $\NeqFour$  Poincar\'e
supergravity  has  ultraviolet divergences  already at the one-loop supergravity level 
 (the relevant counterterm is proportional to the square of the matter-field  
 stress tensor \cite{fischler,FT})
%%%%%%%
%At this loop level we can therefore focus on the part of the effective
%action that depends solely on the fields in the
%%%%%%%
we may restrict consideration to a finite sector with all external states
belonging to the supergravity multiplet and track down the contribution of 
the matter multiplets to the $U(1)$  anomalous amplitudes. The $U(1)$ charges
of matter fields are related to those of the fields of  the supergravity multiplet 
by the $SO(6,n_{\rm v})$ symmetry of matter-coupled $\NeqFour$ supergravity. 
%%%%%%%
%symmetry breaking to
%both to its anomalous part as well as to its matrix elements -- i.e.
%to the anomalous scattering amplitudes.
%%%%%%%

The construction of one-loop amplitudes in $\NeqFour$ supergravity with and without matter
multiplets is made straightforward by the use of the duality between color and kinematics of 
super Yang-Mills theory amplitudes, and of the corresponding double-copy construction of 
(super)amplitudes in related supergravity theories, uncovered in \cite{BCJ, BCJLoop}.
%%%
\iffalse
%
The construction of one-loop amplitudes in $\NeqFour$ supergravity with and without matter 
multiplets is made straightforward by the use of the duality between color and kinematics 
of super Yang-Mills theory amplitudes, uncovered  in \cite{BCJ, BCJLoop}, and the corresponding 
double-copy construction of (super)amplitudes in related supergravity theories \cite{BCJLoop}. 
%
\fi
%%%
According to the  color/kinematics  duality conjecture, the integrands of super
Yang-Mills theory amplitudes can 
be organized in terms of graphs with only cubic vertices such that there is a one-to-one correspondence 
between the Jacobi identities obeyed by the graphs' color factors and their kinematic numerator factors.
Whenever such a representation is available, the (super)amplitudes of a related supergravity theory are  obtained 
by simply replacing the color factors with the kinematic numerator factors of a second-factor 
gauge theory. 
The validity of the construction can be easily confirmed through the evaluation of the $D$-dimensional 
unitarity cuts.

\

This  paper is organized as follows. In section~2 we will  discuss  the $U(1)$
anomaly of the Poincar\'e supergravity from the
perspective of the superconformal formulation of the theory.  We will then 
construct in detail the anomaly-induced terms in
the one-loop  effective action in the ``unitary-gauge'' 
 formulation of the theory with manifest
$SU(4)$ R-symmetry. We shall  assume  a general reparametrization-invariant 
regularization scheme and 
%%%%%%%
%AT:   they are not added, they are there 
%if all   ef action (not just anom term) is computed in susy preserving way
% which  may  not preserve supersymmetry and 
%%%%%%%
will  comment on the structure of  other 
(parity-even, non-anomalous,  finite) terms that should also  appear in the effective action 
 % need to be added to the effective action to restore 
to maintain  supersymmetry.  
We shall  discuss  a 
consistent assignment of  the $U(1)$ charges to extra  matter
multiplets  coupled to  $\NeqFour$  Poincar\'e supergravity  and
%%%%%%%
%Together
%with the charges of the on-shell supergravity multiplet states 
%in the $\NeqFour$ supergravity multiplet,  this will allow us to
%%%%%%% 
identify a certain $U(1)$ symmetry of asymptotic states such that their
corresponding charges are the same as their
charges with respect to $U(1)\subset SU(1,1)$.

In section~3, after a general discussion on the structure and
double-copy  construction of the scattering
amplitudes in matter-coupled $\NeqFour$ supergravity, we will proceed to
compute  explicitly, through the double-copy construction and the generalized 
unitarity method,  the three-, four- and five-point amplitudes that break the 
asymptotic-state $U(1)$ symmetry previously identified
and which  are forbidden in the ${\cal N} \ge 5$ supergravities. These amplitudes
are  UV finite and have  rational dependence on external momenta. 
As usual in scattering
superamplitude calculations, the result preserves manifestly the
supersymmetry of the asymptotic states. We also discuss in
some detail the same  amplitudes in matter-coupled $\NeqFour$ supergravity.

In section~4 we will analyse the soft scalar limits of these amplitudes
and use them  to   present a
well-motivated conjecture for the all-multiplicity one-loop
superamplitudes with all fields in one of the two $\NeqFour$
on shell supergravity multiplets.  These  amplitudes   correspond to a particular 
term   in the  one-loop
effective action  containing  two gravitons and having 
holomorphic scalar-field dependence.  We compare this effective action term with
the general form of the anomaly-induced effective action found in section 2. 
%%%%%%%
%AT
% As expected, the difference   is only  in   finite, local terms related 
% to the fact that  supersymmetry was not manifestly preserved 
% in the   construction of the latter. 
%%%%%%%

%%%%%%%
%whose presence can be
%ascribed to the 
%non-supersymmetric regularization
%scheme used in the construction of the anomaly-induced effective action.
%%%%%%%

We close in section 5 with remarks on nontrivial contributions of the one-loop anomalous 
amplitudes to higher-loop amplitudes.   We also discuss the existence of an analogous 
$U(1)$ (electric/magnetic) duality symmetry in certain matter-coupled supergravity theories, 
with fewer supercharges, which can be obtained by truncation from $\NeqEight$ supergravity. 
This symmetry appears to be broken (via a mechanism similar to the one described above) by 
finite graviton-matter amplitudes.

\iffalse
We close in section~5 with remarks on  nontrivial contributions of
the one-loop anomalous amplitudes to higher-loop amplitudes
as well as on the existence of an analogous $U(1)$ (electric/magnetic)
duality symmetry  in matter-coupled supergravity theories with fewer supercharges
which, for certain (non-zero) numbers of matter multiplets,  can be obtained by truncation from 
$\NeqEight$ supergravity.  
This symmetry appears to be broken (via  a mechanism similar to the one described  above) 
by finite graviton-matter amplitudes.
\fi

In Appendix A    we shall comment on the  vector field contribution to the 
chiral anomaly on a gravitational   background.  In Appendix B  we shall  discuss the local 
parity-even    scalar-curvature-curvature term  which is a natural superpartner of the local 
part of the  parity-odd  anomalous   term  in the effective action.
Appendices C, D and E contain details on the construction and evaluation of 
$n=3,4,5$ -point %3-, 4-  and 5-point 
anomalous amplitudes.

%%%%%%
% it  is an open question if the presence of such anomalous one-loop terms actually influences 
% 3-loop  divergent  lowest-order S-matrix elements at all ...
%
% \nf PSG is one-loop  finite so it makes sense to discuss its one-loop effective action and S-matrix.
% Once  extra  matter  multiplets   are added  there are  already one-loop divergences like square of 
% the stress tensor of matter multiplets $T_{\m\n} T^{\m\n}$  \cite{fischler,FT}.  We shall formally
% ignore this issue  concentrating on finite sectors of S-matrix  related to  $U(1)$ anomalous 
% terms in the effective action. 
%%%%%% 

%%%%%%%%%%%%%%%%%%%%%%%%%
%\section{Formulations of ${\cal N}=4$ supergravity}
%%%%%%%%%%%%%%%%%%%%%%%%%
%\bldraft{ 2) Various formulations of  Poincar\'e N=4 supergravity:     SU(4) vs SO(4);     anomalies in the two formulations??    }  \
%\draftnote{decide whether or not to keep this section}
%%%%%%%

%%%%%%%%%%%%%%%%%%%%%%%%%%%%%%%%%%%%%
\renewcommand{\theequation}{2.\arabic{equation}}
 \setcounter{equation}{0}
%%%%%%%%%%%%%%%%%%%%%%%%%
\section{$U(1)$  anomalies in conformal and Poincar\'e   $\N=4$ supergravities} \la{sec3}
%%%%%%%%%%%%%%%%%%%%%%%%%

Our aim here will   be to    study  the structure of  the $U(1)$ anomaly in  \nf  Poincar\'e supergravity (PSG) \cite{dfg,mar}
(with or without additional  $\N=4$   matter  multiplets) 
and its consequences for the corresponding scattering amplitudes.
By $U(1)$  anomaly here we mean the  anomaly  of an auxiliary    gauge symmetry 
in covariant formulation in which a global $SU(1,1)$   duality symmetry is realized linearly or, 
equivalently, the anomaly of a particular global $U(1)\subset SU(1,1)$  symmetry
in a ``physical gauge'', where  the  local $U(1)$ symmetry is  gauge-fixed.

As   was shown   in \cite{roo},  pure \nf   PSG     theory  \cite{cre}  may be interpreted as a  spontaneously
 broken (and gauge-fixed)
version of \nf  conformal   supergravity (CSG) \cite{bdw}  coupled to six \nf vector multiplets (assuming 
the  higher-derivative ``kinetic'' term of the   CSG   theory  is not  included).  
A similar formulation also holds  for $\NeqFour$ PSG coupled to 
  $n$  matter \nf vector multiplets:  one is to start with  $n+6$  vector multiplets  
(with six vector multiplets having ``wrong''  sign of kinetic term) coupled to  conformal supergravity. 
  
  Given  that  
  (i)  the anomalies of spontaneously broken and unbroken  
  phases of CSG may be  expected to be the same (as suggested by the classical equivalence of the two phases), 
  and also that  
  (ii)  anomalies are usually controlled  by  lowest  derivative terms (i.e.  they are unchanged by 
  addition of higher-derivative terms with same symmetry), 
  one may conjecture that  the $U(1)$ anomaly of \nf  PSG \cite{mar} can be  understood  in 
  terms of the corresponding anomaly \cite{rvn} of the system of higher-derivative CSG  coupled to $\NeqFour$ 
  vector multiplets. 
 
 We shall demonstrate that this is so in sections \ref{sec31} and \ref{sec32} below. 
 Then, we shall discuss  the detailed structure of  the  corresponding anomalous effective action 
 of \nf  PSG in section~\ref{sec33}.

%%%%%%%%%%%%%%%%%%%%%%%%%%%%%%%%%%%%%%
\subsection{$U(1) $ anomaly in $\NeqFour$ conformal supergravity \label{sec31}}
%%%%%%%%%%%%%%%%%%%%%%%%%%%%%%%%%%%

Let us  begin by recalling  the  field content of \nf  CSG \cite{bdw,ber}.
%The story of $U(1)$ symmetries   in $\NeqFour$  conformal supergravity (CSG) \cite{bdw} is 
%rather  confusing (see, e.g.,  comment 3 on p. 293 in \cite{rvn}  )  and is clearly discussed in \cite{ber}.
In a ``unitary gauge''  formulation  
the scalar sector  is   described
 by a  complex  scalar $C$  parametrizing the coset $SU(1,1)/U(1)$. 
The fields   with non-zero chiral $U(1)$ weights  $c$ are then:\foot{
In this   paper we shall  use   Minkowski signature   notation  with  $*$ being 4d duality, 
$(T^{\m\n})^* = \ha \epsilon^{\m\n\k\l} T_{\k\l}$, \ \ 
$T^\pm_{\m\n} \equiv \ha ( T_{\m\n} \pm  i  T^*_{\m\n})$ and similar 
definitions for field strength $F^\pm _{\m\n}$   and   curvature tensor $(R^\pm)_{\m\n}^{\l\r}$.
Note also that the 
 CSG   action involves both $T^-$ and $T^+$, i.e. depends on real $T_{\m\n}$  with six
independent components. The position of $SU(4)$ indices on fermions is 
such that they all can be considered left-handed Weyl fermions.}

\hspace{-5truemm}
\begin{tabular}{lll}

1 complex scalar $C$ (-2)&  4 left spinors $\Lambda_i\ (- \trd)$&  10 complex scalars $E_{(ij)}$ (-1) \cr 

6  (anti)self-dual tensors $T^{-}{}^{ij}_{\m\n}$ (-1)& 
 20 spinors $\chi_k^{[ij]}\ (- \ha)$&  4  left-handed gravitini $\psi^i_\m\ (-\ha)$
 
\end{tabular}
%
% complex scalar $C$ (-2); 4 left spinors $\Lambda_i\ (- \trd)$; 10 complex scalar $E_{(ij)}$ (-1); 
%
% 6  (anti)self-dual tensors $T^{-}{}^{ij}_{\m\n}$ (-1);
% 20 spinors $\chi_k^{[ij]}\ (- \ha)$;
% 4  left gravitini $\psi^i_\m\ (-\ha)$.

\noindent
In the ``covariant''  formulation with manifest (linearly-acting)  $SU(1,1)$   symmetry 
  \cite{bdw}  the scalar   $C$ is replaced by a doublet  of complex scalars 
$\P_\a$ 
with 
\be \la{ppp}   \eta^{\a\b} \P^*_\a \P_\b\equiv \P_1^* \P_1-\P^*_2 \P_2    =\P^\a\P_\a=1, \ \ \ \ \ \ \ \ \a= 1,2\ , \ee
by adding an extra  $U(1)$  gauge symmetry. Then
   $\Phi_\a$ transforms under the global $SU(1,1)$ (with  matrix $ U_{\a}^\b$) 
as well as under the local $U(1)$ (with parameter $\g(x)$)
as follows: $\Phi'_\a = e^{-i \g (x) }   U_{\a}^\b\Phi_\b $. 
The field  $\Phi_\a$  is assigned  the  $U(1)$ chiral weight $-1$  while other fields have the same weights
as above. 
This  assignment is consistent with supersymmetry transformations at full non-linear level as given in
\cite{bdw} (with the parameter $\epsilon_i$  of Poincar\'e supersymmetry having weight 1/2). 
 Only the scalar doublet $\Phi_\a$ transforms under $SU(1,1)$, but all other fields   transform 
under the  local $U(1)$ (with the corresponding weights). 
That means that all the  fields   with derivative couplings and non-zero chiral weights
 couple to the   composite $U(1)$   gauge field   $a_\m$ 
  through covariant derivatives  (we ignore the fermionic term $ { i \ov 2} \bar \Lambda^i \gamma_\m \Lambda_i $ in $a_\mu$)
 \be\la{1} 
D_\m  = \del_\m     + i a_\m    \ , \ \ \ \ \ \ \ 
      a_\mu=  { i \ov 2} (\Phi^\a \del_\mu \Phi_\a - \Phi_\a \del_\mu \Phi^\a) = i \Phi^\a \del_\mu \Phi_\a \ . 
 \ee 
The composite real field   $a_\mu$  transforms under $U(1)$  (by a gradient)  and 
 is invariant  under $SU(1,1)$.\foot{Our definition of $a_\m$ differs by a factor of $i$ from the one in \cite{bdw,roo}.}

 For example, in the  particular $U(1)$    gauge
  \be \P_1 = \P_1^* \ , 
  \la{gre}
  \ee
  one may  parametrize  $\P_\a$   in terms of the above  complex scalar $C= \P_2/\P_1$, i.e.   
  \be \la{ga}
  \Phi_1 = (1- |C|^2)^{-1/2}\ , \ \ \ \ \ \ \ \ \ \  \ \Phi_2 = C(1- |C|^2)^{-1/2} \ . 
 \ee
 An $SU(1,1)$ transformation requires a 
 compensating local $U(1)$ transformation to preserve the gauge \rf{gre}. 
 Then  
 \be 
 a_\mu =  { i \ov 2} (1 -  |C|^2)^{-1}   (  C  \del_\mu C^* -C^* \del_\mu C)   \ ,   \la{1.2} \ee
 with  field   strength   $ da  \sim  dC \wedge dC^* + ..., \   $,  $ da \wedge da =0$. 
 This  $a_\m$ is no longer a  singlet of redefined $SU(1,1)$; 
 indeed, 
   under $SU(1,1)$ acting (non-linearly) on $C$  the  field  $a_\mu$ is shifted by the gradient of 
  field-dependent function multiplied by a rigid $SU(1,1)$ parameter, i.e. by an induced $U(1)$ 
  transformation with field-dependent parameter.
An anomaly in the latter $U(1)$ symmetry implies that the finite part of the effective action breaks the rigid $SU(1,1)$
symmetry.\foot{A similar anomaly of continuous $SU(1,1)$ appears  in 10d type IIB  supergravity   \cite{gr}.}  

%%%%%%%%%%%%%%%%%%%%%% 
 \def \rk {{\rm k}}
 %%%%%%%%%%%%%%%%%%%%%%
 
 The anomalies on CSG   were  discussed in \cite{rvn} using   this ``unitary  gauge''   formulation. 
 The   anomaly of the rigid $U(1)$  corresponding to  the above $U(1)$ weights    
  comes from the  (chiral) spinors and the self-dual tensors  only.
 They are coupled to $a_\mu$ and the gravitational  connection  but  since $da \wedge da$ here is zero, 
 only the gravitational anomaly   is present. 
 We can normalize it to the anomaly of a single  left-handed  fermion  having the  chiral weight   $c=+1$  
 \be \la{2}
 A_{1/2} = \del_\m j^\m = -{ k \ov 24 (4 \pi)^2}   R  R^* \ ,  \ \ \ \ \ \ \ \ \ \ \     k=c=1  \ . \ee
 where $(R^*)_{\mu\nu\rho\sigma} = \frac{1}{2}\epsilon_{\rho\sigma\k\l}R_{\mu\nu}{}^{\k\l}$ 
 is the dual curvature tensor. 
 %
 %$k=1$ here corresponds to   chiral weight -1  contribution \bldraft{*}.  
 For a   massless   spin (helicity)  $s$  field  with   {\it  standard}   kinetic term 
and chiral weight $c_s$,   the anomaly of the corresponding axial current 
 obeys \rf{2}  with    ($ \rk_s$ is related   \cite{duf} to  the  difference of the number of left- and right- handed 
 zero modes of  the corresponding Laplace operators, with ghost contributions properly accounted for) 
 \be 
 k = c_s \rk_s \ , \ \ \   \ \ \ \ \ \     \rk_s=   (-1)^{2s}  4 (  2 s^3-s )    \ , \la{an}
 \ee
 so that  the  spinor:vector:graviton anomalies are  related as 1:4:(-21). 
 In general, for a collection of fields with  chiral weights $c_s$  and  multiplicities $m_s$   we   get 
 \be  \la{an9}
 k= \sum_s  c_s m_s \rk_s    \ . \ee
 Note that 
since $ j^\m =   { \delta \Gamma \ov \delta a_\mu}$,   
 this anomaly   means that  the corresponding  term in the finite part of the 
 one-loop effective action  is\foot{The effective action $\G$  is given  by the logarithm of  the  
 determinant of a  chiral operator constructed in terms of $a_\m$ and gravitational Lorentz connection 
 $\omega_\m$. The  finite anomalous part of $\G$ depends on both the longitudinal part of $a_\m$  
 and the curvature of $\omega_\mu$. The  Minkowski one-loop  effective action  is 
  defined  as $ e^{i \G} =\langle e^{i S} \rangle$ and thus the anomalous term  in  is real.}
 \be 
 \la{3}
\Gamma_{\rm an}  =   \ka  \int  RR^*\nabla^{-2} \nabla_\mu a^\mu  \ , 
\ \ \ \ \ \ \ \ \ \ \ \ \ \ \   \ka=   {    k \ov 24 (4 \pi)^2}  \ .
\ee 
We shall discuss  in more detail below the structure  of this and related terms in the effective  action.

 The chiral $U(1)$ anomaly of the  {\it conformal} gravitino was found in \cite{fra}
 to be $-20A_{1/2}$  in terms of the  anomaly of a single chiral  fermion. This anomaly is related\foot{The
  two anomalies are simply related \cite{fra}
  by adding gauge-fixing terms, taking into account the relevant ghosts and their
 chiralities,  etc. More explicitly, 
 the axial anomaly depends only  of $\gamma^\m D_\m$   factor in $D^3$-type  conformal gravitino 
  operator:
 taking into account chiralities (or keeping track of $\gamma_5 a_\m$  coupling in covariant derivatives)  one has 
 $D^+  D^- D^+$  so that  only one spinor helicity contributes. 
 That  means gauge  
 (or $SO(4)$  Lorentz connection)  gravitational anomalies 
  of  the gauge-fixed   operators  of the  conformal  ($D^3$)  gravitino  and of the standard ($D$)
   gravitino  actually  coincide, and are the same as  the anomaly of 4   chiral spinors 
  (as  the  gravitino has an extra  4-vector index). 
  In the case of the gravitational anomaly there is additional ``non-minimal''
   curvature contribution to anomaly that happens to be -24  times the chiral spinor anomaly.
   Thus the total  gravitational anomaly from the gauge-fixed conformal or standard gravitino 
   is the same $4-24=-20$.
  The difference  between the two  gravitini 
  comes   from the ghost   sector.  In the conformal  gravitino case 
  the total   contribution of $\gamma^\m D_\m$'s from  all (FP and NK)   ghosts, with chiralities 
  properly taken into account, happens to vanish  \cite{fra}.
  Thus the  
  gauge-field anomaly of the conformal gravitino is the  same as of 4 chiral spinors \cite{rvn}, while 
  its gravitational
  axial anomaly  is $4-24=-20$ of the single spinor value \cite{fra}. 
  In the  standard   gravitino case  the  FP  
  ghosts  have the same  chirality as the gravitino but the opposite statistics, while 
  the  NK    ghosts  have the opposite chirality  (and their contribution is 1/2 of the FP ghost one), 
  so that the final count of the gauge anomaly is 
  $4- 2 +1= 4-1=3$ times $A_{1/2}$ . 
  This agrees  with the expectation that  the gauge  anomaly  should be proportional to the  helicity, 
  implying that there is a factor of 3 between the standard gravitino and the  single spinor gauge 
  anomaly  \cite{mar}. 
  In the gravitational anomaly case one is  still to add the non-minimal term
  contribution leading to $(4- 24)-1  = -20 -1 = - 21$  coefficient in \rf{an}. 
  It should be noted that  a count of anomalies is  of course 
  not correlated with a count of physical degrees of freedom (or overall 
  power of  box operator in the partition function which is 8 in the case of conformal
   gravitino and  2= 2(4-2-1) in the case of the standard gravitino)
   as there  one counts  all derivative operators   without taking into account 
   the corresponding  chiralities. 
  }
  to the anomaly of the standard (Poincar\'e)  gravitino \cite{grav,gra,aw}    which is  $-21 A_{1/2}$ 
  (in agreement with \rf{an}  with $s=3/2$).
  
   The  chiral anomaly of the  selfdual tensor $T^{-}_{\m\n}$   can be found,  e.g., by 
 replacing it with a complex  transverse  vector $\zeta= \xi + i \eta$ 
as  $T= d\xi + *d \eta $  (see sect.  3.2 in \cite{ftr})
 and using   the result \cite{aw}  for the anomaly of the chiral  vector in four dimensions.\foot{In the CSG action one
  has a $DT^{-} DT^{+}$ kinetic term but extra $D^2$ that 
 appear  in the kinetic terms  for chiral   vectors will  not influence the chiral anomaly.} 
 An alternative way \cite{rvn}  is to replace $T$   with a symmetric product of two 
  two-component left-handed spinors, finding that  
% and thus express it as $2 + 2$  spinor anomaly, i.e.
 the anomaly of a single $T^-$ field is   $4 A_{1/2}$. 
 
 Taking into account the chiral  weights 
 $c_s$ 
 %$w_i$  
 and the  number 
 %$N_i$ 
$m_s$ of components  of each field,  the total 
 $U(1)$ anomaly  count $\sum_s c_s m_s  \rk_s A_{1/2}  $  of $\NeqFour$ CSG is then  \cite{rvn}:\foot{We 
 use labels $(c)$ and $(s)$   to  distinguish the conformal and the standard gravitino anomalies.}
 \bea
 A_{\rm CSG}= A_{\L_i}  +  A_{T}   +  A_{\chi}  +  \AC_{\psi^i_\m} \qquad\qquad \qquad\qquad\qquad\qquad\qquad\qquad
\qquad\qquad\qquad  \no\\  
 = \Big[ (- \trd) (4)\times 1     + (-1) (6)\times  4   
 + (-\ha) (20)\times 1     
+ (-\ha) (4) \times  (-20) \Big] A_{1/2} =0 \ , 
 \la{4}
 \eea
 i.e. the pure \nf CSG  theory  has no ``external''  $U(1)$ anomaly.\foot{This $U(1)$ symmetry    has   no associated 
 dynamical gauge field  in \nf CSG \cite{bdw}.}
 This  would  formally   imply  that  the global $SU(1,1)$   symmetry is preserved  in the  corresponding quantum theory.
 However,  pure  \nf CSG has  conformal \cite{ftr} and $SU(4)$ \cite{rvn}   gauge  anomalies and thus is  inconsistent 
(e.g. non-unitarity) at the quantum level. 
 
 \
 
 Let  us consider  now the system of  \nf CSG  coupled  to  $n $    $\NeqFour$ vector multiplets (VM)
  with fields  $(A_\m,\psi^i, \phi^{ij})$.
  Following 
 the conventions of \cite{roo} we shall assume that 
 {\it left}  fermions $\psi^i$ of VM  have  $U(1)$ chiral weight +$\ha$, i.e.   
 opposite to the one of 
   $\chi ^{ij}_k$ or  the  conformal gravitino $\psi^i_\mu$.\foot{The chiral weight of a {\it right}-handed
    fermion   $\psi_i$  is the same as that of  the gravitino $\psi^i_\m$ but we are counting the anomaly relative to a 
    left-handed spinor.} This gives  the  additional  VM   contribution  to the anomaly as  
 \be 
 A_{\rm  VM}=  n (+\ha) (4)   A_{1/2} =  2 n A_{1/2} \ , \ \ \ \ \   \ \ \ \ \ \ \qquad     A_{\rm  CSG+  VM} =   A_{\rm  VM} \ .   \la{6}
 \ee 
The total  anomaly  is thus 
\be   A_{\rm  CSG+  VM} =   A_{\rm  VM}=  2 n A_{1/2} \ , \la{ta}  \ee
and  it  never vanishes for any non-zero number of VMs.  In particular, 
   the system of  $\NeqFour$  CSG   plus  $n=4$  VM   that has  no conformal \cite{ftr} and $SU(4)$ \cite{rvn} gauge 
 anomalies. Yet, it still has  a $U(1)$ gravitational anomaly, 
 implying \cite{dfg} 
the  breaking  of the  rigid $SU(1,1)$  symmetry  in the combined   system  at the quantum level.
Since $SU(1,1)$  here  acts on the vectors from VM as an on-shell   duality rotation, 
it  is present only on equations of motion (unless one gives up manifest Lorentz symmetry and 
uses doubled formulation)  and  in any case it does not survive    generalization to  non-abelian VM case. 

 %This may not be  too  bad 
 %as  $SU(1,1)$    here in any case   is manifest only on-shell (unless we  give up  
 %manifest Lorentz invariance)  and also given that it does not survive   generalization to SYM case. 

 %%%%%%%%%%%%%%%%%%%%%%%%%%%%%%%%%%%%%%
\subsection{$U(1)$ anomaly %${\cal N}=4$  vector multiplet coupled to conformal supergravity multiplet
in   ${\cal N}=4$  Poincar\'e supergravity  \label{sec32}}
%%%%%%%%%%%%%%%%%%%%%%%%%%%%%%%%%%%

 Let us now consider the $\NeqFour$ system of $n=6$  VM coupled to CSG   multiplet but without 
 adding the higher-derivative (Weyl tensor  squared +...) action of the CSG itself.
  As was argued in \cite{roo}, in the spontaneously-broken phase 
 in which (at least some of) the VM scalars have constant vacuum values,  
 this theory  is  classically equivalent (upon Weyl  and S-supersymmetry  gauge fixing  and solving for the auxiliary fields 
 $E,T,V$  and $\chi, D$  which are  not propagating here) to the standard $\NeqFour$  Poincar\'e supergravity (PSG). 

 Anomalies of $\NeqFour$ PSG   without any reference to this  conformal supergravity construction
 were discussed in \cite{mar}. Our aim  below  will  be  to explain how  they can be  understood 
  using   the above results \cite{rvn}   about  the  anomalies of the CSG + VM system.
  
  In general, 
one  may expect that there  should be anomaly matching between spontaneously broken and 
  unbroken phases. In the unbroken phase, anomalies of CSG and VM can be counted
 separately as all massless modes are obvious  -- only fermions from the VMs
 then contribute to the anomaly (gravitino, vectors, and $\L_i$  do not have free kinetic terms) because the 
 reality of the VM precludes assigning a nonzero $U(1)$ charge to the vectors \cite{roo}.
 In the broken phase we get kinetic terms for the  metric, vectors, gravitino and $\Lambda$ 
 from the  VM part, meaning they explicitly  contribute. In the broken phase there is  some  mode 
 rearrangement but at the end anomalies should match.

In \nf PSG  the contributions to the gravitational 
 $U(1)$ anomaly (the non-gravitational anomaly is  zero because $da \wedge da=0$) come from the 
 gravitini, spinors and vectors. 
  The  chiral weights  of  the gravitino, $T_{\m\n}$  and $\Lambda_i$    in the CSG setting were 
  $-\ha, -1,  - \trd$  ($\Lambda_i$  that had $D^3$ kinetic term in the CSG 
  becomes the physical  spinor of $\NeqFour$ PSG).
   These are   the same weights used in \cite{mar},  up to overall 
  rescaling by -2. Since the gravitino is now standard (not conformal)  its anomaly 
  is  $(-21) A_{1/2}$  (rather than $(-20) A_{1/2}$). Finally,  the contribution 
  of vectors (or self-dual tensor contribution in \cite{mar})  is the  same as 
   that of 
   %v2
   the  self-dual   tensor in CSG.\foot{There we had two transverse 
  vectors equivalent to $T$. 
  Explicitly (cf. eq.3.49 in  \cite{ftr}), 
   we have \cite{nuf,roo} a vector-tensor
   coupling of the type $ FT + TT$. If we add  the $\del T\del T$  terms in  CSG  action, we will get 
   $\zeta^* \del^2 ( \del^2 + m^2) \zeta$ type action for the complex transverse vector
   $\zeta= \xi + i\eta$. It is important  also to note that while in CSG  case the   $T$-tensor couples to
the scalar connection  $a_\m$ directly via covariant
derivatives  and thus    the corresponding determinant 
contains   scalar contributions, there is no such coupling in the PSG  context:
 here the 
 %v2
 scalar couplings   enter originally as prefactors in the kinetic terms of the vectors. 
 However, the   consistent phase space  formulation requires   specific  measure factors 
 that are  cancelled if we first redefine  the  vector fields   to absorb 
 these scalar  couplings (cf. two-dimensional sigma models). 
 Then we get  instead  derivative couplings  of (doubled) vectors to scalar   connection
 (related remark appeared in \cite{bhn}).}
   %The massless   part of the resulting determinant still contributes to
   %anomaly.}
  Thus the total $U(1)$ anomaly count is   
   (cf. \rf{4}) 
  \bea 
 A_{\rm PSG}&=& A_{\L}  +  A_{A}  +  \AS_{\psi_\m}  \no \\
 &=&  (- \trd) (4)  \times 1 A_{1/2}  + (-1) (6) \times 4  A_{1/2}  
 + (-\ha) (4)  \times (-21)  A_{1/2} =  12 A_{1/2}  \ .  \la{7}
 \eea
 This  is the same (up to  an  overall $-2$ factor due to different normalization of anomaly)
 as in %the second equation on p.386 of 
 \cite{mar}. 
% (there the order of terms and the order of factors is  opposite to ours). 
 This is also exactly the same as the   anomaly of  $n=6$    $\NeqFour$ VM coupled to CSG  in  \rf{6}
 which should be the anomaly in the unbroken phase.

We can understand  this    in detail    by restoring the CSG 
  multiplet contribution (which  is zero)  and then  tracking how the anomaly 
  rearrangement takes place in the  $\NeqFour$ PSG  derived following \cite{roo}, i.e.
   by assuming the broken phase
  and gauge fixing and solving for the non-dynamical fields. 
  In the broken phase we  have:\  (i) 
   $\L_i$ and $T^{ij}_{\m\n}$   contribute   as before;  (ii)
   gravitino contribution  is reduced  by -1  from -20 to -21  due to breaking of S-supersymmetry
  (and solving for
  %v2
   non-dynamical fields);\foot{In the spontaneously broken phase the 
  gravitino  should be absorbing  the massless  
  Goldstino of S-supersymmetry  to become the   standard gravitino.}
   (iii)   20 fermions  $\chi^{ij}_k$  of CSG and the rest of % (5 sets) of 
  $\psi^i_{I}$ fermions 
   of 6 VMs  should  no longer contribute  (they are integrated  out or 
    set to zero in the spontaneously broken phase).\foot{Note that 
  since  $ \psi^i$   has opposite chirality to the gravitino one, 
 adding its contribution is equivalent to subtracting the contribution 
 of the same-chirality fermion,  as was the case in the   gauge anomaly count for the standard
 gravitino (4-2+1=3  due to FP and NK ghosts).}

 The count of anomaly in the broken phase then proceeds as follows:
 \bea 
&&(A_{\rm CSG} + A_{\rm VM})_{\rm br.\, ph.} =
A_{\L}  +  A_{T}   + 0 \times   A_{\chi}  +  \big[ A_{\psi_\m}  +    A_{\psi^i}\big]
+   0 \times (6-1)  A_{\psi^i} 
 \no \\
&& =  (- \trd) (4)  A_{1/2}  + (-1) (6) 4  A_{1/2} 
 +  \big[ (-\ha) (4)  (-20)    + (+\ha) (4)    \big] A_{1/2}
 =  12 A_{1/2}   \la{77}
 \eea
 where the terms in the  square brackets  represent  the  standard  gravitino   contribution, 
 in perfect agreement with direct   $\NeqFour$ PSG  anomaly  count \rf{7}.\foot{Let us add a few remarks.
One could try to argue that  adding   the CSG action   to the VM action should 
 not change anomalies as this is like adding a  higher-derivative regulator  
 %v2
 (e.g.,  for the fermions   $ D \to   D (1  + M^{-2} D^2 ) $). 
 Indeed, one may think that as anomalies  may appear only at  one-loop  they  thus 
  cannot depend   on the  coupling  constant 
 $a$ in  the   combined  Lagrangian   $ L=  a  ( C^2_{\m\n\kappa\lambda} +  ...)  +  g^{-2} (F^2_{\m\n} + ...)$. 
However, the    anomalies depend on which operator is used in regularization and this has 
   to be the  kinetic operator in the action: the anomalies 
    are related to  finite anomalous parts of the one-loop determinants.
%\foot{Fijikawa-type argument should always be  viewed from a physical 
% point of view  -- one cannot generate anomaly from non-derivative couplings}
%
  While for $a=0$  the   antisymmetric tensor $T$   enters only algebraically, 
  it mixes with the vectors  of the VMs  and that suggests its  definition 
 in terms of  derivatives of a complex vector from the   start, getting a  non-trivial kinetic operator.
 For $\L_i$ and $\psi^i_\m$ there is also no problem as  one gets  their kinetic operators 
 (in the broken phase) as 
 $a D^3 + g^{-2}D$. Anomalies are the  same  for any value of the coefficient $a$
 (the $D^2 + M^2$  factor    
 does not contribute  to the anomaly). 
 However,   in the case  of the   fermions $\chi$  we have  a discontinuity:  
 they  have no kinetic term if $a=0$  and thus  contribute for $a\not=0$ but do  not contribute 
 for $a=0$. 
 The relation between anomalies of the CSG +VM and PSG systems 
 is thus not  completely 
 straightforward as  we need to  discard some contributions 
 (of $\chi$ and  also of extra fermions of 6  vector  multiplets)  in the broken phase. 
 }

%%%%%%%%%%%%%%%%%%%%%%%%%
 %This is at least a heuristic explanation, one may wonder if the argument can be made in a more
 %detailed form taking into account the ghost  contributions more explicitly. 
 % That means instead of $(-\ha) (20) + 6 (-\ha) (4) $ %deficit
% we should just have $  (-\ha) (4) $
% which is  the difference between the contributions of 4 conformal  $(-\ha) (4)  (-20)$
 %and 4 standard  $(-\ha) (4)  (-21)$   gravitini.
% ????????????
 %%%%%%%%%%%%%%%%%%%%%%%%
 
 The conclusion is therefore that the  $U(1)$ anomaly of $\NeqFour$ PSG  can be understood 
 as the  anomaly in the superconformal phase and that  gives an alternative  
 justification of  the~claim~of~\cite{mar}.
 
 \ 
 
 Let  us conclude this   section with  some comments on the $SU(4)$ anomalies. 
 While  the local $SU(4)$ symmetry is   gauge-fixed in the unitary gauge  
by the condition  completely  fixing \cite{roo}   the  VM scalars  to be  constant
 or zero in the broken phase,  
 one may (as in \cite{mar}) still   formally  consider the ``external'' 
 axial anomaly of the $SU(4)$ current.\foot{To preserve the gauge-equivalence, 
   one would need to add a local counterterm 
  depending on compensating scalars  if one does not fix the ``unitary'' gauge \cite{gdw}.}
 The conclusion of \cite{mar}   is that this anomaly  is non-vanishing (but has 
  no real consequences).\foot{If we  solve for $V^{i}_{j\m}$  at the classical level, as it enters 
  the action only algebraically  (see  \cite{roo}), we find:
$V^i_{\mu j} \sim \phi^{Iij} \del_\mu \phi_{Iij} + \bar \psi ^{Ii} \g_\m \psi _{Ii} + \bar 
\Lambda^i \g_\m \Lambda_i$. After  gauge-fixing $SU(4)$ by imposing a condition on scalars,
we are left only with global $SU(4)$   and  as long as there are no scalars left transforming 
non-trivially under this $SU(4)$  there are  no immediate consequences of the $SU(4)$ anomaly. 
%%%%%%%%%
%In \cite{mar}    above eq 12 there is a   comment on $SU(4)$ anomaly  and his claim is that 
%``if there are no scalars, as in the $SU(4)$ of $\NeqFour$ supergravity,  the Maurer-Cartan   
%equation is simply $F=0$  so presumably  the  anomalies   should be irrelevant.''
%%%%%%%%%
One reason for  looking at the $SU(4)$ anomaly is that it  should be in 
%v2
the same multiplet with the Weyl 
anomaly  and thus,  understanding why it is not relevant or how it is cancelled by 
a local counterterm  depending on compensator multiplet fields in the case of $n=6$ VM system 
may be   relevant   in a  more general  context.} 
 Let us see  how to  relate the $SU(4)$ anomaly count in $\NeqFour$ Poincar\'e supergravity \cite{mar}    
 to the discussion of  the $SU(4)$ anomaly in the $\NeqFour$ conformal supergravity \cite{rvn}, viewing  
 \cite{roo} the $\NeqFour $   PSG  from the superconformal point of view.

In the pure $\NeqFour$ CSG   the count of the non-abelian  $SU(4)$ anomaly goes  as follows \cite{rvn}: normalizing the
anomaly to the $d_{abc}$ symbols in the fundamental representation, $d^\rep{4}_{abc}$, 
a left-handed spinor $\psi^i$ (or Majorana spinor whose left-handed part transforms in $\rep{4}$ of $SU(4)$)
 contributes $+1$; then, the CSG spinor $\Lambda_i$ contributes $-1$;  the left-handed gravitino  $\psi^i_\m$ contributes  
 $+4$; $T^{ij}_{\m\n}$ does not contribute as the two-index antisymmetric representation of $SU(4)$  is  real; 
$\chi^{[ij]}_k$ 
gives\foot{One may use the relation between the $d_{abc}$ symbols in a mixed-symmetry representation and in the 
fundamental representation, $d^{\chi^{[ij]}_k}_{abc}=\frac{1}{2}(N^2-7N-2)d_{abc}^\rep{N}\big|_{N=4} = -7 d_{abc}^\rep{4}$.}
 $-7$, and thus the  total  axial gauge anomaly is 
\bea
&&A^{\rm (gauge)}_{\rm CSG}= A_{\L}   +  A_{\chi}  +  \AC_{\psi_\m}  =
  (-1)  A_{1/2}  + (-7) A_{1/2}  + 4  A_{1/2} =  -4 A_{1/2} \la{9}\ , \\
  &&\ \ \  A_{1/2}= { 1 \ov 24 (4\pi)^2}  \Tr ( FF^*) \ . \la{99}
 \eea
If we add coupling to $n$  of  $\NeqFour$ VMs   with the left-handed spinor $\psi^i$   being   in $\rep{4}$ representation 
of $SU(4)$ (like the gravitini),  then 
\be 
A^{\rm (gauge)}_{\rm VM}= n  A_{1/2}  \ , \ \ \ \ \ \ \  \ \ \ \ \ \ \ \ 
A^{\rm (gauge)}_{\rm CSG} + A^{\rm (gauge)}_{\rm VM}= (n-4)  A_{1/2}\ . \la{10} 
\ee
Thus  for $n=4$ we have the cancellation  of the $SU(4)$ 
anomaly  while   for $n=6$   we have the anomaly equal to $2  A_{1/2}$.

Let us now compare this with  the    count    of  the $SU(4)$ anomaly 
 in $\NeqFour$ PSG  \cite{mar} interpreted as a superconformal system in the  broken phase \cite{roo}  with all extra 
gauge symmetries fixed.
Here  we have the  same four spinors $\Lambda_i$  in the fundamental representation of $SU(4)$ and the same 
gravitino  $\psi^i_\m$,  but no spinors $\chi$. 
The standard gravitino  contribution to  the gauge anomaly is 
proportional  to the helicity     so it 
should have a relative factor of 3  compared to the chiral spin 1/2 fermion. This  implies
 \be 
A^{\rm (gauge)}_{\rm PSG}=   A_{\L}    +  \AS_{\psi_\m}  = (-1)  A_{1/2}   + 3  A_{1/2} =  2 A_{1/2} \la{11}\ . 
 \ee
This reproduces the count of the $SU(4)$ anomaly in conformal supergravity in \rf{10}  for $n=6$.

  %%%%%%%%%%%%%%%%%%%%%
\subsection{$U(1)$ anomaly in Poincar\'e supergravity coupled to $\nv$ vector multiplets \label{U1nv}}
%%%%%%%%%%%%%%%%%%%%%%

In the case when  CSG is coupled to  $n= 6 + \nv$   rather than   just  six vector multiplets,   i.e.   to  extra  $\nv$ 
``matter'' multiplets,   the anomaly  relation \rf{ta}  does not directly apply.  Indeed,  the   matter  VMs  surviving 
 as  dynamical fields in the broken  phase may  acquire different $U(1)$ charges than
the ones  assumed  in the unbroken phase.\foot{
%v2
 In the presence of matter multiplets, the $U(1)\subset SU(1,1)$
symmetry is a combination of the $U(1)$ symmetry in the matter-free theory and $U(1)$ duality symmetries of the 
matter VMs. It is also possible that, in the process of fixing the conformal (super)symmetry, the matter vector fields 
absorb some power of a CSG field that was  charged under the $U(1)$ symmetry  and thus acquire different 
charges than in eq.~\rf{6}.}
As the chiral  weights   should be consistent also with the supersymmetry
of the PSG   theory, a way to fix them  is to  use  
  that  for  $\nv=6$   
   the resulting  \nf PSG + matter 
 theory  can be interpreted as a  truncation   of $\NeqEight$ supergravity.

 Namely, let us   decompose 
the  $\NeqEight$ graviton multiplet 
 into  the $\NeqFour$ components following the embedding
\be
SU(8)\supset SU(4)\times SU(4)\times U(1) \ .
\label{decomposition}
\ee
This decomposition encodes the double-copy structure of supergravity theories, first realized in the 
string-theory KLT relations~\cite{KLT}, which
imply that the spectrum of the theory and its on-shell interactions may be represented in 
terms of two copies of the $\NeqFour$ sYM theory, each of which has an $SU(4)$ symmetry.
%
%The reason for this identification of $U(1)$ charges is that the way the $\nv$ vector multiplets are added in the 
%double-copy construction of scattering amplitudes matches, for $\nv=6$, the way the 
%six vector multiplets enter the  decomposition of the $\NeqEight$ supergravity. 
%
The lone $U(1)$ in (\ref{decomposition}) will be identified with the $U(1)$ symmetry of $\NeqFour$ supergravity (at least up 
to conjugation by $SU(1,1)$ elements)
and the first $SU(4)$ will be identified with the R-symmetry of $\NeqFour$ supergravity. 
This is possible because the charges of the PSG and matter multiplet asymptotic states are the same under the two 
symmetries.

The decomposition of the   $SU(8)$  representations  appearing in   $\NeqEight$ theory 
in representations of $SU(4)\times SU(4)\times U(1)$, denoted by $(SU(4), SU(4)')^{U(1)}$,  is:
\bea
\rep{1}&=&(\rep{1},{\rep1})^0
\no\\
\rep{8}&=&(\rep{4},{\rep1})^q\oplus (\rep{1},\rep{4})^{-q}
\no\\
\rep{28}&=&(\rep{6},{\rep1})^{2q}\oplus (\rep{1},\rep{6})^{-2q}\oplus (\rep{4},\rep{4})^{0}
\la{26} \\
\rep{56}&=&({\bar{\rep{4}}},{\rep1})^{3q}\oplus (\rep{1},{\bar{\rep{4}}})^{-3q}\oplus (\rep{6},\rep{4})^{q}
\oplus (\rep{4},\rep{6})^{-q}
\no\\
\rep{70}&=&(\rep{1},{\rep1})^{4q}\oplus (\rep{1},\rep{1})^{-4q}
\oplus ({\bar{\rep{4}}},\rep{4})^{2q}\oplus (\rep{4},{\bar{\rep{4}}})^{-2q}
\oplus(\rep{6},\rep{6})^{0} \ .
\no \eea
Here $q$ is the normalization of the $U(1)$ charge  which will be fixed below.  
%All components that are invariant under the second $SU(4)$ group form the $\NeqFour$ supergravity multiplet:
All the components invariant under the second $SU(4)$ group form the $\NeqFour$ supergravity multiplet: 
\be
(\rep{1},{\rep1})^0,\,(\rep{4},{\rep1})^q,\,(\rep{6},{\rep1})^{2q},\,({\bar{\rep{4}}},{\rep1})^{3q},\,(\rep{1},{\rep1})^{4q}, (\rep{1},\rep{1})^{-4q}
\ .
\label{PSGmultiplet}
\ee
We can also identify four $\NeqFour$ gravitino multiplets, transforming in the $\rep{4}$ of the second $SU(4)$ group,
\be
(\rep{1},\rep{4})^{-q},\, (\rep{4},\rep{4})^{0},\,(\rep{6},\rep{4})^{q},\, ({\bar{\rep{4}}},\rep{4})^{2q},\,(\rep{4},{\bar{\rep{4}}})^{-2q} \ ;
\ee
as well as six $\NeqFour$ vector multiplets transforming in the $\rep{6}$ of the second $SU(4)$ group,
\be
(\rep{1},\rep{6})^{-2q},\,(\rep{4},\rep{6})^{-q},\,(\rep{6},\rep{6})^{0} \ .
\label{Vmultiplet}
\ee
The conjugate representations (i.e. asymptotic states with opposite helicity) have opposite $U(1)$ charges.
We note that, as expected from the discussion in the beginning of this section, the charges of the fields of the matter vector 
multiplets under the $U(1)$ 
are different from the $U(1)$ charges in the unbroken phase of CSG coupled to $n=6+\nv$ 
vector multiplets.

A further argument for the identification of the supergravity $U(1)$ symmetry with the  $U(1)$ symmetry appearing in  
\rf{decomposition} is that they have similar consequences on scattering amplitudes. Indeed, the supergravity $U(1)$ 
symmetry acts on vector fields as electric/magnetic duality rotation; as such it implies (see e.g. \cite{ros}) that scattering 
amplitudes of vector fields of the same flavor vanish identically unless they have an equal number of positive and negative 
helicity fields.
If scattering amplitudes preserve the $U(1)$ symmetry of eq.~\rf{decomposition}, then they must carry vanishing charge.
Restricting, as above, to the scattering of a single type of vector field it immediately follows that the amplitude vanishes 
unless one scatters an equal number of positive helicity and negative helicity fields. 
Thus, the two symmetries have the same consequences on scattering amplitudes.

With the $U(1)$ charges in eq.~\rf{PSGmultiplet} we can compute the anomaly contribution of the $\NeqFour$ graviton 
multiplet following \rf{an}, \rf{an9} as
\be\la{ghm}
k_\text{PSG} = 4\rk_{3/2} (q)+6\rk_1 (2q)+4\rk_{1/2}(3q) = -24 q =12 \ ,  
\ee
where we   have   chosen the chiral weight of the gravitino to be $q=-1/2$, i.e.  the 
same as in \rf{4} and \rf{7},  to  reproduce the  anomaly  coefficient  12 in eqs.~\rf{7} and \rf{77}. 

With   this  choice the anomaly contribution of one $\NeqFour$ vector multiplet is:
\be
k_{\rm v} = \rk_1(-2q)+4\rk_{1/2}(-q)=-12q = 6  
 \ , \la{vv} 
\ee
which is different from the anomaly \rf{6} of a VM in the conformal phase.\footnote{One may understand this and the fact 
that the vector fields in the supergravity and matter multiplets have the same $U(1)$ charges as a consequence 
of the $O(6,n_{\rm v})$ symmetry of matter-coupled $\NeqFour$ supergravity.}
The total anomaly coefficient for the $\NeqFour$ supergravity coupled with $\nv$ vector multiplets  is  then 
\be\la{vvv}
k_{\text{PSG+VM}}=k_\text{PSG} +\nv k_{\rm v}  = - 12 q (2+\nv) =   6(2+\nv) \ .   
\ee
This, along with eq.~\rf{3},  implies  that the corresponding anomalous part of the effective action is
\be
\Gamma^{\NeqFour, \nv}_\text{an}=\frac{1}{2}(2+\nv)\Gamma^{\NeqFour, \nv=0}_\text{an}
= \frac{2 +\nv}{4(4\pi)^2}\int RR^*\nabla^{-2} \nabla_\mu a^\mu\ .  %*\frac{1}{\Box} \partial^\mu a_\mu \ .
\label{nv_anomaly}
\ee 
In secs.~3 and 4 we will reproduce this dependence on the number of vector multiplets from scattering 
amplitude calculations.

\def \an {{\rm an}}
\def \inv {{\rm inv}}
 \def \aa{{\rm a}}
 
 %%%%%%%%%%%%%%%%%%%%%%%%%%%%%%%%%%%%%%%%%%%%%%%%%%%%%%%%%%%
 \subsection{Structure of  the anomalous part of effective action}
 \label{sec33}
 %%%%%%%%%%%%%%%%%%%%%%%%%%%%%%%%%%%%%%%%%%%%%%%
 
 Let us now   comment further  on 
 the meaning of  the above  $U(1)$ anomaly and the the related  breaking of 
 the global $SU(1,1)$  symmetry of \nf PSG theory in the context of 
 the  one-loop  supergravity effective action  in  an  external 
 scalar   and gravitational background.

 \subsubsection{General comments}

 Let us  first not fix a  $U(1)$ gauge, so that  the  composite $U(1)$   gauge   field 
 $a_\mu$  in \rf{1}  transforms by a gradient 
 under a  chiral rotation  of $\P_\a $.
 Suppose  we consider the one-loop effective action for a  Majorana fermion coupled  
 to  gravity (though the Lorentz connection) and chirally (i.e. with $\g_5$) to $a_\m$. 
  If     we  split  $a_\m$ into the longitudinal and transverse parts, 
 $a_\m = a^{||}_\m + a^\perp_\m$,\  $\nabla^\m  a^\perp_\m=0$ then, 
integrating the $U(1)$ anomaly to obtain the corresponding effective action \rf{3}, we find that the latter
may be written as\foot{For a real fermion 
 there is no gravitational (i.e. local  Lorentz)  anomaly, so
 the local  Lorentz symmetry is unbroken  and the effective action depends  on the metric $g$
 rather than on the vierbein.}
 \be 
 \G[a;g] = \G_{\an}[a^{||}; g] + \G_{\inv}[a^{\perp}; g] \ , \ \ \ \ \ \ 
 \G_{\an}[a^{||}, g]
 = \ka \int RR^*\nabla^{-2} \nabla^\mu a_\mu  \ , \la{1a} \ee
 where $\ka$ stands for the overall coefficient in \rf{3}. 
   As  both $a^{||}_\m$ and $ a^\perp_\m$ are  separately  $SU(1,1)$  invariant, 
   the same applies to $\G_{\an}$  and  $\G_{\inv}$.

 Let us parametrize the scalar doublet $\P_\a$    as   
 \be 
 \P_1 = \sqrt{ 1+r^2} \  e^{i( \aa - b) }\ ,\ \ \ \ \ \ \ \ \ \   \P_2 = r \  e^{ i(\aa + b) }
 \ , \la{1b}
 \ee
 where $r,\aa,b$ are three real  fields.  Then only $\aa$ is transforming
 under  the local $U(1)$ symmetry 
  (by a shift), while all the three fields transform under the $SU(1,1)$.
   The connection  \rf{1} and the PSG scalar Lagrangian take  the form:\foot{An alternative
   parametrization is ($r= \sinh \r$):\ \ 
   $ \P_1 = \cosh \r \  e^{i( \aa - b) }\ ,\ \    \P_2 =\sinh \r  \  e^{ i(\aa + b) }$, 
    with $a_\m = - \del_\m \aa  + \cosh 2 \r \  \del_\m b$ and 
    ${\cal L}= -   ( \del_\m \r)^2  - \sinh^2 2 \r\ ( \del_\m b)^2$.}
 \bea 
&& a_\m = - \del_\m \aa  + (1 +2 r^2)  \del_\m b  
 \ , \la{222}\\
 && {\cal L} = D_\m \P_\a  D^\m \P^\a = |D_\m \P_1|^2 - |D_\m \P_2|^2
 = - {( \del_\m r)^2 \ov 1 + r^2 }   -  4(1 + r^2) r^2 ( \del_\m b)^2  \ . 
        \la{1d}
 \eea
 Both $a_\m$ and ${\cal L}$ are  $SU(1,1)$ invariant, while ${\cal L}$ is also invariant under  
 local $U(1)$ transformations.  Then \rf{1a} implies 
 \bea 
&& \G= \G_{\an}[\aa,b,r; g] + \G_{\inv}[b,r; g] \ , \ \ \ \ \no \\
&& \G_{\an}
 = - \ka \int RR^* \, \aa    
 +    \ka \int RR^*\nabla^{-2}  \nabla^\mu \big[( 1+ 2 r^2) \ \del_\m b \ \big]   \ ,  \la{1c} \eea
 so that while $\G$ is $SU(1,1)$ invariant, it is not invariant under the local $U(1)$ transformations
 because it depends on the scalar field $\aa$. This anomalous term, however, is {\it local}, and thus can be
 cancelled by a local counterterm. The important difference   compared to a standard gauge theory 
 (where the  gauge field  is  a fundamental field   and the anomalous term 
 is  nonlocal)  is that here the basic variables in the path integral 
 are the  scalar  fields $\aa,b,r$ rather than $a_\m$.
 Thus, defining the effective action as 
 \bea 
 &&\G'= \G'_{\an}[b,r; g] + \G_{\inv}[b,r; g] \ , \ \ \ \ \
 \  \   \\ && \G'_{\an}= \G_{\an} + S_{\rm c.t.} \ , \ \ \ \ \ \ \ \ \ 
 S_{\rm c.t.}=\ka \int RR^*  \big[ \aa  +  f(b,r) \big]   \ , 
 \la{1f}  \eea
 we   may  restore the local $U(1)$   invariance, i.e. the  absence of dependence on 
 the gauge degree of freedom  $\aa$.\foot{The ``minimal''
 counterterm, $  \int \, RR^*\, \aa$,
  may be written in terms of the original $\P_\a$ as 
  $ { 1 \ov 4}  \int RR^* \ln { \P_1 \P_2 \ov  \P^*_1 \P^*_2 }$, 
  illustrating its non-invariance under $SU(1,1)$.}
 Here    we  introduced $f(b,r)$ to parametrize possible ambiguity in choice 
 of a local counterterm.\foot{In general, a  local counterterm  may also contain many other terms, 
 including derivative-dependent and curvature-independent ones that should be constrained by the
 requirement of  preservation of  desired symmetries of the theory (i.e.  satisfaction of  the respective Ward
 identities). Note, however, that \ (i) we cannot introduce local  derivative terms without extra powers of
 a dimensionful parameter,  and \ (ii)   we cannot   construct an $SU(1,1)$ invariant using just algebraic 
 functions of $\aa,b,r$, i.e.  adding $f(b,r)$ cannot restore the $SU(1,1)$  symmetry.}
 For example, we may choose to also cancel the local $b$-dependent term in $\G_{\an}$
 in \rf{1c}  ending up with 
 $\G'_{an} = \ka \int RR^*\nabla^{-2}  \nabla^\mu \big( 2 r^2  \ \del_\m b \ \big)$.
 However, being nonlocal, $\G'_{\an}$   cannot
 be completely   eliminated by a local counterterm. 
 Furthermore, since $S_{\rm c.t.}$   is not $SU(1,1)$ invariant, 
 the resulting effective action $\G'$ is not invariant under $SU(1,1)$ transformations. 
 This is thus  an illustration of a general 
 ``compensator'' mechanism discussed, e.g.,
 in \cite{gdw}.

 Equivalently, the $U(1)$  anomaly implies that starting  with a classical  
 theory  in two different physical 
 $U(1)$ gauges one finds  two different  quantum  effective actions, but 
  the possibility to cancel the anomaly by a
 local counterterm means that these  two effective actions
 differ only by  local terms. 
  We may then choose  a particular $U(1)$ gauge from the start and interpret 
  $\G_{\an}$ as the part of the effective action that breaks $SU(1,1)$ invariance 
  (this is the framework used in \cite{mar}).
  
 For example,   the gauge where 
 $\aa=b$  corresponds to \rf{ga}, i.e. the gauge in which  $\P_1$  is real and 
 $C= {r \ov \sqrt{ 1 +r^2} } e^{2 i b}$. 
Another  gauge is, e.g., $\aa=0$  where $\P_1 \sim \P_2^*$, i.e. 
 $\P_1 = \sqrt{1 +r^2} e^{-i b} , \  \P_2 = r e^{i b} $.
 In these two cases  we find 
 \bea \la{1e}
&& (\G_{\an})_{\aa=b}=     
     \ka \int RR^*\nabla^{-2}  \nabla^\mu \big(2 r^2  \ \del_\m b \ \big) \ , \ \ \ \ \ \ \\ &&
   (\G_{\an})_{\aa=0}=  \ka \int RR^*\,  b    +
     \ka \int RR^*\nabla^{-2}  \nabla^\mu \big(2 r^2  \ \del_\m b \ \big) \ . \la{1ee}\eea   
The two effective actions  differ   by a local term  which is linear in the scalar 
field $b$  (implying, e.g.,  that in the first  gauge  the ``anomalous''  S-matrix describes 
the scattering of a smaller number of external states).
 One  may of course rule out this difference by requiring that we  add a 
 local counterterm as part of the definition of the theory.\foot{As already mentioned above, 
 extra counterterms may be required   to make   the anomaly  consistent   with supersymmetry
 (see  \cite{gdw, Bossard:2012xs}).}

 Such a finite local  counterterm also appears in the relation between  the $SO(4)$  \cite{cre}  
 and  $SU(4)$ \cite{su}  formulations of $\NeqFour$ supergravity: 
 the map between the two theories requires a local chiral redefinition of  spinors and a duality 
 rotation of vectors (resulting in a local contribution of the type $\int RR^*   f(b,r) $ to the anomaly)
 as well as  a local reparametrization of the scalar fields in the $\aa=0$ gauge. 

\subsubsection{The case of $SU(4)$ invariant   version  of 
 $\NeqFour$ Poincar\'e  supergravity \label{U1psg}}

Let us recall that  there are two ``unitary-gauge'' 
 formulations of \nf PSG: one with $SO(4)$ symmetry \cite{cre}  and 
one with $SU(4)$ symmetry \cite{su}. 
They are related  by a field redefinition of the scalar fields, a chiral rotation of  fermions and a duality 
rotation of vectors.
They may also be understood as  corresponding to two different  $U(1)$ gauges in the 
superconformal formulation \cite{roo}.  Because  of  the duality anomaly the corresponding 
quantum  effective action should be different, but  only by local  terms (explaining the puzzle
of  ``quantum-inequivalence''  found in  \cite{gri}). 

 Let us discuss in detail the $SU(4)$ invariant  version of $\NeqFour$   supergravity \cite{su}.
It corresponds to the following  $U(1)$  gauge \cite{rw,FKP}:
\be  
\Im \Phi_1 = \Im  \Phi_2 
   \ .   \la{ag}  
\ee 
Introducing the  two  independent scalar fields  as\foot{The field   redefinition 
relating $\tau$ to  the complex scalar $C$ of the $SO(4)$ invariant  formulation 
(corresponding to the gauge \rf{gre}) 
is $\tau =  i { 1 + C^* \ov 1-  C^*}$,  with 
${\del^\m \tau  \del_\m \bar \tau  \ov (\Im \tau )^2}=   4 {\del^\m \bC \del_\m \bC^* \ov (1 -\bC\bC^*)^2  }$.}
\be
   \tau \equiv   \cC+ie^{-\vp}
=  i { \Phi_1^* + \Phi_2^* \ov \Phi^*_1- \Phi_2^*} %=   i { 1 + C^* \ov 1-  C^*}
 \ ,\ \ \ \ \ \ \   
\la{22}\ee
we find that the composite connection and the scalar kinetic term are
\bea 
&&a_\m =-  { \del_\m (\tau +\bar \tau)  \ov  4\, \Im \tau } =   - {1 \ov 2} e^\vp \del_\m \cC    \ , \la{hho}\\
&&
- 4 \DD^\m \P^\a \DD_\m \P_\a % =  4 {\del_\m \bC \del_\m \bC^* \ov (1 -\bC\bC^*)^2  }
 =  {\del^\m \tau  \del_\m \bar \tau  \ov (\Im \tau )^2}
= (\del_\m \vp)^2 + e^{2 \vp} (\del_\m \cC)^2 \la{vb}
  \ , \eea
while the vector-scalar interaction terms are  ($F^\pm = {1\ov 2} (F \pm i F^*)$) 
\be 
L=  {1 \ov 4} i \tau\, F^+_{\m\n}F^+_{\m\n}      %\f{\phi^1-\phi^2}{\Phi}
-  {1 \ov 4} i  \bar \tau\, F^-_{\m\n}F^-_{\m\n}  %\f{\phi_1+\phi_2}{\Phi^\ast}
=-   {1 \ov 4}   e^{-\vp}  F_{\m\n}F_{\m\n}  -   {1 \ov 4}   B  F_{\m\n}F^*_{\m\n}  \ . \la{aca}
\ee
%%%%%%%%%%%
%Recall that  a pure gradient term in $a_m$  leads to a local term in the   effective  action, 
%and  formulations  of $\NeqFour$   supergravity   differing by a  field   redefinition 
%and also a duality transformation (as $SO(4)$ and $SU(4)$ formulations do \cite{su}) will 
%have effective action differing by  such local counter terms (in the present case 
%$\int RR^*  \ln { 1 + e^{-\vp}   - i \cC  \ov  1 + e^{-\vp}   +  i \cC  }  $). 
%%%%%%%%%%%
To recall,  we started  with a formulation with  manifest linearly realised  
$SU(1,1) $   and the  $U(1)$ gauge symmetry  (with $ a_\m$   being  $ SU(1,1)$
  invariant);   
  once   we fixed  a  $U(1)$  gauge,  $a_\m$   starts  transforming by  a gradient  under a 
  subgroup of $SU(1,1)$ which is broken in this gauge.
   Then,  at the quantum level,  we find the anomaly of that subgroup, whose precise embedding into  $SU(1,1)$ depends 
   on a the particular gauge choice.
In the gauge \rf{ag}  the $SU(1,1)$ symmetry  becomes $SL(2,R)$ acting on $\tau$  in the usual way, through 
M\"obius transformations, 
\be \tau'= { a \tau + b \ov c \tau + d}   \ , \ \ \ \ \   \ \ \ \      ad - bc=1 \ .
\la{sl} 
\ee
In $SL(2,R)$ language we have  3 subgroups of the duality group acting on the scalars (and vectors 
and fermions):
%AR
%\footnote{They correspond, respectively, to the transformations with ghost fields $e, h$ and $f$ discussed
%in \cite{Bossard:2012xs}. {\bf WHY we need to say that here ??}: 
%
\begin{enumerate}
\item  shift of  $\tau$    
 by a real constant (shift of $B$  only) 
 :  $\tau' =  \tau + b$ \ ($c=0,\,  a=d=1$)
\item  rescaling of $\tau$  (and  vectors, in the opposite way):  $\tau' = a^2 \tau$\  ($c=0,\, b=0,\, d= a^{-1}$)
\item   non-linear transformation: $\tau' = {\tau \ov 1 +   \eps  \tau}$ \ ($c=\ep,\, b=0,\, a=d=1$)  
or, in an  infinitesimal form,  $ \delta \tau  =  - \epsilon \tau^2$, 
 with vectors transforming as  $\delta F = \epsilon  F^*$   and fermions  rotating chirally,   $\delta \psi \sim  \epsilon \gamma_5 \psi$,  see   \cite{su}. 
\end{enumerate}
The $U(1)$ connection \rf{hho}   
is  invariant under first two subgroups and transforms 
under the  third   one    by   a gradient, 
$\delta a_\m = - \ha   \eps    \del_\m (\tau - \bar \tau) $.
Since   $\tau - \bar \tau  = 2i e^{-\vp} $
 the anomalous term in the effective  action \rf{3}
thus transforms as 
$\delta \Gamma \sim  \epsilon      \int   RR^* \, e^{-\vp} 
$.

\

The above relations  imply   that   in this gauge   
we find the  following anomalous term \rf{3}  in the effective action  (cf.\rf{1c},\rf{1f})
\bea
&&  \Gamma_{\an} \  =   \ka     \int RR^* \nabla^{-2} \nabla^\m  ( e^\vp \del_\m \cC ) 
\qquad \qquad  \no    \la{anb} \\
&&  \qquad  =\ka 
    \int RR^* \cC     +   \ka  \int RR^*  \nabla^{-2}    \nabla^\m  \Big[ \big( \vp  +  {1 \ov 2} \vp^2 + ... \big)  \del_\m \cC \Big]
         \la{ann}\ , \\
         &&    \ka=- {1 \ov 2} \times 12 \times   { 1 \ov 24(4 \pi)^2}        =- {  1 \ov  4 ( 4 \pi)^2}  \la{jjj}\ . 
\eea
Here the extra factor $ - {1 \ov 2}  $ in $\ka$  comes  from  \rf{hho}   and  12 is   the  total  value of coefficient $k$  in \rf{3}
corresponding to PSG   in  \rf{7} or \rf{ghm}.

Recall that the total coefficient $12$  in \rf{7} includes the contribution $-24= (-1) 6 \times 4$ 
of six vectors  with chiral weight $-1$.  We  can then check the normalization of $\G$  in \rf{jjj}   by  relating the 
above    discussion to the known anomalous correlator \cite{dolg} (see also \cite{end,reu,erd})
  of the quantum   Maxwell   theory  in  a  
gravitational background   
\be 
\langle F^{\m\n}F^*_{\m\n}  \rangle = {1  \ov 3 (4 \pi)^2}  RR^*    \ , \la{ff}
\ee
which also   implies  the presence in the effective action of a vector contribution which is  anomalous 
under the duality rotation $\delta  F_{\m\n} = \epsilon F^*_{\m\n} $ (see  Appendix A).\foot{This conclusion may be 
reached by relating the discussion above to the quantum anomaly of the chirality 
current $K_\m = \epsilon^{\m\n\l\r}   A_\n \del_\l A_\r, \   \del^\m  K_\m =  {1 \ov 2}  F^{\m\n}F^*_{\m\n}$.} 
   Expanding the   effective action corresponding to the  classical action 
   \rf{aca} in powers of the scalar fields, we then  conclude   that   the term linear   in $B$   in the corresponding effective action 
   is  given  by\foot{We note in passing that the  result  for the vector  loop contribution  to the corresponding term 
   in the effective action found   by a diagrammatic method  in \cite{gri}  disagrees  with this  by an extra 3/2 factor
   (which should  be due to the use of a regularization which is not reparametrization-invariant).}  
   \be 
   \Gamma^{(\rm vec)}_{\an}  = - { 1 \ov 4}   \int %d^4 x \sqrt g \ 
      \langle (F^{\m\n}F^*_{\m\n}) (x)   \rangle   B(x)   + ...  
   =    - { 1 \ov 12  (4 \pi)^2 }   \int \,  %d^4 x  \ 
     RR^*\,   B(x)   + ...   \ .  \la{ab} 
   \ee 
   Recalling  again   that 
 according to \rf{7}   the anomalous contribution  of the full  $\NeqFour$   PSG  multiplet 
 should be   3  times a  single vector contribution (i.e. 4, if vector has  weight +1)
  we conclude that \rf{ab} is indeed in agreement with \rf{ann}, \rf{jjj}.

 % (vector:spinor:gravitino  contributions to gravitational chiral  anomaly are 
%1: 1/4: (-21)/4;  their numbers in $\NeqFour$ supergravity are 6:4:4    and their relative weights are 
%1: 3/2:1/2). 

While   the quantum vector field  contribution to $\int RR^* B$ term   in \rf{ann} 
follows   directly from \rf{ff}, one can  find  similar spinor contributions   from the form 
of the corresponding $\gamma_5 \del_\m B$ covariant   derivative couplings in \cite{su}
(producing  $\int RR^* B$ terms  in $\G$   via  fermion triangle loop diagram).
It is worth emphasizing that  here 
 one need  not go through  the duality group anomaly discussion  of \cite{mar}  -- one is 
  simply  computing certain  leading terms in the one-loop effective action.\foot{Note 
  also that while one   might   argue that  the local  $\int RR^* B $ term is ambiguous as we can eliminate it 
by adding a local counterterm,  doing this would  effectively drive us away   from the 
$SU(4)$    version of  supergravity  -- once we have fixed the  $U(1)$ gauge \rf{ag}, 
and  thus decided   about which subgroup of the  $SU(1,1)$ duality group is anomalous 
(with shifts of $B$ and rescalings being non-anomalous)   we should not add 
further local  parity-odd counterterms.}

\

With an expression for the anomalous part of the effective action in hand we may consider extracting 
the corresponding contribution
to various  one-loop scattering amplitudes by expanding in powers of the scalar fields,
\bea  &&   \G_{\an} =\G_1 + \G_2 + ... \ , \ \ \ \ \ \qquad 
   \Gamma_1  =       \ka  \int RR^*   B \ \la{nnb}  ,\\  \ \ \ \ \ \ \ 
 && \Gamma_2  =    \ka     \int RR^* \nabla^{-2}  ( \del^\m  \vp   \del_\m B   + \vp \nabla^2  B  ) 
      \ ,   \la{bnn}
\eea
and constructing tree-level amplitudes with the resulting effective vertices. 
These 1-PI   vertices give particular   contributions to the S-matrix  with at least 2 gravitons  on external lines. 
For example, we get $hhB$  {\it  3-point }  amplitudes and $hh \vp B$   {\it  4-point }  amplitudes.
 
 As follows from the $\NeqFour$   supergravity Lagrangian (here we set $K= \ha$ in \cite{su};\  $r=1, ..., 6$)
 \be \la{laga}
 {\cal L}= - R - {1 \ov 2 } ( \del^\m \vp \del_\m \vp  +    e^{2 \vp} \del^\m   B  \del_\m   B)  
   -   {1 \ov 4}   e^{-\vp}  F^r_{\m\n}F^r_{\m\n}  -   {1 \ov 4}   B  F^r_{\m\n}F^{r*}_{\m\n}  + ... \ , 
   \ee
we get the following  linearized equation of motion  for $B$: 
 $ \nabla^2  B =    {1 \ov 4}    F^r_{\m\n}F^{r*}_{\m\n} + ...$. 
Thus connecting  the one-loop  $ RR^*   B$  vertex   and the  tree-level $B  F^r_{\m\n}F^{r*}_{\m\n}$ 
vertex by the $B$-field   propagator we get also   $hhAA$  {\it  4-point}  amplitude 
represented by the $ \int RR^*   \nabla^{-2}  F^r_{\m\n}F^{r*}_{\m\n}$ term  in the generating functional.

It should be noted    that these are not the complete S-matrix elements  with a given  choice of external states:   there are 
obviously other  nonlocal terms  in the effective action not directly related to the  duality  anomaly term \rf{ann}   
discussed above. In particular, the 4-point matrix element $h^{++}h^{++}A^+A^+$, while receiving contributions 
from the $B$ intermediate state, can be shown to vanish identically when all the contributions are included, in agreement 
with the consequences of supersymmetry Ward identities.

Indeed,  in the above discussion we concentrated just on a particular set of terms  in the effective action 
which are directly related    to the anomaly.  The effective action  should   contain, of course,  many other local and nonlocal terms. 
Supersymmetry may require that some of them be natural partners, being parts  of the same superinvariants.
The  contributions to  scattering amplitudes  coming just from the anomalous   terms in \rf{3} 
may  therefore appear to break supersymmetry  (separating the field configurations mentioned  above 
in helicity components one may  notice that some of them are forbidden by the 
analysis of \cite{Grisaru:1977px}). 
Since  we do not expect  supersymmetry to be anomalous, the effective action should  thus contain also  other terms 
related by supersymmetry transformations to the  anomalous ones in \rf{nnb},\rf{bnn}.
For example, we  may expect   to find  also   the 
$\int R R \vp$   term \cite{gri,gil}   as a partner of $\int RR^* B$  
as well as an $\int R^{\m\n\l\r} F_{\m\n} F_{\l \r} $  term 
corresponding  to the  graviton-vector-vector    amplitudes.
We will 
  return to this discussion  in sections ~\ref{amplitudes} and \ref{eff_action}  below 
  where we will evaluate 
the  relevant scattering amplitudes
in a scheme that manifestly preserves supersymmetry at the linearized level.

%%%%%%%%%%%%%%%%%%%%%%%%%%%%%%%
\renewcommand{\theequation}{3.\arabic{equation}}
 \setcounter{equation}{0}
%%%%%%%%%%%%%%%%%%%%%%%%%
\section{$U(1)$ symmetry violating scattering amplitudes \label{amplitudes}}
%%%%%%%%%%%%%%%%%%%%%%%%%

Let us now   address the question  of  how the $U(1)$  anomaly  of  the  $\NeqFour$     Poincar\'e  supergravity 
 is  reflected in the S-matrix of the theory.  
As we have seen above,  the anomaly implies  that the effective  action for  gravitons and scalars 
should contain anomalous   terms   which, in turn, should  correspond  to  particular   
$U(1)$ symmetry violating one-loop  scattering amplitudes. 

To construct the S-matrix we will not start directly from 
the PSG action but instead use the  (generalized) unitarity method \cite{UnitarityMethod,BCFUnitarity}, 
color/kinematics duality and the double-copy construction to express it 
in terms of the S-matrix of the $\NeqFour$ supersymmetric YM theory and pure YM theory coupled 
to scalar fields \cite{BCJ, BCJLoop}.

\subsection{Generalities on gauge theory and supergravity scattering amplitudes}
\label{general_amplitudes}

The ($D$-dimensional) generalized unitarity method together with the KLT \cite{KLT} relations, 
determining the tree-level amplitudes of a (super)gravity theory in terms of the scattering amplitudes 
of two (supersymmetric) gauge theories, provide a sure way for 
constructing one- and higher-loop amplitudes in $\NeqFour$ supergravity  (pure or coupled
to vector multiplets). In this approach the spectrum of $\NeqFour$ supergravity is realized as a tensor 
product of the fields of  $\NeqFour$   and $\N=0$   supersymmetric  YM  (sYM)  theories, i.e. 
as a product of   an $\NeqFour$ vector multiplet and a single vector field (we denote YM or sYM 
 gluons by $g$ 
and supergravity abelian vector fields by $A$):
\be
(g^+, \lambda^{+}_{ABC}, s_{AB}, \lambda^{-}_{A}, g^-)\otimes (g^+, g^-)=
(h^{++}, \psi^{+}_{ABC}, A^+_{AB}, \chi^{+}_{A}, \bartaua)
\oplus
( \taua, \chi^{-}_{ABC}, A^-_{AB}, \psi^{-}_{A}, h^{--})\ .
\label{spectrum1}
\ee
Here and below $(\taua,\bartaua)$ is the  complex  field that labels the external scalar states in 
the supergravity scattering amplitudes; in terms of the two vector fields it is
\be
\taua = g^+_{\NeqFour}\otimes g^-_{{\cal N}=0} 
\quad , \qquad
\bartaua = g^-_{\NeqFour}\otimes g^+_{{\cal N}=0} 
\ .
\label{taua}
\ee
Additional $\nv$  vector multiplets  may be described  as
\be
(g^+, \lambda^{+}_{ABC}, s_{AB}, \lambda^{-}_{A}, g^-)\otimes \phi_p =
(A_p^+, \lambda_{p,ABC}^{+}, s_{p,AB}, \lambda_{p,A}^{-}, A_p^-)
\ee
where $\phi_p$ with $p=1,\dots,\nv$ are real scalar fields.  

While the construction of the scattering amplitudes of the $\NeqFour$ sYM factor is  clear (and will be reviewed shortly), 
this is less clear  for  the bosonic factor  since one may consider several different self-couplings of the scalar 
fields. The correct choice follows from the observation that, on the one hand, half-maximal supersymmetry implies  
unique consistent coupling of vector multiplets to supergravity and,  on the other,  that  the 
models of  $\NeqFour$ supergravity coupled to $\nv=2,4,6$ vector multiplets can be realized as orbifolds of 
${\cal N}=8$ supergravity. This construction implies that the tree-level scattering amplitudes of the bosonic 
factor must be chosen to be the same as those of the $\NeqFour$ sYM theory,
% truncated to its bosonic sector, 
i.e. that the $\nv$ scalars should have  quartic self-couplings.
%v2
 \footnote{Moreover, the scalar contact term is 
also required by color/kinematics duality of the scalar-coupled YM theory.}

The scattering amplitudes of $\NeqFour$ sYM theory manifestly 
preserve linearized $\NeqFour$ supersymmetry. The on-shell 
fields of  this  theory can be combined into  a chiral superfield
\begin{equation}
\Phi(\eta)=
   g^-
+\eta^A\lambda^{-}_A
+\frac{1}{2!}\eta^{A}\eta^{B}\phi_{AB}
+\frac{1}{3!}\eta^{A}\eta^{B}\eta^{C}\epsilon_{ABCD}\lambda^{+,D}
+\frac{1}{4!}\eta^{A}\eta^{B}\eta^{C}\eta^{D} \epsilon_{ABCD}\, g^+\ , 
\label{eq:superfield}
\end{equation}
where $\lambda^+_{ABC}=\epsilon_{ABCD}\lambda^{+,D}$ and $\eta^A$ are four Grassmann variables. 
The scattering amplitudes of component fields are assembled into a superamplitude
\begin{equation}
\label{eq:superspace-tree}
{\cal A}_n = 
%
%g^{n-2}
% 
\frac {i\,\delta^{(4)} (\sum_{i=1}^n \lambda_i
  \tlambda_i) \delta^{(8)} (\sum_{i=1}^n \lambda_i \eta^A_i)}{\langle
  1 2\rangle \langle 2 3\rangle \cdots \langle n 1\rangle} \sum_{r=0}^{n-4}
\mathcal{P}_n^r\ ,
\end{equation} 
where $\lambda_i$ are the spinors corresponding to $i$-th  momentum,
$k_{i,\mu} (\sigma^\mu)_{\alpha{\dot\alpha}} = \lambda_{\alpha i}{\tilde\lambda}_{\dot\alpha i }$
and $\mathcal{P}_n^r$ are degree-$(4 r)$ polynomials in the Grassmann variables
$\eta^A_i$.  The invariance under the $R$-symmetry implies that
$\mathcal{P}_n^r$ are invariant under $SU(4)$ rotations of the
Grassmann variables $\eta_i^A$.  The lowest-order term in the $\eta$
expansion has Grassmann weight $8$, while the highest-order term has
Grassmann weight $4 n - 8$.  CPT conjugation exchanges weight $4
r+8$ with weight $4 n - 4 r - 8$.  The $r=0$ term in
eq.~\eqref{eq:superspace-tree} has $\mathcal{P}_n^0 = 1$ and contains
all the $n$-point maximally helicity-violating amplitudes.

Component amplitudes may be extracted by multiplying the superamplitude 
with the appropriate product of superfields and integrating over all Grassmann
parameters: 
\be 
A_n(k_1, h_1; \dotsc; k_n, h_n)=\int
\prod_{i=1}^nd^4\eta_i \;\prod_{i=1}^n \Phi_{h_i}(\eta_i)\;{\cal A}_n(k_1, \eta_1,
\dotsc, k_n, \eta_n)\,.  
\ee 
The superfields $\Phi_{h_i}(\eta_i)$ have
a single non-vanishing term corresponding to the field with helicity
$h_i$. For example, NMHV $n$-point amplitudes appear inside the superamplitude 
${\cal A}_n$ as
\begin{eqnarray}
  \label{eq:components}
  {\cal A}_n(k_1, \eta_1, \dotsc, k_n, \eta_n) = \cdots &&+ (\eta_1)^4 (\eta_2)^4
  (\eta_3)^4 A_n(-,-,-,+,+,\dotsc,+) 
  \\
  &&+ (\eta_1)^4 (\eta_2)^4 (\eta_4)^4 A_n(-,-,+,-,+,\dotsc,+) + \cdots,
\nonumber
\end{eqnarray} 
where $(\eta)^4$ stands for the $SU(4)$-invariant expression $\tfrac 1 {4!}
\epsilon_{ABCD} \eta^A \eta^B \eta^C \eta^D$.

The (super)amplitudes of ${\cal N}=4$ sYM theory can also be formulated in anti-chiral superspace; they are obtained 
from the chiral superspace expressions by conjugating all spinors (i.e. interchanging $\lambda$ and $\tlambda$ and 
their corresponding spinor products)  and Fourier-transforming all Grassmann variables $\eta_i^A$.
Denoting by ${\tilde \eta}$ its conjugate variable, 
in the corresponding superfield the negative helicity gluon wave function comes multiplied by $({\tilde\eta})^4$ and the 
positive helicity gluon wave function has no ${\tilde\eta}$ factors.

The organization of non-supersymmetric amplitudes is less compact at a generic loop order. Tree-level gluon 
amplitudes may however be obtained from tree-level amplitudes of $\NeqFour$ sYM  by requiring that only gluons appear 
on the external lines. Similarly, for a scalar-coupled YM theory with scalars having the same quartic self-interaction as
the $\NeqFour$ scalars, tree-level scattering amplitudes can be obtained by truncating,  e.g.,  eq.~\rf{eq:superspace-tree}
to gluon and scalar external states. 
Since at  tree level and for such external states 
%AAT
no fermions can appear on the internal lines, the resulting  bosonic amplitudes  formally obey the same 
supersymmetry Ward identities as the  corresponding  $\NeqFour$  sYM amplitudes  with only bosons on external lines.
In this  restricted  sense one may say   that   there is a linearized  (extended)  supersymmetry algebra 
(relating bosons  to bosons)  acting on the asymptotic space of states of the scalar-coupled YM theory.

The double-copy structure of the supergravity amplitudes 
implied at the tree-level by the KLT relations 
was clarified recently  
in~\cite{BCJ}, where it was realized that the gauge theory amplitudes can be arranged in a graph-organized 
representation so as to make manifest a duality between their color and kinematic factors. 
In the same organization,  the $L$-loop amplitude is
\begin{equation}
{\cal A}^{L-\text{loop}}_m\ =\ 
i^L \, g^{m-2 +2L } \,
\sum_{i\in \Gamma}{\int{\prod_{l = 1}^L \frac{d^D p_l}{(2 \pi)^D}
\frac{1}{S_i} \frac {n_i C_i}{\prod_{\alpha_i}{p^2_{\alpha_i}}}}} \  .
\label{LoopGauge} 
\end{equation}
Here the sum runs over the complete set $\Gamma$ of 
$m$-point $L$-loop graphs with only cubic (trivalent) vertices, including all permutations of external 
legs, the integration is over the $L$ independent loop momenta $p_l$  
and the denominator is given  by the product of  all propagators of the corresponding 
graph.
The coefficients $C_i$ are the color factors obtained by assigning to every three-vertex in a  
graph a factor of the structure constant ${\tilde f}^{abc} = i \sqrt{2} f^{abc}=\Tr([T^{a},T^{b}]T^{c})$
while respecting the cyclic ordering of edges at the vertex.
The hermitian generators $T^a$ of the gauge group are normalized so that 
$\Tr(T^a T^b) = \delta^{ab}$.
The coefficients $n_i$ are kinematic numerator factors depending on
momenta, polarization vectors and spinors. For supersymmetric
amplitudes in an on-shell superspace, they will also contain Grassmann 
parameters.
The symmetry factors $S_i$ of each graph remove any overcount
introduced by summing over all permutations of external legs (included by 
definition in the set $\Gamma$), as well as any internal automorphisms of 
the graph,  i.e. symmetries of the graph with fixed external legs.

The color/kinematics duality~\cite{BCJ} is manifest when the kinematic numerators of a graph 
representation of the amplitude satisfy antisymmetry and (generalized) Jacobi relations around 
each propagator -- in one-to-one correspondence with the color-factors.  
That is, schematically for cubic-graph representations, it requires that
\begin{equation}
C_i + C_j + C_k =0 \qquad  \Rightarrow \qquad  n_i + n_j + n_k =0 \, .
\label{BCJDuality}
\end{equation} 
Related supergravity  amplitudes  are then trivially given in the same graph organization but with the 
color factors replaced by another copy of (a potentially different) gauge theory kinematic factors (which 
are not required to satisfy the duality): 
\begin{equation}
 {\cal M}^{L-\text{loop}}_m = i^{L+1} \, \left(\frac{\kappa}{2}\right)^{m-2+2L} \,
\sum_{i\in\Gamma} {\int{ \prod_{l = 1}^L \frac{d^D p_l}{(2 \pi)^D}
 \frac{1}{S_i}
   \frac{n_i {\tilde n}_i}{\prod_{\alpha_i}{p^2_{\alpha_i}}}}} \,.
\hskip .7 cm 
\label{DoubleCopy}
\end{equation}
Here $\kappa$ is the gravitational coupling.  The Grassmann parameters that may appear in the two gauge theory 
factors are therefore inherited by the corresponding supergravity amplitudes.

The organization of $\NeqFour$ supergravity states in on shell multiplets follows directly from eqs.~\rf{spectrum1} 
and  \rf{eq:superfield}. The states in each parenthesis on the right-hand side can be organized in an on shell chiral 
$\NeqFour$ multiplet:
\bea
\Phi^+(\eta)&=&  \bartaua
+\eta^A \chi^{+}_A
+\frac{1}{2!}\eta^{A}\eta^{B}A^+_{AB}
+\frac{1}{3!}\eta^{A}\eta^{B}\eta^{C}\epsilon_{ABCD}\psi^{+,D}
+\frac{1}{4!}\eta^{A}\eta^{B}\eta^{C}\eta^{D} \epsilon_{ABCD}\, h^{++}\ ,~~~~ ~~~~
\label{phi+}
\\
\Phi^-(\eta)&=&   h^{--}
+\eta^A\psi^{-}_A
+\frac{1}{2!}\eta^{A}\eta^{B} A^-_{AB}
+\frac{1}{3!}\eta^{A}\eta^{B}\eta^{C}\epsilon_{ABCD}\chi^{-,D}
+\frac{1}{4!}\eta^{A}\eta^{B}\eta^{C}\eta^{D} \epsilon_{ABCD}\, \taua\ .
\label{phi-}
\eea
They are CPT conjugates of each other\foot{Since both multiplets are written in chiral superspace, a CPT transformation
also assumes a Fourier transform of the Grassmann variables $\eta$.}.

The double-copy construction of $\NeqFour$ supergravity amplitudes thus manifestly preserves the  linearized 
$\NeqFour$ supersymmetry and the resulting amplitudes have manifest $SU(4)$ R-symmetry.  
The relation to $\NeqEight$ supergravity discussed in sec.~\ref{U1nv}  implies the existence, at least at tree level, 
of an additional $U(1)$ symmetry -- the lone $U(1)$ in \rf{decomposition}. 
We can identify the corresponding charges from the double-copy perspective as a
combination of the little-group transformations on the two gauge theory factors:
\be
{\qq}_{{}_{U(1)} }(\Phi\otimes{\tilde\Phi}) = 2q\big(h({\tilde\Phi})-h(\Phi)\big) \ .
\label{2copy_charges}
\ee
Here $\Phi$ denotes a field in the $\NeqFour$ factor, ${\tilde\Phi}$ denotes a field in the non-supersymmetric factor,
$h(\Phi)$ is the helicity of  $\Phi$. \  $\Phi\otimes{\tilde\Phi}$ denotes the supergravity field with helicity 
$h({\tilde\Phi})+h(\Phi)$ and other quantum numbers given by the quantum numbers of  $\Phi$ and ${\tilde\Phi}$.

Half-maximal supersymmetry is substantially less restrictive than maximal supersymmetry. Since, in particular, it
acts only on one gauge theory factor, supersymmetry cannot relate amplitudes that differ in the field content or 
helicity assignment  of  the 
%bosonic gauge theory 
non-supersymmetric factor. For example, the four-graviton amplitudes 
${\cal M}(1^{++}2^{--}3^{++}4^{--})$ and ${\cal M}(1^{++}2^{--}3^{--}4^{++})$, while related by supersymmetry in 
${\cal N}=8$ supergravity, belong to independent superamplitudes
%are no longer related 
in a  half-maximal  ${\cal N}=4$ supergravity theory. 
Because of this, the standard organization of amplitudes in N$^{k}$MHV sectors, following the number of Grassmann variables is 
no  longer sufficiently descriptive. We will instead refer to amplitudes as N$^{k}$MHV$^{(p,q)}$ where $p$ and $q$ 
are the number of $\Phi^+$ and $\Phi^{-}$ external states, respectively,  and the total number of external legs is $(p+q)$.

\subsection{Special classes of anomalous amplitudes \label{special_classes}  }

The fact that one of the gauge theory factors is non-supersymmetric implies that certain amplitudes 
that vanish identically at tree level, are non-vanishing at one-loop level. Restricting ourselves to  external 
gluons only, these amplitudes are the 
so-called all-plus   and single-minus amplitudes: assuming all participating gluons are either incoming 
or outgoing, these amplitudes have all   and  all-but-one gluon of identical helicity, respectively. 
It is the supergravity amplitudes built out of such special non-supersymmetric amplitudes that we shall be focusing on.

As we shall see, through the double-copy construction, these gauge theory amplitudes lead to non-vanishing
supergravity amplitudes with nonzero $U(1)$ charge; they thus break this $U(1)$ symmetry, i.e. it is  
anomalous. We stress that this anomaly cannot be interpreted as a failure of the double-copy construction 
because the results are checked against a direct evaluation of $D$-dimensional unitarity cuts.
% Freedman's question

It is interesting to note that, even from a gauge theory perspective, non-vanishing  all-plus and single-minus 
amplitudes may be interpreted as a consequence of an anomaly of a symmetry of on-shell asymptotic states. 
Indeed, the tree-level scattering amplitudes of pure YM theory as well as of pure YM theory coupled to scalars with 
the same self-interactions as in $\NeqFour$ sYM theory obey the same supersymmetry Ward identities as the 
tree-level amplitudes of $\NeqFour$ sYM theory with only external bosons.  One of the consequences  of these Ward identities 
is the vanishing of the all-plus and single-minus amplitudes. 
%
%AAT
Absence of the fermionic superpartners  in the bosonic theory  
leads to  one-loop amplitudes that  no longer obey supersymmetry Ward identities -- e.g. non-vanishing 
all-plus and single-minus amplitudes --  and thus to a formal breaking of the above 
tree-level  supersymmetry  understood as   acting only on bosonic  asymptotic states. 
  In this sense, the  bosonic tree-level  symmetry of the  YM theory  ``inherited''  from  
  supersymmetry of the sYM theory is anomalous. 
  
Moreover,  %at least at one-loop level, the 
non-vanishing  one-loop  amplitudes arise in dimensional regularization through an   $\epsilon/ \epsilon$ mechanism similar 
to the one leading to   chiral anomalies. For the all-plus amplitudes this property is manifest in their dimension-shifting 
construction \cite{Bern:1996ja} in terms of the MHV amplitudes of $\NeqFour$ sYM theory\footnote{While extensively tested, 
the origin of this rather curious relation remains to be understood. The amplitude  representations obtained this way can be 
easily checked against the results of explicit calculations.}
\be
{\cal A}^{(1);\text{YM}}_{n}(1^+,\dots, n^+) = -2\epsilon(1-\epsilon)(4\pi)^2%\Big[\, 
\frac{1}{\langle ij\rangle^4}
{\cal A}^{(1);\text{ sYM MHV}}_{n}(1^+,\dots,i^-,\dots, j^-,\dots n^+)\Big|_{D\rightarrow D+4} \ .
%\,\Big] 
\label{dimshift}
\ee
Here  ${\cal A}^{(1)}_{n}$  denotes the color-dressed one-loop $n$-point amplitude.
The dimension shift $D\rightarrow D+4$ yields an $\epsilon^{-1}$ UV divergence which is then cancelled by the overall 
%AAT
$\epsilon$ factor above.\foot{One starts with the amplitude in $D=4-2 \epsilon$   and then  performs the shift $D\to D+4$
so  that all integrals over loop momenta are  evaluated in $8-2\epsilon$ dimensions  while the external momenta
remain in $D=4$.} 

To see that the all-plus and single-minus YM amplitudes potentially can lead to amplitudes breaking the $U(1)$ symmetry 
of \rf{decomposition} it suffices to discuss an example. The external legs of an $(n+2)$-point $\overline{\rm MHV}$ amplitude 
are two positive and $n$ negative helicity gluons. Tensoring each external state with the positive helicity gluon of an 
all-plus amplitude and identifying the supergravity states using \rf{spectrum1} and \rf{taua} leads to two positive-helicity 
gravitons and $n$ $\bartaua$ scalars (see \rf{taua}). This amplitude  carries the charge $4qn = -2n$ under the $U(1)$
symmetry in \rf{decomposition} and thus, if non-vanishing, is an anomalous amplitude, i.e. it  breaks this symmetry.

Anticipating the result of the next section that such superamplitudes are non-vanishing, we can classify the supergravity 
superamplitudes that can be constructed on the basis of the anomalous amplitudes of non-supersymmetric gauge theories.
Since there are two classes of such amplitudes, the double-copy construction implies that there are two classes  of the corresponding 
supergravity amplitudes as well.  Denoting the double-copy operation by the tensor product symbol $\otimes$, the 
independent one-loop anomalous supergravity superamplitudes (with rational momentum dependence) can be organized 
in the  two classes,
\bea
\label{type1}
&& {\cal A}_{n}^{(1); \text{ sYM N}^k\text{MHV}} \otimes {\cal A}^{(1);\text{YM}}_{n}(1^+,\dots,i^+,\dots n^+) \ , 
\\
\label{type2}
&& {\cal A}_{n}^{(1); \text{ sYM N}^k\text{MHV}} \otimes {\cal A}^{(1);\text{YM}}_{n}(1^+,\dots, i^-,\dots n^+)   \ ,
\eea
with $ \ k=0,\dots,n-4$.
Since there is no symmetry relating a positive-helicity gluon to a negative-helicity one in pure YM theory
these superamplitudes are unrelated to each other. Two more classes of amplitudes can be constructed by 
conjugation; in the language of the double-copy construction their pure YM components are the all-minus
and single-plus amplitudes.

%\draftnote{The conjugate amplitudes with all minus or all but one minus are...You may either use words here or give the %analogous 2 eqs}

It is interesting to note that, starting with five external legs, it is, in principle, possible that  other supergravity 
amplitudes are also anomalous. A  potential  example is provided  by the supergravity amplitude obtained from 
the one-loop five-point MHV $\NeqFour$ sYM superamplitude and the one-loop five-point ${\overline{\rm MHV}}$ 
pure YM amplitude; for any choice of external helicities this superamplitude carries nonzero $U(1)$ charge. 
Similarly to the amplitudes in eqs.~\rf{type1} and \rf{type2}, this amplitude also vanishes at the tree level. If 
non-vanishing at loop level, it  can have non-rational dependence on momentum 
invariants. While it would be interesting to construct and analyze it, we will not do it here and focus instead 
on superamplitudes in the two classes \rf{type1} and \rf{type2}.

All external states of the amplitudes of the first type, eq.~\rf{type1}, belong to $\Phi^+$ on shell chiral multiplets, 
eq.~\rf{phi+}, whereas those of the CPT-conjugate amplitudes belong to on shell chiral multiplets of the type $\Phi^-(\eta)$, 
eq.~\rf{phi-}. 
It is perhaps natural to use an anti-chiral superspace for one of them. We will later use the anti-chiral superspace 
for the former.
For  $k=0$ it is easy to write a superspace expression for  the contribution of such amplitudes to the effective  
Lagrangian; it consists of a sum of a chiral and anti-chiral superspace integrals
\be
{\cal L}_{\text{eff}} = \frac{1}{(4\pi)^2}\Big ( \int d^8\theta \,\sum_n\, d_n\, W^n 
                                                                        + \int d^8 \bar \theta \,\sum_n\, d_n\, \overline W^n \Big )\ ,
\label{superspace}
\ee
so that the action is hermitian for real coefficients $d_n$. They are to be determined by explicit calculations.
The component expansion of the chiral superfield $W(x,\theta)$ is
\be
W(x,\theta) = \tau+\theta^{\alpha}_{A}\chi_{\alpha}^{A}+\theta^{\alpha}_{A}\theta^{\beta}_{B}F_{\alpha \beta\,CD}\epsilon^{ABCD}
+\theta^{\alpha}_{A}\theta^{\beta}_{B}\theta^{\gamma}_{C} \psi_{\alpha \beta\gamma\,D}\epsilon^{ABCD}
+\theta^{\alpha}_{A}\theta^{\beta}_{B}\theta^{\gamma}_{C}\theta^{\delta}_{D} C_{\alpha \beta\gamma\delta}\epsilon^{ABCD}, 
\label{Wsuperfield}
\ee
and it contains the same states as the momentum space on shell superfield \rf{phi-}.
%AR
Here 
$C_{\alpha \beta\gamma\delta}$ is totally symmetric spinor  component  of  the self-dual Weyl tensor 
which on shell is the same 
as the self-dual   curvature tensor $R^-$,   describing  the negative helicity graviton.  
The Grassmann variables $\theta$ and $\eta$ are formally related by Fourier transform. 
Note that the dimension of $W$ is zero and therefore any power of this superfield has  dimension zero 
so that the one-loop multi-point amplitude has a correct dimension.

%AR
\def  \CC {{\rm C}}

The effective Lagrangian for amplitudes of the second type, eq.~\rf{type2}, is somewhat  more 
involved.
It depends on the anti-chiral superfield $\overline W$ and its space-time derivatives as well as on 
the anti-chiral superfield   $\CC_{\alpha\beta\gamma\delta} (x, \theta, \bar \theta)$  whose first component is the self-dual Weyl tensor $C_{\alpha\beta\gamma\delta}(x)$:\foot{The existence of  a
 linearized anti-chiral  superfield $\CC_{\alpha\beta\gamma\delta} $ is a consequence 
of the fact that the spectrum of $\NeqFour$ supergravity does not contain spin-5/2 states and that the graviton is on shell.
Indeed, assuming that the right-hand side of the second equation in \rf{constraintsCW}
  is non-zero we can decompose it into a completely symmetric 
spin-5/2 term and a term proportional to $D^{i\alpha}\CC_{\alpha\beta\gamma\delta}$ which is set to zero by on-shell conditions.
Absence of spin-5/2 fields sets to zero the first term as well, leading to the second equation in  \rf{constraintsCW}.}
\be
D_\eta ^i\overline W(x, \theta, \bar \theta)=0 \, , \qquad D_\eta ^i  \CC_{\alpha\beta\gamma\delta} (x, \theta, \bar \theta)=0 \ .
\label{constraintsCW}
\ee
Since the dimension of $C_{\alpha\beta\gamma\delta}(x)$ is 2 and its indices must be contracted with some derivative factors, 
a certain amount of non-locality is necessary to write down the effective Lagrangian; because of this it is naturally written in momentum space.
For $k=0$ in \rf{type2} we can write a superspace expression for the contribution of such amplitudes to the momentum 
space effective Lagrangian:
\bea
{\cal L}(p)&=&f_4 {\delta \Big (\sum_{i=1}^4 p_i\Big )  \over s_{12}s_{13}s_{23}} 
\Big ( \int d^8 \bar \theta \,  \CC_{\alpha\beta\gamma\delta} (p_1, \bar \theta) 
 D^{\alpha \dot \alpha} D^{\beta \dot \beta}  {\overline W} (p_2, \bar \theta)  D^{\gamma}{}_{\dot \alpha} D^\delta{}_{\dot \beta} 
 {\overline W} (p_3, \bar \theta){\overline W} (p_4, \bar\theta)    \nonumber\\
 &&  \qquad\quad
 + \int d^8  \theta \,  \bar \CC_{\dot \alpha \dot \beta\dot \gamma\dot \delta} (p_1, \bar \theta)  
 D^{\alpha \dot \alpha} D^{\beta \dot \beta}  W (p_2,  \theta)  
 D^{\dot \gamma}{}_{ \alpha} D^{\dot \delta}{}_{ \beta}  W (p_3,  \theta) W (p_3, \theta)  \Big ) \ .
\label{new2}
\eea
The coefficient $f_4$ can be determined by an explicit scattering amplitude calculation (see sec.~\rf{other_sg_amplitudes}). 
The effective Lagrangian is again a sum of a chiral and anti-chiral superspace integrals and 
therefore,  if $f_4\ne 0$, it breaks  the $U(1)$ symmetry. Eq.~\rf{new2} can be easily extended to capture 
higher-point amplitudes of the type \rf{type2} with $k=0$: one simply multiplies the chiral and anti-chiral integrands 
above by further factors of ${W}$ and ${\overline W}$, respectively. Other generalizations may also be possible, but 
we will not discuss them here.

It is worth mentioning that, while in the discussion above we assumed that one of the  two gauge theory factors 
is the    ${\cal N}=4$ sYM theory,
 the same analysis can be carried out if   ${\cal N}=4$  theory    replaced 
by the ${\cal N}=1$ or the ${\cal N}=2$ sYM theory. 
As argued in the previous section, the resulting supergravity theories
should have a $U(1)$ symmetry analogous to that in eq.~\rf{decomposition} and the scattering amplitudes constructed as
above  should  provide  candidates for anomalous amplitudes breaking it.
The complete classification of the supergravity amplitudes  in this case 
is slightly more involved than just \rf{type1}, \rf{type2} due to the existence of several types of N$^k$MHV amplitudes 
in ${\cal N}=1,2$ sYM theories.

The discussion above can also be extended trivially to a non-supersymmetric YM theory coupled with scalars; we 
will do this in sec.~\ref{Ampnv}. 

Let us now turn to   examples of anomalous amplitudes in pure $\NeqFour$ supergravity.

\subsection{Graviton-scalar amplitudes in $\NeqFour$ supergravity \label{hhscalar_amplitudes}}

There are many graviton-scalar amplitudes in $\NeqFour$ supergravity and  most of them have 
counterparts in $\NeqEight$ supergravity. Here we shall focus on the one-loop amplitudes described in the 
previous section, which carry non-zero $U(1)$ charge. 
All-plus and single-minus amplitudes have been constructed in \cite{Bern:1995db}; the all-plus amplitudes 
can also be obtained by dimension-shifting \rf{dimshift} the color/kinematics-satisfying representations of the 
one-loop four-point \cite{Green:1982sw}, five-point \cite{Carrasco:2011mn} and six- and seven-point 
\cite{Bjerrum-Bohr:2013iza} $\NeqFour$ MHV superamplitudes. The six- and seven-point calculations 
use the general framework \cite{Boels:2013bi} which, in principle, can be used at higher multiplicities.

We will collect here the results for the three-, four- and five-point  supergravity superamplitudes which contain 
component amplitudes with two positive-helicity gravitons and one, two and three scalar fields $\bartaua$.
Relegating some of the details of the calculations to Appendices~\ref{4pt_oneminus}, \ref{4pt_allplus} 
and \ref{5pt_allplus}, we find:
\bea
{\cal M}_3^{(1);\,  {\cal N}=4\,\text{PSG}}(1,2,3)&=&
\frac{i}{(4\pi)^2}\left(\frac{\kappa}{2}\right)^3\delta^{(8)}(\sum_{i=1}^3{\tilde \eta}_{i,A}{\tilde \lambda}_i)\ , 
\label{anomalous_3pt}
\\[2pt]
{\cal M}_{4}^{(1);\,  {\cal N}=4\,\text{PSG}} (1,2, 3,4)&=&
\frac{i}{(4\pi)^2}\left(\frac{\kappa}{2}\right)^4\delta^{(8)}(\sum_{i=1}^4{\tilde\eta}_{i,A}{\tilde\lambda}_i) \ ,
\label{anomalous_4pt}
\\[2pt]
{\cal M}_5^{(1);\,  {\cal N}=4\,\text{PSG}}(1,2,3,4,5) &=& 
\frac{2i}{(4\pi)^2}\left(\frac{\kappa}{2}\right)^5\delta^{(8)}(\sum_{i=1}^5{\tilde\eta}_{i,A}{\tilde\lambda}_i) \ .
\label{examples}
\eea
Since they are proportional to eight powers of ${\tilde\eta}$ we may refer to them as 
${\overline{\rm MHV}}{}^{(n,0)}$ superamplitudes, with $n=3,4,5$, because all their external legs 
are in the $\Phi^+$ multiplet. 
The MHV$^{(0,n)}$ amplitudes, with negative-helicity gravitons and scalars 
$\taua$ are obtained by conjugation (i.e. through ${\tilde\lambda} \rightarrow\lambda$, ${\tilde\eta}\rightarrow\eta$).

It should be mentioned that, as discussed in sec.~\ref{4pt_oneminus},  the three-point superamplitude 
is not constructed directly via the double-copy relation \cite{BCJLoop}. This is because, at least naively, the 
three-point (one-)loop amplitudes vanish identically in $\NeqFour$ sYM theory while the three-point 
all-plus amplitude is formally singular \cite{Bern:2005hs}. As a means of regularizing this $0/0$ situation 
we construct\footnote{We thank Z.~Bern and L.~Dixon for sharing their unpublished notes on this calculation.}
this superamplitude as the soft-graviton limit of the $k=0$ four-point one-loop amplitude of the  type \rf{type2}.

For completeness, we mention that the four- and five-point superamplitudes \rf{anomalous_4pt} and \rf{examples} 
are of type \rf{type1} with $k=0$ and $k=1$, respectively.

These amplitudes are forbidden in $\NeqEight$ supergravity by the $SU(8)$ on-shell R-symmetry. 
In particular, since they are by construction invariant under the $SU(4)$ R-symmetry of one $\NeqFour$ sYM theory 
factor and the external states are inert under the $SU(4)$ R-symmetry of the other $\NeqFour$ sYM theory factor, 
it is the $U(1)$ that appears in the decomposition \rf{decomposition}, $SU(8)\supset SU(4)\times SU(4)\times U(1)$,
that forbids the amplitudes \rf{anomalous_3pt},  \rf{anomalous_4pt}, \rf{examples} in the $\NeqEight$ theory.  
We therefore see explicitly that this symmetry, while present at tree-level in $\NeqFour$ supergravity,  is 
broken at the one-loop level.

To extract component amplitudes we multiply the superamplitude 
 by the appropriate on-shell superfields and integrate over all 
$\eta$ variables. The terms in the superamplitude that are proportional to $({\tilde \eta}_1)^4({\tilde \eta}_2)^4$ are 
the component amplitudes ${\cal M}(1^{++}2^{++}3^{\bartaua})$, ${\cal M}(1^{++}2^{++}3^{\bartaua}4^{\bartaua})$ 
and ${\cal M}(1^{++}2^{++}3^{\bartaua}4^{\bartaua}5^{\bartaua})$, respectively. Presence of external scalar fields
is,  however,  not a requirement as among the superpartners of,  e.g.,  ${\cal M}(1^{++}2^{++}3^{\bartaua})$ we can find
the graviton-vector-vector amplitude ${\cal M}(1^{++}2^{+}3^{+})$. This term can also be found in the superspace 
expression \rf{superspace}.

\subsection{Nonlocal amplitudes in $\NeqFour$ supergravity \label{other_sg_amplitudes}}

The characteristic property of the amplitudes discussed in the previous section is that they are local. As such, 
while still anomalous, they may be adjusted or even eliminated completely by simply {\em defining} the theory 
to contain finite local counterterms that simply set them to zero. The same cannot be immediately said about 
nonlocal amplitudes; it is therefore of interest to see whether there exist such amplitudes which are also anomalous.

Perhaps the simplest such amplitude is ${\cal M}(1^{--}2^{++}3^{++}4^{\bartaua})$, whose soft graviton limit 
leads to \rf{anomalous_3pt}. The calculation in Appendix~\ref{4pt_oneminus} implies that, in anti-chiral superspace,
this superamplitude is 
\bea
{\cal M}_4^{(1);\,  {\cal N}=4\,\text{PSG}}(1,2,3,4)=-\frac{i}{(4\pi)^2}\frac{1}{[31][14]}\frac{[32][24]\spa2.1}{[21]}
\delta^{(8)}(\sum_{i=1}^4{\tilde \eta}_{i,A} {\tilde \lambda}_i) \ .
\label{4pts_nonlocal_amp}
\eea
Extracting component amplitudes (see Appendix~\ref{4pt_oneminus} for details) does not remove the apparent 
nonlocality of this expression. Note that the spinor product ratio prefactor can also be written as 
$\langle 1|{k\llap/}_2 {k\llap/}_3|1\rangle^2/(stu)$ which reproduces the momentum dependence of the four-point 
superamplitude following from the superfield expression~\rf{new2}. This implies that, as anticipated, $f_4\ne 0$.
%%%%%%%%%
% \draftnote{Note also that the superfield expression for the non-local 4-point amplitude in \rf{new2} can 
% be shown to be proportional to the expression \rf{4pts_nonlocal_amp}.}
%%%%%%%%%

It is also possible to construct the five-point supergravity superamplitudes of type \rf{type2}; the result appears 
to be nonlocal and rather unwieldy and we will not present it here.

The five-point nonlocal superamplitude of type \rf{type1} can also be easily constructed: it corresponds to choosing 
$k=0$ in \rf{type1}. The result of  its  calculation is (see  Appendix~\ref{5pt_allplus})
\bea
\label{5ptsuperamp_final}
&&
{\cal M}_5^{(1);\,  {\cal N}=4\,\text{PSG}}(1,2, 3,4,5) 
= i \big(\frac{\kappa}{2}\big)^5 \,\frac{\delta^{(8)}(Q_5)}{3\,(4\pi)^2}  \,
\sum_{S_5} \,  \frac{1}{4}\frac{({\widehat\gamma}_{12})^2}{s_{12}} \no
\\
&& 
= i\big(\frac{\kappa}{2}\big)^5 \,\frac{\delta^{(8)}(Q_5)}{(4\pi)^2}  \,
\Bigl[
\frac{{\widehat\gamma}_{12}^2}{s_{12}}+\frac{{\widehat\gamma}_{13}^2}{s_{13}}
+\frac{{\widehat\gamma}_{14}^2}{s_{14}}+\frac{{\widehat\gamma}_{15}^2}{s_{15}}
+\frac{{\widehat\gamma}_{23}^2}{s_{23}}+\frac{{\widehat\gamma}_{24}^2}{s_{24}}
+\frac{{\widehat\gamma}_{25}^2}{s_{25}}+\frac{{\widehat\gamma}_{34}^2}{s_{34}}
+\frac{{\widehat\gamma}_{35}^2}{s_{35}}+\frac{{\widehat\gamma}_{45}^2}{s_{45}}
\Bigr]   
\label{final5ptsuperamlitude}
\eea
with 
\be
{\widehat\gamma}_{12}=\frac{\spb{1}.{2}^2\spb{3}.{4}\spb{4}.{5}\spb{3}.{5}}
{\spa{1}.{2}\spb{2}.{3}\spa{3}.{5}\spb{5}.{1}-\spb{1}.{2}\spa{2}.{3}\spb{3}.{5}\spa{5}.{1}} 
\ee
and the other ${\widehat\gamma}_{ij}$ obtained by relabeling. $Q_5$ is defined in eq.~\rf{gdelta}. 
A component S-matrix element contained in this superamplitude is ${\cal M}_5^{(1)}(h^{++},h^{++}, \bartaua, A^+, A^+)$.

It is interesting to note that, up to a trivial numerical coefficient, the square bracket in eq.~\rf{final5ptsuperamlitude}
is also the divergence of the five-point amplitude of ${\cal N}=8$ supergravity in $D=8-2\epsilon$ \cite{Carrasco:2011mn}. 
The collinear limits have the expected behavior dictated by the fact that the amplitudes evaluated here and their 
four-point counterparts vanish exactly at tree level,   implying that only the tree-level splitting amplitudes are necessary.
These features suggest that, as in the case of ${\cal N}=8$ supergravity, the amplitude 
\rf{5ptsuperamp_final} is related to the covariantization of a four-point term in the effective action.

\subsection{Amplitudes in ${\cal N}=4$ supergravity coupled to $\nv$ vector multiplets \label{Ampnv}}

We may explore further the relation between the matrix element\foot{We use the definition  $R^\pm = \ha (R \pm i R^*)$   for the   ``self-dual''
 curvature components.}
 $\langle (R^+)^2\, \bartaua^n\rangle$
and the anomaly of the $U(1)$ symmetry in \rf{decomposition}  by coupling $\NeqFour$ supergravity with 
$\nv$ vector multiplets and comparing the S-matrix elements with the
corresponding  anomalous term in the effective action.

As discussed above,  in the KLT-based generalized unitarity approach to supergravity scattering amplitudes 
 as well as in the double-copy construction, the scattering amplitudes in $\NeqFour$ supergravity coupled to $\nv$ 
 vector multiplets   may be found  by supplementing the non-supersymmetric gauge theory factor by  $\nv$ real scalar fields,
\be
(\NeqFour \ \text{sYM})\otimes \big[ ( {\cal N}=0\ \text{sYM} ) \oplus \nv \, \text{real scalars}\big] \ .
\label{Neq4_nv}
\ee 
The $\nv$ scalars couple to the ${\cal N}=0$ gluons though the standard minimal coupling and also 
have quartic self-couplings similar to the scalar fields of $\NeqFour$ sYM theory.\footnote{Since we are 
interested only in scattering amplitudes of ${\cal N}=0$ gluons the scalar self-couplings 
need not be specified.}

At one-loop level all fields of a gauge theory make independent contributions to gluon scattering amplitudes.
Moreover, for the finite helicity amplitudes, supersymmetric Ward identities imply that the contributions of particles 
of different spin circulating around the loop are related, $A_n^{(1);s=1}= -A_n^{(1);s=1/2} = A_n^{(1);s=0}$, where
the spin zero particle is a complex scalar.
Thus, we may express the all-plus and single-minus pure YM amplitudes  in the previous section 
in terms of the contribution of the scalar fields circulating in the loop. 
We can do this directly by using $A_n^{(1);s=1}= A_n^{(1);s=0}$ or by writing the spectrum of free YM theory
in terms of the spectra of supersymmetric YM theories:
\be
({\cal N}=0\ \text{sYM}) = (\NeqFour\ \text{sYM}) - 4(\NeqOne\ \text{chiral multiplet}) +2\ \text{real scalars} \  .
\ee
Here the second term on the right-hand side cancels the contribution of the $\NeqFour$ fermions and the third 
term cancels the contribution of the $\NeqFour$ and $\NeqOne$ scalars. Due to supersymmetry, the contribution 
of the first two multiplets to the all-plus and single-minus amplitudes vanishes identically and one concludes that 
these amplitudes are given by two real scalars (or one complex scalar) running in the loop. 

Thus, in the theory $({\cal N}=0 \ \text{sYM})   \oplus  (\nv\ \text{real scalars})$ \  the all-plus and single-minus 
one-loop gluon amplitudes  are  given by $\nv+2$ real scalars running in the loop and can therefore be obtained 
by multiplying the amplitudes in pure YM theory by $\ha (\nv+2)$. 
Proceeding to the $\NeqFour$ supergravity coupled to $\nv$ vector multiplets, since in the  double-copy approach 
the integrands of the anomalous S-matrix elements  are proportional  to the all-plus or single-minus 
YM amplitudes, they can be obtained from those of pure $\NeqFour$ supergravity by multiplication by $\ha (\nv+2)$.

The $\nv$-dependence of  the overall coefficient matches the one  in eq.~(\ref{nv_anomaly}); this suggests
that the $U(1)$ anomaly captured by the scattering amplitude calculation is that of the $U(1)$ subgroup 
appearing  in~\rf{decomposition}.

%%%%%%%%%%%%%%%%%%%%%%%%%%%%%%%%%%%
\renewcommand{\theequation}{4.\arabic{equation}}
 \setcounter{equation}{0}
%%%%%%%%%%%%%%%%%%%%%%%%%
\section{Effective action from scattering amplitudes \label{eff_action}}
%%%%%%%%%%%%%%%%%%%%%%%%%

As reviewed in sec.~\ref{intro}, it was shown in  \cite{ArkaniHamed:2008gz} that the tree-level amplitudes 
of $\NeqEight$ supergravity vanish identically in the single soft scalar limits (i.e. the limit in which the 
momentum of a single scalar field  goes to  zero).  Moreover, it was  argued that this is a manifestation of the 
$E_{7(7)}$ duality symmetry of the theory and that, in the absence of $E_{7(7)}$ anomalies, this should 
extend to all-loop orders.
One may also understand the vanishing of the soft scalar limit in $\NeqEight$ 
supergravity as a consequence of the $SU(8)$ symmetry which, if unbroken,  requires that 
all amplitudes corresponding to field configurations that are not $SU(8)$ invariant vanish identically.
To see this, let us  consider a general $n$-point scattering amplitude that is $SU(8)$ invariant and has 
at least an external scalar. Then, since 
the scalar fields transform under $SU(8)$, the field configuration obtained by dropping this scalar field
cannot be $SU(8)$ invariant and thus the corresponding $(n-1)$-point amplitude 
should vanish identically. Thus  the soft scalar limit of the $n$-point amplitude  should be  zero if $SU(8)$ 
is not anomalous.

Let us study the single soft-scalar limits of the anomalous amplitudes constructed in the previous section.
We will find that, while vanishing at the tree level, the soft scalar limits are non-vanishing at the one-loop level. 
We will then use them to   find  
which ``anomalous''  terms   in the effective action 
correspond to amplitudes breaking the $U(1)$ symmetry. 

\subsection{Soft scalar limit and the single-multiplet soft scalar function}

Since $\NeqFour$ supergravity is a consistent orbifold projection of $\NeqEight$ supergravity  \cite{cgr},
its tree-level amplitudes are a subset of those of $\NeqEight$ supergravity and therefore should also have vanishing 
soft scalar limits at tree level. This is a consequence of the $U(1)$ symmetry \rf{decomposition} because
the scalar field labeling the $\NeqFour$ supergravity amplitudes is not charged under the $SU(4)$ R-symmetry.
%%%%%%%%%%%%%%%%%%%%%%%%%
%As we saw in the previous section, there exist one-loop amplitudes that break 
% this symmetry, which is therefore anomalous.  
%The fact that these amplitudes  break the $U(1)$ symmetry inherited from $\NeqFour$ supergravity \rf{decomposition} 
%implies that their soft scalar limits   are   non-vanishing.
%%%%%%%%%%%%%%%%%%%%%%%%%
The fact that there exist sequences of loop-level amplitudes that break this $U(1)$ and differ only in the number of external 
scalars  implies that their soft scalar limits   are   non-vanishing.
Below we will construct the resulting soft scalar function and use it to
find all one-loop terms in the effective action with holomorphic dependence on $\taua$.

Scattering amplitudes have universal factorization properties in the limit in which the momentum of an external 
particle (say, the $n$-th one)  is soft. In general, the $L$-loop scattering amplitudes behave as 
\cite{Berends:1988zp, Weinberg:1965nx, Weinberg:1964ew, BernGrant}
\be
{\cal M}^{(L)}_n(1,2,\dots n-1, n) \stackrel{k_n\rightarrow 0}{\longrightarrow } \,\frac{\kappa}{2}
\sum_{\ell=0}^{L}{\cal S}^{(\ell)}_n\,{\cal M}^{(L-\ell)}_{n-1}(1,2,\dots n-1)  \ .
\label{soft_general}
\ee
In this expression the  coupling constant $\kappa/2$ is assumed to be placed in vertices. It has 
been argued in \cite{Bern:1998sv} that, in pure supergravity theories, the soft scalar function 
does not receive loop corrections; therefore,  the expression above collapses to a single term
\be
{\cal M}^{(L)}_n(1,2,\dots n-1, n) \stackrel{k_n\rightarrow 0}{\longrightarrow } \,\frac{\kappa}{2}
{\cal S}^{(0)}_n\,{\cal M}^{(L)}_{n-1}(1,2,\dots n-1)  \ .
\label{singleterm}
\ee
For color-ordered amplitudes in a gauge theory the soft scalar factor depends on the label of the 
soft leg (here $n$) and the legs adjacent to it (i.e. $1$ and $(n-1)$). In supergravity (as in any unordered 
theory) the soft factor ${\cal S}_n$ depends on all the external legs. In tree-level (super)gravity 
it is typically given by
\be
{\cal S}_n=\sum_{i=2}^{n-2} s(1, i, n-1, n) \ .
\label{structure_soft_fct}
\ee
Despite appearances, ${\cal S}_n$ is symmetric under the interchange of legs $1$ and $(n-1)$ with the others.
This expression captures the fact that, in an unordered amplitude,  some fixed external leg can formally be ``adjacent" to any two other legs.

The examples of amplitudes in eqs.~\rf{anomalous_3pt}, \rf{anomalous_4pt} and \rf{examples} are sufficient
to determine the soft scalar functions for the case when all particles belong to the same $\NeqFour$ multiplet 
as the soft scalar. We should stress that the functions $s(1, i, n-1, n)$ determined from them may not be 
correct when any of the external legs $1, i, (n-1)$ belongs to a different multiplet than the soft scalar leg, $n$.

Regardless of the arguments of \cite{Bern:1998sv} on the non-renormalization of soft functions, 
the fact that the anomalous amplitudes vanish identically at tree level implies that only the leading order
soft scalar function is important at one-loop level.\footnote{It is interesting to also consider soft scalar limits
of one-loop amplitudes that do not vanish at tree-level, i.e. of amplitudes with vanishing total $U(1)$ charge. 
The fact that the anomalous $U(1)$ is a symmetry of 
tree-level amplitudes implies that the structure of the amplitude in the soft scalar limit is that in eq.~\rf{singleterm}
with $L=1$ and with ${\cal M}^{(1)}_{n-1}$ being a one-loop anomalous amplitude. It seems therefore that we can 
expect that the one-loop amplitudes that have a non-zero tree-level counterpart (and thus are  
allowed in $\NeqEight$ supergravity) also have non-zero soft scalar limits.} 
Extracting the component amplitudes and fitting the soft scalar limit onto eq.~(\ref{singleterm}) and with 
${\cal S}_n^{(0)}$ of the form (\ref{structure_soft_fct}) implies that
\bea
{\cal S}^0_n =
\sum_{\text{s=2}} 0
+
\sum_{\text{s=3/2}} \frac{1}{4}
+
\sum_{\text{s=1}} \frac{1}{2}
+
\sum_{\text{s=1/2}} \frac{3}{4} 
+ 
\sum_{s=0} 1
= \sum_{i=2}^{n-2} \frac{1}{2}(2-h_\text{external, i})\ , 
\label{soft_scalar_fct}
\eea
where the sums run over all the external legs of the amplitude except for the one on which the soft scalar
limit is taken and $h_\text{external, i}$ is the helicity of the $i$-th external leg. We stress that, due to its derivation, 
this expression holds {\em{only}} for amplitudes in which all external legs are in the same on-shell $\NeqFour$ 
multiplet.\footnote{We have also successfully tested this expression for the spin-2 fields in the conjugate 
$\NeqFour$ on-shell multiplet.}
By construction, this soft scalar function allows one to obtain the superamplitudes 
\rf{anomalous_3pt}-\rf{examples} from each other.

The soft scalar function \rf{soft_scalar_fct} may also be written in superspace. To this end one selects the $n$-th external 
leg  of the superamplitude and chooses its Grassmann parameter dependence such that it corresponds to a scalar field 
(i.e. one multiplies the anti-chiral superspace superamplitude by $({\tilde\eta}_n)^4$ and integrates over ${\tilde\eta}_n$) 
and then  sets $k_n=0$. The result is:
\be
{\cal S}^{(0)}_n = n-3 \ .
\label{superspace_soft_fct}
\ee
We can use this soft scalar function to construct higher-point superamplitudes with all fields belonging to 
the same multiplet by starting from the lower-point ones. This is analogous to the inverse soft limit used at tree 
level to construct all tree amplitudes. 
In that case a detailed analysis was necessary to identify the origin of the soft limit and the relevant 
BCFW shifts. A similar analysis is not needed here because there is no momentum dependence associated 
with the additional scalar leg.

%%%%%%%%%%%%%%%%%%%%%%%%%%%%%%%%%%%
%\subsection{Inverse soft scalar limit and single-multiplet MHV $(n+2)$-point superamplitudes}
%%%%%%%%%%%%%%%%%%%%%%%%%%%%%%%%%%%

\subsection{Inverse soft scalar limit and the MHV$^{(0, n+2)}$ and ${\overline{\rm \bf MHV}}^{(n+2,0)}$ 
%$(n+2)$-point 
superamplitudes}

The general form of an $(n+2)$-point ${\overline{\rm MHV}}^{(n+2,0)}$ amplitude 
(i.e. the ${\overline{\rm MHV}}$ with all fields belonging to the $\Phi^+$ multiplet, \rf{phi+}) is 
\be
{\cal M}_{n+2}^{(1);\,  {\cal N}=4}(1,2,\dots,n+2) = c_{n+2}\delta^{(8)}(\sum_{i=1}^{n+2}{\tilde\eta}_{i,A}{\tilde\lambda}_i) \ ,
\label{np2pts}
\ee
where $c_n$ can be a function of momentum invariants which has no multi-particle poles.  The analogous 
MHV$^{(0, n+2)}$ amplitude, containing the matrix element $\langle h^{--} h^{--}\,(\taua)^{n}\rangle$, is equally simple and
is naturally written in chiral superspace: it is simply obtained by the transformation $\lambda_{\alpha i}\leftrightarrow {\tilde\lambda}_{{\dot\alpha}, i} $ and $\eta\leftrightarrow {\tilde\eta}$.
These superamplitudes are related by soft scalar limits. Using the superspace form of the soft scalar function 
(\ref{superspace_soft_fct}) and accounting for the absence of multi-particle poles,
it is not difficult to see that all the coefficient functions are constant
\be
c_3= \frac{i}{(4\pi)^2} \left(\frac{\kappa}{2}\right)^{3} \ , \ \ \ 
\qquad
c_{n+2} = \frac{i}{(4\pi)^2}(n-1)! \left(\frac{\kappa}{2}\right)^{n+2}\quad \text{for} \quad n\ge 2 \ , 
\label{coef}
\ee
where for $n=1$ we used eq.~(\ref{anomalous_3pt}). Thus, the $(n+2)$-point ${\overline{\rm MHV}}^{(n+2,0)}$ 
superamplitude 
%with all fields belonging to the same $\NeqFour$ multiplet 
is
\bea
\label{nptsuperamp}
{\cal M}_{n+2}^{(1);\,  {\cal N}=4}(1,2,\dots,n+2) 
&=& \frac{i}{(4\pi)^2}(n-1)! \left(\frac{\kappa}{2}\right)^{n+2}\delta^{(8)}(\sum_{i=1}^{n+2}{\tilde\eta}_i^A{\tilde\lambda}_i) \ ,
\eea
and its graviton-scalar component is\footnote{It may be useful to compare our approach with other inverse 
soft-limit loop-level constructions which, in contrast to what have done, use soft-functions visible at tree-level. 
Using the tree-level graviton soft function, e.g., the rational terms in the one-loop $n$-graviton amplitudes in 
$\NeqFour$ supergravity were recently constructed in \cite{Dunbar:2012aj}. }
\bea
{\cal M}_{n+2}^{(1);\,  {\cal N}=4}(h_1^{++},h_2^{++},{\bartaua}_{3}\dots,{\bartaua}_{n+2}) 
&=& \frac{i}{(4\pi)^2}(n-1)! \left(\frac{\kappa}{2}\right)^{n+2}[12]^4 \ .
\label{np2pts_components}
\eea

Let us now construct the graviton-scalar effective action that reproduces the amplitudes in eq.~\rf{np2pts_components}. 
We assume that the classical  supergravity action is normalized as \cite{Bern:1993wt}
\be
S=\frac{1}{2} \left(\frac{\kappa}{2}\right)^{-2}\int d^4 x \, \sqrt g  R+\dots \ ,
\label{einstein}
\ee
i.e.  $\kappa$ is the gravitational coupling. 
The term in the effective action that contributes to the matrix element $\langle h^{++} h^{++}\,(\bartaua)^{n}\rangle$ 
in eqs.~(\ref{np2pts_components}) and has the highest number of fields is  
($R^\pm = \ha (R\pm i R^*)$)
\be
\Gamma^{(1)}_{n+2} =  \int d^4 x\  s_n\,  (R^+)^2\, \bartaua^n + {\rm c.c.}\ .
\label{effectiveactionterm}
\ee
It is not difficult to construct the contribution to the two-graviton--($n$-scalar) 
S-matrix element of such an effective action term:
\be
{\cal M}_{\Gamma_{n+2}^{(1)}}(h_1^{++}, h_2^{++}, {\bartaua}_3\dots 
{\bartaua}_{n+2}) =  2i \, s_n \, n!\,\left(\frac{\kappa}{2}\right)^{n+2} [12]^4
\label{np2Gammamatelem}
\ee
where $k_1$ and $k_2$ are the momenta of the two gravitons and the overall $n!$ factor  accounts for the 
nonstandard normalization of an (effective) action term containing the $n$-th power of a scalar 
field (the overall factor of 2 has a similar origin). 

\iffalse
%%%%%%%%% FIGURE %%%%%%%%%%%%%%%%%%
\begin{figure}[t]
\begin{center}
\includegraphics[width=0.7\textwidth]{non1PI_hhtbn.eps}
\end{center}
\caption{\small The two diagrams that appear in the five-point one-loop amplitudes.}
\label{non1PI_hhtbn}
\end{figure}
%%%%%%%%%%%%%%%%%%%%%%%%%%%%%%%%
\fi

To determine the numerical coefficient $s_n$ we compare  \rf{np2pts_components}  
with the same S-matrix element computed from the effective action; the latter is given by the sum of  \rf{np2Gammamatelem} 
and the contribution of  Feynman graphs with one vertex from the lower-point effective action and 
the other vertices from the tree-level Lagrangian.
It is not difficult to see that the tree-level part of such Feynman graphs is forbidden 
by supersymmetry if all its external lines are on shell. 
Since in Feynman graphs the internal lines are off shell, these tree-level Green's functions give a contribution 
proportional to the square of the momentum of the internal 
line, i.e. they are proportional to the tree level equation of motion for the off-shell leg.\footnote{For example, using the 
supergravity action, one can see that an amplitude with a single $\tau$ (the full supergravity field) and any number 
of ${\bar\tau}$  fields  is indeed proportional  to $\Box\tau$. Also, it 
%appears 
is impossible to draw tree-level graphs 
with the field configuration $h^{++}h^{--}{\bar\tau}^m$ because all vertices contain both $\tau$ and ${\bar\tau}$. The 
matrix elements of such operators are set to zero by the supersymmetry Ward identities.}
Such Green's functions can be set to zero by a local field redefinition in the corresponding effective action.

Thus \rf{np2Gammamatelem} represents  the complete contribution of $\Gamma^{(1)}$ to the one-loop 
two-graviton--($n$-scalar) S-matrix element. Comparing it with \rf{np2pts_components} we find that the 
coefficient $s_n$ is
\be
s_n = \frac{1}{2(4\pi)^2} \frac{1}{n} \ .
\ee
Therefore, the one-loop effective action with two gravitons and any number of scalar fields is
\be
\Gamma_{hh}^{(1)} = \sum_{n=1}^\infty\Gamma^{(1)}_{n+2}= 
 \frac{1}{2(4\pi)^2}\int d^4 x \,  (R^+)^2\, \Big(\bartaua+\sum_{n\ge 2} \frac{1}{n} \bartaua^n\Big)+ {\rm c.c.}\ . 
\ee
This expression may  be summed up as 
\bea
\Gamma^{(1)}_{hh} &=&  -\frac{1}{2(4\pi)^2} \int d^4 x \,  (R^+)^2\ln(1-\bartaua)+ {\rm c.c.}  \ .
\la{fff} 
\eea
While  bosonic, it  was constructed from scattering amplitudes that manifestly preserve 
linearized (asymptotic-state) supersymmetry and thus may be promoted to a superspace expression in terms of 
analogs of the  superfields \rf{Wsuperfield} whose fields are identified with the double-copy fields \rf{spectrum1}.
Extracting the two-graviton component of \rf{superspace} it is easy to see that the coefficients $d_n$ in that equation
are given by
\be
d_{n+2} = \frac{1}{(n+2)(n+1)}s_{n+2} =\frac{1}{2(4\pi)^2}\,\frac{1}{n(n+1)(n+2)} \ .
\ee
The resulting resumed one-loop effective action in linearized superspace 
has a relatively simple but unilluminating expression and we will not present  it here.

The derivation above can be repeated with minimal changes in the presence of additional $\nv$ $\NeqFour$ vector 
multiplets. The discussion in sec.~\ref{Ampnv} implies that the analogs of eqs.~\rf{nptsuperamp} and \rf{np2pts_components}
pick up an overall factor of $(2+\nv)/2$. Consequently, the only change to $\Gamma_{hh}^{(1)}$ and the $d_{n+2}$
coefficients is a multiplicative factor of $(2+\nv)/2$.

% R JHEP
Interpreting eq.~\eqref{fff} as the local off-shell extension of the local on-shell effective action one may proceed to 
construct the nonlocal part of the on-shell effective action from 
non-local amplitudes, some of which were described in sec.~\ref{other_sg_amplitudes}. Given a particular field configuration,
one should first compute the one-loop amplitude with that external field configuration and then subtract from it the corresponding 
tree-level amplitude with exactly one vertex from the local one-loop effective action and all the others from the tree-level
action. Whenever nonzero, this difference represents a new contribution to the on-shell effective action. While potentially 
non-local, such terms are dependent on the local effective action being {\it chosen} not to contain terms proportional to the 
classical equations of motion; relaxing this requirement will modify the non-local effective action by further local terms which 
are not necessarily proportional to the classical equations of motion.

%%%%%%%%%%%%%%%%%%%%%%%%%%%%%%%%%%%%%%%%%%%%%%%%
\subsection{Comparison with the anomaly-induced term  in the effective action }
%%%%%%%%%%%%%%%%%%%%%%%%%%%%

In general, the local part of the effective action depends on the regularization scheme used to construct it. Assuming that
the scheme selected by the double-copy construction preserves the shift symmetry of the supergravity axion field $B$ 
\eqref{22}, we can compare \rf{fff} with the anomaly-induced effective action \rf{ann}. Locality of $\Gamma_{hh}^{(1)}$ implies  
that  it can be compared only with the first local term  $\int RR^* B$ in eqs.~\rf{ann}, \rf{nnb}.
%%
% R JHEP
%To compare \rf{fff} with the anomaly-induced effective action \rf{ann} we first notice that, since the effective 
%action $\Gamma_{hh}^{(1)}$ is local, it can be compared only to the first local term  $\int RR^* B$ in eqs.~\rf{ann}, \rf{nnb}.
%%
%
Furthermore, while $\Gamma_{hh}^{(1)}$ was derived in a chiral superspace 
framework   which is   manifestly supersymmetric,   the parity-odd anomalous term in \rf{ann} 
needs to  be supplemented by other  parity-even terms to make  the resulting effective action 
consistent with supersymmetry. One  obvious   candidate   for such an extra term 
is
$ \int RR \vp$   which  should indeed be present in the  one-loop effective action $\G$  as discussed in  Appendix~\ref{ap2}. \foot{Note that 
   up to terms proportional  to equations of motion (i.e. Ricci tensor terms)  $RR$  is equal to $R^*R^*$
   which is a total derivative in four dimensions and thus does not contribute to the S matrix. This also implies that, in four dimensions, 
   the term $\int RR\varphi$ contributes to the S matrix only terms proportional to the momentum of the scalar field.} 

Since the  parity-odd anomalous term \rf{nnb}  is linear in $B$, it  is  natural  to  expect  that   
its  extension  which is consistent with supersymmetry   should be   linear in $\tau = B + i e^{-\vp}$, 
namely
\be
\Gamma =\frac{i}{4(4\pi)^2}\int d^4 x \, \big[ (R^+)^2 {\bar\tau}- (R^-)^2 \tau \big] 
= -  \frac{1}{4(4\pi)^2}\int d^4 x \, \big(  RR^* B   - RR\, e^{-\vp} \big) \ . 
\label{linear_in_B}
\ee
Indeed, if one  makes  a  general ansatz like 
\be
\Gamma = \int d^4 x\;  (R^+)^2 \sum_{n,m}  a_{nm}  \tau^n {\bar\tau}^m + \text{c.c.}\ , \la{tet}
\ee
  imposes  the linearized supersymmetry restrictions on 
 the resulting scattering amplitudes\footnote{At the linearized level, setting 
 $\tau=i+c\,\taua + O(\taua)$  (with some  constant $c$), 
 supersymmetry requires that the only non-vanishing matrix elements are
\bea
&&
(R^+)^2 {\bar\taua}^{1+n}\ , 
\qquad
(R^-)^2 {\taua}^{1+n}
~~
\text{with}
~~
n\geq 0 \ , 
\cr
&&(R^+)^2{\bar\taua}^{n+2}\taua^m\ , 
\quad
(R^-)^2{\taua}^{n+2}{\bar \taua}^m 
~~
\text{with}
~~
n\geq 0, ~m\geq 1 \ .
\nonumber
\eea
} 
and requires that for $\tau= B + i e^{-\vp} $ as in   in eq.~(\ref{22}) 
 this   effective   action $\Gamma$  is linear in $B$,    one is led to  \rf{linear_in_B}.
 
 Terms containing other fields, such as $R^+F^+F^+$ and others which are required by
  linearized supersymmetry and identified at linearized level in the amplitude calculation  
  are to be determined separately (or found by  expanding in components the supersymmetric 
  anomaly-related action constructed  in \cite{Bossard:2012xs}).

Comparing (\ref{linear_in_B}) and $\Gamma^{(1)}_{hh}$ in eq.~(\ref{fff}) and adjusting the normalization 
of the Einstein term \rf{einstein} to match the one in eq.~\rf{laga} (i.e. absorbing a factor of $\sqrt{2}$ in each 
graviton wave function and thus producing an extra factor of $1/2$ in $\Gamma_{hh}^{(1)}$),   we see that 
they  match  provided one makes  the identification
\be
\tau=i-i\ln(1-\taua) \ . 
\la{tat}
\ee
Here the constant term accounts  for the fact that we assume that  $B=\varphi=0$
should correspond to $\taua=0$. 
%
% R JHEP
%
Thus, the assumption that the regularization scheme employed in amplitude calculations
preserves the shift symmetry of the axion implies that this symmetry is realized as a shift and a rescaling of $\taua$.

Assuming \rf{tat}, the  $U(1)$ transformation of $\taua$ does not translate into a simple transformation of the 
supergravity field $\tau$. At the  linearized level,  however, when $\tau-i=i\,\taua$, an infinitesimal $U(1)$ 
transformation of $\taua$ is the same as an infinitesimal transformation of $\tau$ under the anomalous $U(1)$ 
subgroup identified  below eq.~\rf{sl}. 

In the presence of additional $\nv$ vector multiplets $\Gamma$  in \rf{linear_in_B} 
 must be multiplied by a factor of $(2+\nv)/2$
as in \rf{nv_anomaly}.

%%%%%%%%%%%%%%%%%%%%%%%%%%%%%%%
\renewcommand{\theequation}{6.\arabic{equation}}
 \setcounter{equation}{0}
%%%%%%%%%%%%%%%%%%%%%%%%%
\section{Summary and concluding  remarks}
%%%%%%%%%%%%%%%%%%%%%%%%%

In this paper we first discussed in detail the $U(1)$ anomaly of  the $\NeqFour$ Poincar\'e supergravity
realized as conformal supergravity coupled to vector multiplets. In the $SU(4)$-invariant formulation of the theory
we identified the anomalous $U(1)$ subgroup of the duality group $SU(1,1)$. 
We have also identified a $U(1)$ symmetry acting on the on-shell asymptotic states under which all the 
fields carry the same charges as under the anomalous $U(1)\subset SU(1,1)$.

Then, using the double-copy construction, we computed particular one-loop supergravity scattering amplitudes whose 
tree-level counterparts  vanish identically. These amplitudes break the asymptotic-state $U(1)$ symmetry. Interestingly,  
this breaking is related to an anomaly in an asymptotic-state tree-level  bosonic symmetry of pure YM theory
which, in the context  of the  $\NeqFour$   super YM theory,   follows from  supersymmetry. 

The dependence of the symmetry-breaking amplitudes (and of the symmetry-breaking effective action) on the number 
of additional vector multiplets matches the dependence of the anomaly of the $U(1)$ subgroup of the duality group.

We have shown that the soft-scalar limits of the anomalous amplitudes are non-vanishing, 
which is in line with the expectation that a non-anomalous duality symmetry requires  that this type of limit vanishes. 
The soft scalar function extracted this way is momentum-independent and can be used to construct a class of higher-point 
one-loop anomalous amplitudes of any multiplicity.

%AR
Symmetries  analogous to the anomalous asymptotic state $U(1)$ symmetry discussed here  are  present in other 
supergravity theories which, for certain matter content, can be obtained though  factorized orbifolding  
(consistent truncation) from $\NeqEight$ supergravity. 
Examples   include   ${\cal N}=2$ supergravity with $(1+\nv)$ vector multiplets (for $\nv=2,4,6$) and 
${\cal N}=1$ supergravity with one chiral and $\nv$ vector multiplets (for $\nv=2, 4, 6$).  
From a double-copy perspective these theories are realized
as $(\NeqTwo \text{ sYM})\otimes (\text{pure YM}\oplus \phi_p)$ and $({\cal N}=1 \text{ sYM})\otimes (\text{pure YM}
\oplus \phi_p)$ (with the $U(1)$  charges of  fields  determined as in \rf{2copy_charges}). The four-point anomalous 
amplitudes can be   found  using the color/kinematics-satisfying representations of the one-loop 
four-gluon amplitudes in $\NeqOne$ and $\NeqTwo$ sYM theory together with the color/kinematics-satisfying 
representations of the one-loop all-plus amplitudes obtained,  e.g.,  by dimension shifting (see Appendix~\ref{4pt_allplus}). 
They are:
\be
{\cal M}_4^{(1); {\cal N}}(h_1^{++}, h_2^{++},\bartaua_3, \bartaua_4)=\frac{i}{(4\pi)^2}\left(\frac{\kappa}{2}\right)^4[12]^4
\Big[1+(4-{\cal N})\Big(1-\frac{tu}{3s^2}\Big)\Big] \ ,
\ee
where ${\cal N}=1,2$, the first term in parenthesis is the contribution of the $\NeqFour$ supergravity multiplet 
and the second term represents the subtracted contribution of two ${\cal N}=2$ gravitino multiplets (for $ {\cal N}=2$)
or  three ${\cal N}=1$ gravitino and three ${\cal N}=1$ vector multiplets (for ${\cal N}=1$); $s, t$ and $u$ are the usual
Mandelstam variables. 
The scalar fields $\taua$ 
belong to a vector multiplet for an ${\cal N}=2$ theory or a chiral multiplet for an ${\cal N}=1$.
As in the case  considered   in sec.~\ref{Ampnv}, the dependence on the number of vector multiplets enters through  
an  $(\nv+2)/2$ multiplicative factor.

It would be interesting to understand whether the quantities defined by eq.~\rf{2copy_charges} correspond to 
the charges of a physical symmetry even if both gauge theory factors appearing in the double-copy construction are 
supersymmetric.
%, because it may hint towards the mechanism responsible for the appearance of U-duality from gauge theory.  
For a symmetric (i.e. with identical gauge theory factors) construction with $\NeqOne$ sYM factors, the 
charges $\qq(\Phi\otimes{\tilde\Phi})$ of the fields in the supergravity multiplet appear to be  given by a combination of the  
$U(1)$ R-symmetry groups of the two gauge theory factors which can be identified with the  Cartan 
generator of the SU(2) R-symmetry group of $\NeqTwo$ supergravity.
For a symmetric construction with $\NeqTwo$ sYM factors, which realizes $\NeqFour$ supergravity with two vector 
multiplets \cite{cgr, Tourkine:2012vx}, the $\qq(\Phi\otimes{\tilde\Phi})$ of fields in the supergravity multiplet are such 
that they combine into representations of  $SU(4)\supset SU(2)\otimes SU(2)\otimes U(1)_{\text{eq.\;\rf{2copy_charges}}}$.
The scalar field in the supergravity multiplet is uncharged under this $U(1)$ and therefore the anomalous amplitudes
discussed in this paper,  while non-vanishing, do not break it.
The charges of the matter vector multiplets are different from the expected ones, suggesting that for them 
eq.~\rf{2copy_charges} assigns charges corresponding to a linear combination of the $U(1)\subset SU(4)$
R-symmetry and the duality symmetry of each vector multiplet.
It would be useful to clarify this structure    in more detail as well as understand  the meaning of the quantities defined 
by eq.~\rf{2copy_charges} for asymmetric double-copy constructions.

%%%%%%%%% FIGURE %%%%%%%%%%%%%%%%%%
\begin{figure}[t]
\begin{center}
\includegraphics[height=0.155\textwidth]{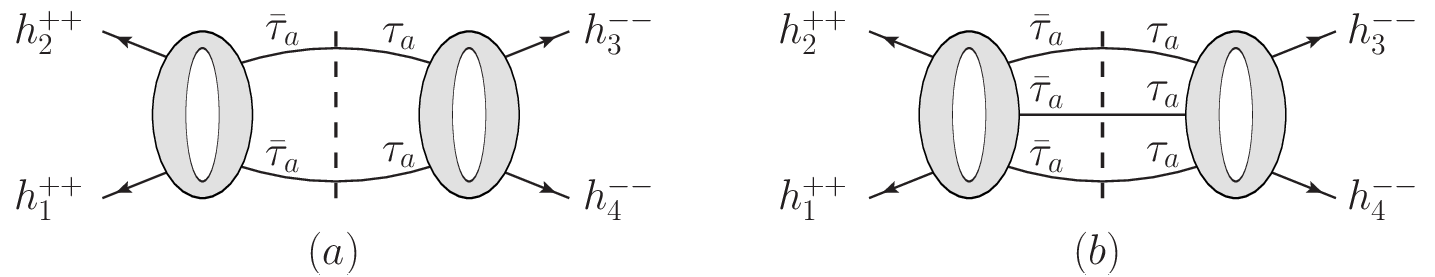}
\end{center}
\caption{\small Cuts of the three- and four-loop four-graviton amplitude in $\NeqFour$ Poincar\'e 
                           supergravity which reveal the contribution of anomalous amplitudes.}
\label{3loopcut}
\end{figure}
%%%%%%%%%%%%%%%%%%%%%%%%%%%%%%%%

The one-loop  $U(1)\subset SU(1,1)$ anomaly  discussed  above  is reflected  also 
in higher-loop scattering  amplitudes  in $\NeqFour$ supergravity (as well as in all other  
theories with fewer supercharges that exhibit this $U(1)$ symmetry at the tree level). 
Indeed, an inspection of the two-particle cuts shows that, beginning at three loops,  amplitudes
that do not a priori  break the $U(1)$ symmetry (such as the four-graviton amplitude) receive 
nontrivial contributions from  anomalous matrix elements like \rf{anomalous_4pt}, \rf{examples}, 
cf. fig.~\ref{3loopcut}$(a)$.  Higher-loop amplitudes receive contributions from higher-point 
one-loop anomalous matrix elements as well as from higher-loop four-point ones, 
cf. fig.~\ref{3loopcut}$(b)$. 
It would undoubtedly be interesting to consider higher-loop
anomalous amplitudes. Based on available gauge theory amplitudes
\cite{Bern:2000dn} it should not be too difficult to find the two-loop counterpart of
\rf{anomalous_4pt}. Unlike the one-loop all-plus YM amplitude, the two-loop 
all-plus YM amplitude is UV-divergent. This divergence will cancel in the supergravity 
amplitude,   presumably  through the mechanism discussed in \cite{Bern:2012gh}.

A full understanding  of the implications of the $U(1)$ anomaly  for  the ultraviolet properties of 
the theory remains an open question, cf. \cite{bhn}.
It is nevertheless interesting to note that, due to the expression for  the anomalous amplitudes in \rf{anomalous_4pt},  
the super-cut in fig.~\ref{3loopcut}$(a)$ is independent of the cut momenta.\foot{Up to irrelevant numerical factors 
this cut is just $[12]^4\langle 34\rangle^4$. Up to a further numerical factor, the cut in fig.~\ref{3loopcut}$(b)$ has 
the same expression.} This implies that the part 
of the three-loop amplitude detected by this cut is effectively (i.e. after the integrals of each of the one-loop amplitude 
factors are evaluated leading to \rf{anomalous_4pt}) a one-loop bubble integral and thus is divergent in the UV.  
%rr
The result of \cite{Bern:2012cd}
then implies  that, in the complete amplitude, this apparent divergence is cancelled by the contribution of other
intermediate states not related by supersymmetry to a two-scalar state as well as by the contribution of other cuts, neither
one of them being obviously related to the $U(1)$ anomaly. This might hint at the existence of a larger symmetry in 
$\NeqFour$ supergravity.

It would also be interesting  to construct higher-point $\NeqFour$ supergravity amplitudes which carry
a non-zero $U(1)$ charge (and therefore  also break $U(1)$ invariance) and  which are obtained 
from gauge theory amplitudes with non-rational momentum dependence.  As already mentioned in 
sec.~\ref{special_classes}, if such amplitudes are non-vanishing, they potentially have nonlocal dependence 
on external momenta. The first candidate has five external legs. We note that such amplitudes
first appear in unitarity cuts of four-point $U(1)$-preserving amplitudes at four-loop order (cf. fig.~\ref{3loopcut}$(b)$ 
for a different field assignment to the cut legs). Whether such anomalous amplitudes affect the UV behavior of the theory 
(which, at four loops, will be unambiguously determined by an explicit calculation currently in  progress 
\cite{4loopNeq4toappear}) remains an open question.

\

%%%%%%%%%%%%%%%%%%%%%%%%%%%%%%%%%%%%%%%%%

\section*{Acknowledgments }
%%%%%%%%%%%%%%%%%%%%%%%%%%%

We would like to thank Z.~Bern, L.~Dixon and H.~Johansson for useful discussions and to Z.~Bern for comments on the draft.
RR thanks M.~G\"unaydin for discussions on symmetries of $\NeqFour$ and $\NeqEight$ supergravity 
theories. AAT  thanks  B.~Eden and G.~Korchemsky   for useful discussions.
The work of JJMC and that of R.K. is supported by SITP,  the NSF grant PHY-0756174 and a grant from the John Templeton Foundation.
The work  of RR is supported by the US DoE under contract DE-SC0008745.
The work of AAT is supported by the ERC Advanced grant No.290456
%``Gauge theory -- string theory duality'' 
and also by the STFC grant
ST/J000353/1.

\newpage
%%%%%%%%%%%%%%%%%%
 \appendix

\refstepcounter{section}
\def\theequation{A.\arabic{equation}}
\setcounter{equation}{0}

\section*{Appendix A:  On vector contribution to gravitational axial anomaly \la{ap1}}

\addcontentsline{toc}{section}{Appendix A:  On vector contribution to  the gravitational axial anomaly}

%\vspace{-6truemm}

The  contribution of vectors to the $U(1)$ duality anomaly  was originally  not included  in \cite{dfg}
and  once  accounted for   in \cite{mar}  led to the conclusion  that  the  duality anomaly 
does not cancel  in \nf PSG  but cancels  in $\N=8$  PSG.  
Since the duality  symmetry   acts 
only  on-shell (unless  one  gives up  Lorentz symmetry and considers a doubled   formulation as, e.g.,  in \cite{bhn})
the vector field contribution to the anomaly may look unfamiliar  and  it was not  derived  explicitly  in  \cite{mar}.
%%%%%%%%
%\foot{For a recent discussion of this contribution  in doubled formalism see \cite{bhn}.}
%There are, however,  other derivations  of  vector contribution to axial  anomaly  that help demystify the claim of \cite{mar}.
%We review  some history of this subject below. 
%%%%%%%%
Below we  shall make  few  clarifying comments and  mention some relevant references. 

Let us start with quantum Maxwell theory in  a curved    background and  consider  a (nonlocal) 
transformation $\delta A_\m= \eps K_{\m\n} A^\n$   where  $K_{\m\n}$   is a  differential operator such that   the corresponding field strength transforms as  $\delta F_{\m\n} = \eps F^*_{\m\n} $, i.e. as in the  duality transformation. 
Then the corresponding Noether current is given by  $ j^\m={\del L \ov \del \del_\m A_\n } \delta A_\n$ 
and its  divergence  is  equal to $ \del_\m j^\m = F^*_{\m\n} F^{\m\n}   +   O(\del_\m F^{\m\n})$. 
Thus   its  quantum expectation value  is  (after omitting the equations of motion term under path  integral)
\be 
 \langle  \del_\m j^\m \rangle =   \langle  F^{\m\n}   F^*_{\m\n}     \rangle   \ . \la{1ab} \ee
Naively,  the expectation  value $ \langle  F^{\m\n}    F^*_{\m\n}      \rangle $ of a parity-odd operator in the parity-even 
Maxwell  theory   should vanish.  However, as in the case of the chiral spinor  current anomaly 
in a theory  of a real spinor, a reparametrization-invariant regularization  leads to the conclusion 
that  this   correlator is   proportional to  the parity-odd curvature tensor contraction  
$RR^*\equiv \ha \ep^{\m\n\k\l}   R^{\r\s }_{\ \ \m\n}   R_{\r\s\k\l}$.

To find the proportionality coefficient  one may use, e.g., the standard perturbation theory by expanding near flat  space 
to  second order in $h_{\m\n} = g_{\m\n} - \delta_{\m\n}$. The problem reduces  then to the computation of the correlator 
 $ \langle (FF^*)(x_1)   T_{\m\n}(x_2)  T_{\k\l}(x_3)  \rangle $  (where $T_{\m\n}$ is the  stress tensor of the Maxwell  field) 
 in the flat-space free  abelian vector  theory.  
 This correlator   vanishes  at separated   points but  receives a  contact  term contribution: 
 in momentum space it corresponds to a one-loop triangular diagram  that  gives a  non-vanishing  finite 
 contribution   if  computed in a way consistent with reparametrization invariance (e.g. in dimensional regularization). 

Alternatively, one  may   determine  the integrated   value of $ \langle FF^*\rangle\ $ in a curved background 
using topological anomaly  considerations, i.e. by relating it  to the  difference between the number 
of self-dual and anti-self-dual
zero modes of the differential operator acting on  (anti) self-dual rank-2 tensors.  
It can then be related to the spectral index (using $\zeta$-function  regularization) 
as in \cite{gra,duf}  and \cite{rvn}\foot{Using 
that $\langle FF^* \rangle\  = \langle F^2_+     -   F^2_- \rangle\    $  
this  correlator can be expressed as   $  \tr ( P_+   G  - P_-  G )  $  where $G$ is  the field strength Green's function  
(i.e. $\langle F_{\m\n} (x_1) F_{\k\l}(x_2)  \rangle\  $) at coincident  points  
and $P_\pm$ are projectors onto (anti)self-dual components, i.e. it is   related to the index of 
the second-order order operator  acting on $F$. This operator   is found  by starting   with 
$D_\m F^{\m\n} =0$ equation    (which in the standard Lorentz gauge  leads   to the   $ - D^2 A_\m  + R_{\m\n} A^\n $
operator) 
and  acting with another derivative $D_\k$. The resulting equation 
$D_\k D^\m F_{\m\n}   -    D_\n D^\m F_{\m\k}   =0 $ can be simplified using 
$\epsilon^{\m\n\k\l }D_\n F_{\k\l }=0 $   leading to the  operator 
$ ( - D^2   +    X R ) F=0 $
where  $R$ is the  full curvature and $X$ is  proportional to  the  generator of $SO(4)$ 
in the vector representation  (see, e.g.,  \cite{duf}).
The final result for the correlator  is then  given     by  $\tr (  P_\pm     X R X R$) 
where the trace is over the corresponding representation  (the familiar analog in spinor 
case is  $\tr ( \gamma_5 \gamma_{\m\n} \gamma_{\l\r})     R^{\m\n \r\s} R^{\l\r}_{\ \ \r\s}  $).}
\foot{Here $\tau$ and $P$ are the Hirzebruch signature and the Pontryagin number, respectively and the first equal sign is the statement of the Hirzebruch signature theorem.}
\be 
\la{rr}
n(1,0) - n(0,1) =  \tau={ 1 \ov 3} P=    { 1\ov 3 (4\pi)^2}  \int  d^4x\    RR^*  
\ , \ee 
which  is  equivalent to the integrated form of    the local  relation in  eq.~\rf{ff}.  
To compare, for the chiral spinors  one gets 
$n(\ha,0) - n(0,\ha) =  - {1 \ov 24} P  $  which is consistent with eq.~\rf{2}.

The anomaly of a spin 1 field can also be  found  as follows   \cite{rom,rvn}:
%%%%%%%%
%Romer, van N in PLB 162, 1985, 290 cite on p 292  Romer   PLB 83, 1979, 172   for anomaly 
%of  antisymmetric selfdual tensor:  since
%%%%%%%% 
since a self-dual tensor can be counted as  a direct product of two  chiral spinors, 
its anomaly is $(2+2)$ times the  anomaly of a single spinor, i.e.  there  is  a factor of  $4$ 
difference  between its 
%v2
anomaly 
 and the anomaly of a single chiral spinor (an extra factor of $2$ 
comes  in the case of a full vector field).
%%%%%%%%
%(extra 2  comes   from us talking about full vector or full tensor, not just selfdual part).
%, so  there  is  $  {1 \ov 4}$ factor  difference  compared to a single vector.
%In general,  for spin $s$   selfdual  forms  one finds \cite{rom,duf}  (cf. \rf{an})
%\be \la{sss}
%n(s,0) - n(0,s)=  (-1)^{2s} (  2 s^3- s )  \, \tau \ ,  \ee
%  which gives familiar values for  $s= 1/2, 1, 3/2$. 
%The  result for spin 1 case is in disagreement with computation of \cite{gri}.
%%%%%%%%
To apply this count to $\NeqFour$ supergravity as in \rf{7} one needs also to take into 
account the chiral weights of fields, which need  not be same.

The same (``topological'')  count of the vector field  contribution to the $U(1)$ chiral  anomaly 
($4$ times that of a  spinor)  was used in \cite{mar}.
%%%%%%%%
%where  it was used implicitly 
% that   under the duality transformation $\delta F_{mn} = \eps F^*_{mn}$  one 
% finds  $\langle F_{mn} F^*_{mn} \rangle\  $ 
 %representing the divergence of the corresponding current. 
% This  fact was  where   
%%%%%%%%
 The  vector    contribution to  the chiral  gravitational anomaly was   
 made explicit in \cite{dolg}, 
 where  it  was  computed   directly   by expanding near a 
 flat background   and using  a diagrammatic unitarity-based  method. 
 The equivalent  result   was  found using standard covariant methods  in \cite{end,reu}
%%%%%%%%
%\foot{An advantage  of   keeping the background metric general  is 
%that  the  gravitational covariance is manifest  (assuming one also uses  say a proper time or zeta-function 
% prescription) while in  a 
% near flat expansion  in diagrammatic approach one needs
% to maintain it  by carefully  choosing the  regularization prescription, see \cite{gra}.}
%%%%%%%%
and also  by computing the  correlator  $ \langle FF^*   T_{\m\n} T_{\k\l} \rangle\  $    in 
coordinate space in  \cite{erd}.  A computation   of the vector contribution to the $U(1)$ duality  
anomaly using  the ``first-order''  doubled  formalism  was   given  in \cite{bhn}. 
 
%%%%%%%%%%%  
%(see comments in Nielsen et al  sect 2). But it would still be nice to see how that comes out of 
% a direct QFT  computation in near-flat expansion....
%%%%%%%%%%%

Let us  mention  again  that   viewing the $U(1)$  anomaly from the effective action point of view 
where  it  corresponds, in particular,  to the presence of a  local term  $\int RR^* B $ in \rf{ann}
(which for the vector field contribution  is proportional   $\int  \langle FF^* \rangle\  B $)
removes the mystery  related to the  vector field anomaly count in \cite{mar}.

\def \p {\phi} 
\def \L {\Lambda} 

\refstepcounter{section}
\def\theequation{B.\arabic{equation}}
\setcounter{equation}{0}

\section*{Appendix B:   On the $%\displaystyle{\int}
 \int 
R^*R^*\vp$  contribution to the effective action \la{ap2}}

\addcontentsline{toc}{section}{Appendix  B:   On the $\int R^*R^*\vp$  contribution to the effective action
} %A:  On vector contribution to  gravitational axial anomaly}

Starting with a supersymmetric theory   with a  classical  action containing the bosonic terms in \rf{laga} 
one  should  expect that the  effective action  should also be supersymmetric so that 
the anomalous terms  in \rf{ann}   should be also accompanied by other terms   that are parts of the same 
superinvariant. 
This supersymmetric aspect of the duality anomaly was  discussed,  e.g., in \cite{gdw}  and explained 
in detail in the present context in \cite{Bossard:2012xs}.

The  anomalous terms  in \rf{ann}  should thus  have their  ``supersymmetry partners''
in the full effective action. 
In particular,  the presence of  the local  anomalous parity-odd  term  $\int RR^* B $ in \rf{ann}   
implies that there should be also another local  parity-even term $\int RR \vp $.
Below  we shall   discuss  how  one  can directly  find such term in the  one-loop effective action. 
%%%%%%%%%%%
%Being local,  this term is of course sensitive to a choice of the   regularization prescription. 
%%%%%%%%%%%

 Consider   the Lagrangian ${\cal L}= k(x)(  \del \p)^2  = -   k(x)  \p \Delta \p + ...  $   where   $\p$ is a quantum field on a curved
  background and $k$ is  a background
 field. If we integrate over  $\p$  we get  a  complicated   dependence on the derivatives of $k$, but one may wonder if the 
 effective action contains  also a contribution that survives  if $k$ is constant, such as  $\int \sqrt g \,  \ln k \,  RR\, $.  
 This question depends of course on the choice of regularization scheme and path integral measure: 
 since the term we are interested in is local it  can be  changed by adding a local counterterm. 
 Indeed, if   we   make a local field redefinition $\p' = k^{-1/2}   \p $ then the remaining dependence on $k$ will be only through its derivatives. 
 The resulting Jacobian contribution to the effective action  can be regularized as follows:\foot{An alternative argument is 
 based on counting only 0-mode contributions, assuming $\delta(0)$  term with sum over all modes is set to zero.}
 \be \la{a1}
 \G_1 = \ha \Tr \ln (   k^{-1/2} e^{ - \L^{-2} \Delta})  \ , \ \ \ \ \ \ \ \ \  \L \to \infty \ . 
 \ee
 Using the asymptotic expansion of the heat kernel in four dimensions 
 $  e^{ - \L^{-2} \Delta } = \L^{4}  b_0  + \L^{2}  b_0   +  b_4    + O(\L^{-2}) $ 
 and assuming that all  cut-off dependent terms cancel between different quantum fields (or are removed  by 
 adding divergent counterterms or by some further regularization  prescription like $\zeta$-function regularization) 
   we are then left with  $\G'_1 = - { 1 \ov 4}  \int d^4 x \sqrt { g} \, \ln k  \,  b_4(x,x) $ 
 where   $b_4$ is the familiar conformal anomaly coefficient. 
 
 In the case  of a vector field   in four dimensions  with $k=e^{-\vp}$, i.e. with a Lagrangian
  \be
  {\cal L}= -\fo  e^{-\vp}   F_{\m\n} F^{\m\n} 
  \ee  
  as in  \rf{laga}, one  can show (by similar arguments as in the 
  two-dimensional case \cite{st})  that the effective action   found by integrating out $A_\m$   satisfies   \cite{gil} 
 \be \la{a2}
 \G[\vp, g] - \G[- \vp, g] = -   { 1 \ov 64 \pi^2 }   \int  d^4 x \sqrt g\,   R^*R^*\, \vp  \ ,  \ee
 where   $R^*R^* = R^2_{\m\n\k\l} - 4 R^2_{\m\n} + R^2$ is   the Euler number density. 
 This implies that
 \be \la{a3} 
 \G[\vp, g] = -   { 1 \ov 128 \pi^2 }   \int  d^4 x \sqrt g\,   R^*R^*\, \vp   +   O\big( (\del \vp) ^2\big) \ ,  \ee
 where all other  terms should be   even in $\vp$ (as follows from the fact that a vector-vector 
 duality   transformation  reverses the   sign of $\vp$)  and  should  depend only  its derivatives. 
  
 Note that the    coefficient of the local term in \rf{a3}   disagrees with the one found in \cite{gri} 
 where  a different regularization was used. This   illustrates   again that this coefficient  is, in general,  scheme-dependent. 
 In a  supersymmetric theory one should use a regularization  preserving  supersymmetry (which was not a priori the case in \cite{gri}). 
 
 To compute  the coefficient of this    leading   $\int  RR \vp$  term\foot{Here we do not  distinguish 
 between $R^*R^*$ and $RR$ as they differ  by  Ricci-tensor   terms that do not contribute to on-shell scattering amplitudes.} 
 such that it is consistent with the anomalous term discussed in Appendix~\ref{ap1}
  one may expand in powers of $\vp$  and 
 near-flat metric $h_{\m\n}= g_{\m\n}-\delta_{\m\n}$    so that   the leading  contribution will be determined by the correlator
 $ \langle F^2_{\m\n}(x_1)    T_{\k\l}(x_2)  T_{\r\s}(x_3 ) \rangle\ $ in  free   Maxwell theory. 
 This  correlator vanishes at separated points    but  will  contain  a local  $\delta$-function term  consistent 
 with \rf{a3}  under a particular regularization. 
 
Apart from the vector-scalar kinetic coupling of the type discussed above, the  $\NeqFour$ supergravity  
Lagrangian \rf{laga} also contains the scalar-scalar term  $e^{2 \vp} \del^\m B \del_\m B$   in \rf{vb}.
Its   contribution was also   computed in \cite{gri}   but 
again  depends on a particular regularization used. 
For example, given that  \rf{vb}   is a  sigma model (based on the $SL(2,R)/U(1)$  coset) one 
may   assume that the corresponding  path integral measure  should contain the usual $\sqrt G$   factor  
cancelling the effect of  redefinition of $B$ by $e^{ \vp}$ and thus   suggesting 
   that there should be  no  quantum  scalar $B$  contribution to \rf{a3}.\foot{At the same time, 
 having no contribution  from fermions  (assuming that they  have canonically normalized kinetic terms)  
 appears to be in conflict with supersymmetry --  spin 3/2 and 1/2  fermions   did contribute to 
  similar $\int RR^* B$ term in \rf{ann}.}
%%%%%%%%%%%
%\foot{One may wonder if   fermions    may have non-trivial $e^{\vp}$  scalar    factors in their 
%kinetic terms  if   $\NeqFour$   supergravity  is  taken in a   form as it   appears  from a more  symmetric theory, 
% e.g.,  from a superconformal one or  from 10-dimensional one. 
% This is  not the case for the formulation of \cite{roo}   but  dimensional reduction from 10 or 11 dimensions
%may indeed   bring such factors,   depending on how fermionic fields are defined.} 
%%%%%%%%%%%
 
One may wonder if   the  coefficient   of the local term in  \rf{a3}   in $\NeqFour$   supergravity is controlled by 
the total  ``conformal anomaly''  coefficient $b_4$.  Indeed,  the sum of $b_4$    coefficients 
in the supergravity multiplet can be written 
as  \cite{duf,Gri}\foot{Here $ {1 \ov 180} RR$, etc., terms  in $b_4$ cancel out between different spin contributions.}
\be 
  b_4{}_{\rm tot} =  { 1 \ov 32 \pi^2} \,     a_{\rm tot} \,   R^* R^* \ , \ \ \ \ \ \ \ \ \ \ \ 
a_{\rm tot}= { 1 \ov 24} (   58 n_2  - 17 n_{3/2}   - 2 n_1   - n_{1/2}   + n_0) \ , \la{a4}\ee
where   $N_s$ are the numbers of spin $s$ fields. In   $\NeqFour$  PSG 
\be 
n_2=1, \ \ \ n_{3/2}= n_{1/2} =4, \ \  \   n_1=6, \ \  \ n_0=2 \ . \ \ \ \ \ \ \ \ 
a= -1  \ , \la{a5}\ee
Thus, if all  fields were coupling to $\vp$ with the same weight as the vector fields, then the resulting 
coefficient $a_{\rm tot}$  would be  twice the one in \rf{a3}, i.e. 
$ -{ 1 \ov 64 \pi^2 } $,   which is, incidentally,  the same as the coefficient of $\int RR^*B$ term  in \rf{ann}.  

%%%%%%%%%%%
%The bottom line  is that to  compute   the coefficient   of  both $RR^*B$ and $R^*R^* \vp$ terms  
%consistently with supersymmetry one   
%is to choose a regularization that preserves supersymmetry.  
%Nonlocal terms   are not (or  at least   less)   sensitive  to regularization, so
%studying  them provides  a  more reliable way of
% extracting one-loop   corrections  to effective action and thus also to the corresponding S-matrix. 
%%%%%%%%%%%

%%%%%%%%%%%%%%%%%%%%%%%%%%%%%%%%%%%%%%%%%%%%

\refstepcounter{section}
\def\theequation{C.\arabic{equation}}
\setcounter{equation}{0}

\section*{Appendix C:  \ The one-loop four-point superamplitude containing ${\cal M}_4^{(1)}(h^{--},h^{++}, h^{++},\bartaua)$ \label{4pt_oneminus}}

\addcontentsline{toc}{section}{Appendix C:  \ The 1-loop 4-point superamplitude containing ${\cal M}_4^{(1)}(h^-,h^+, h^+,\bartaua)$ 
}

In this appendix we shall use the double-copy construction \cite{BCJLoop} reviewed in 
sec.~\ref{general_amplitudes} to compute the superamplitude containing the three-graviton-scalar amplitude. 
Because the $\NeqFour$ sYM amplitude has loop-momentum-independent numerator factors of its one-loop 
four-point amplitude,  we can use the formulation  \cite{Bern:2012gh} of the double-copy construction which 
expresses the four-point supergravity amplitude as
\bea
\label{Neq4_4ptamps_0}
&&{\cal M}_{4}^{(1);{\cal N}=4\,\text{PSG}} (1,2, 3,4)
\\
&&\qquad
=i\left(\frac{\kappa}{2}\right)^4 s_{12}s_{23} {\cal A}_{\NeqFour}^{\rm tree}(1,2,3,4)
\left( A^{(1)}_{{\cal N}=0}(1,2,3,4)+A^{(1)}_{{\cal N}=0}(1,3,4,2)+A^{(1)}_{{\cal N}=0}(1,4,2,3)
\right)
\nonumber
\eea
where $A^{(1)}_{{\cal N}=0}$ is a pure or matter-coupled YM one-loop color-ordered 
amplitude. The helicities of supergravity
fields are determined in the usual way, by adding the helicities of the individual gauge theory fields, 
cf. sec.~\ref{general_amplitudes}. 

The only $\NeqFour$ sYM amplitude is the MHV one, which we choose to write in anti-chiral superspace
\be
A^{\rm tree}_{{\cal N}=4}(1,2,3,4) = i\frac{1}{[12][23][34][41]} \delta^{(8)}(\sum_{i=1}^4{\tilde \eta}_{i,A}{\tilde \lambda}_i)\ .
\label{4pttreesuper_conj}
\ee
For the field configuration we are interested in, ${\cal M}^{(1)}(h^-,h^+, h^+,\bartaua)$, the relevant ${\cal N}=0$  
gauge theory amplitude is the single-minus amplitude \cite{Bern:1995db}
\be
A_{{\cal N}=0}(1^-,2^+,3^+,4^+) = \frac{i}{3(4\pi)^2}\frac{\spa2.4[24]^3}{[12]\spa2.3\spa3.4 [41]} \ .
\label{oneminus}
\ee
It is not difficult to see that it is symmetric under permutations of 2,3,4, as Bose symmetry requires. In the presence of
$\nv$ scalar fields this amplitude acquires an extra factor of $(2+\nv)/2$.

Therefore, using (\ref{Neq4_4ptamps_0}), the ${\cal N}=4$ supergravity amplitude with the ${\cal N}=0$ amplitude 
factor being (\ref{oneminus}) is
\bea
{\cal M}_4^{(1);{\cal N}=4\,\text{PSG}}(1,2,3,4)&=&-\frac{i}{(4\pi)^2}\frac{\spa2.4[24]^3}{[12]\spa2.3\spa3.4 [41]}
\frac{s_{12}s_{23}}{[12][23][34][41]} 
\delta^{(8)}(\sum_{i=1}^4{\tilde \eta}_{i,A}{\tilde \lambda}_i)\no 
\\
&=&-\frac{i}{(4\pi)^2}\frac{1}{[31][14]}\frac{[32][24]\spa2.1}{[21]}
\delta^{(8)}(\sum_{i=1}^4{\tilde \eta}_{i,A}{\tilde \lambda}_i) \ .
\label{4ptnonlocal_app}
\eea
This is the superamplitude quoted in eq.~\rf{4pts_nonlocal_amp}.

We may extract several instances of ${\cal M}_4^{(1)}(h^{++},h^{++}, h^{--},\bartaua)$, which superficially 
differ only by their momentum assignment. Two examples are \footnote{This result was obtained independently 
by Z. Bern and L. Dixon \cite{BernDixon}.}
\bea
{\cal M}_4^{(1);{\cal N}=4\,\text{PSG}}(1^{--},2^{++},3^{++},4^{\taua}) = 
-\frac{i}{(4\pi)^2}\frac{1}{[31][14]}\frac{[32][24]\spa2.1}{[21]}[23]^4 \ ,
\label{v1}
\\
{\cal M}_4^{(1);{\cal N}=4\,\text{PSG}}(1^{--},2^{\taua},3^{++},4^{++}) = 
-\frac{i}{(4\pi)^2}\frac{1}{[31][14]}\frac{[32][24]\spa2.1}{[21]}[34]^4 \ .
\label{v2}
\eea
In the limit in which the momentum of the scalar field is soft the second expression (\ref{v2}) vanishes 
identically since it scales like some positive power of scalar momentum $k_2$. The first expression appears 
to give a finite expression. However, since the 3-point amplitude with two positive-helicity gravitons is 
${\overline {\rm MHV}}$,
momenta should be continued such that the products $[ab]\ne 0$ while $\spa{a}.{b}=0$. The right-hand side of 
(\ref{v1}) is proportional to $\spa1.2$ and therefore vanishes in the soft limit as well, as required by consistency
with supersymmetry and (\ref{v2}).

Upon use of the graviton soft function \cite{Bern:1998ug}
the soft graviton limit $k_1\rightarrow 0$ together with the fact that in ${\cal N}=4$ antichiral superspace
the negative helicity graviton wave function contains no factors of ${\tilde \eta}$, we find the answer \cite{BernDixon}
quoted in eq.~\rf{anomalous_3pt}:
\be
{\cal M}_3^{(1);{\cal N}=4\,\text{PSG}}(1,2,3)=\frac{i}{(4\pi)^2}
\delta^{(8)}(\sum_{i=1}^3{\tilde \eta}_{i,A}{\tilde \lambda}_i)
\ee
In the presence of $\nv$ scalar fields coupled to the 
YM theory or, equivalently, in the presence of additional $\nv$ vector multiplets coupled to $\NeqFour$ supergravity,  
both this superamplitude and the one in eq.~\rf{4ptnonlocal_app} acquire  an extra factor of $(2+\nv)/2$.

\refstepcounter{section}
\def\theequation{D.\arabic{equation}}
\setcounter{equation}{0}

\section*{Appendix D: \ The one-loop four-point superamplitude containing 
${\cal M}_4^{(1)}(h^{++},h^{++}, \bartaua,\bartaua)$ \label{4pt_allplus}}

\addcontentsline{toc}{section}{Appendix D: \ The 1-loop 4-point superamplitude containing 
${\cal M}_4^{(1)}(h^{++},h^{++}, \bartaua,\bartaua)$ }

To compute the superamplitude containing the 
two-graviton--two-scalar amplitude we can use again the double-copy construction 
in the form \cite{Bern:2012gh}:
\bea
\label{Neq4_4ptamps}
&&{\cal M}_{4}^{(1);{\cal N}=4\,\text{PSG}} (1,2, 3,4)
\\
&&\qquad
=i\left(\frac{\kappa}{2}\right)^4 s_{12}s_{23} {\cal A}_{\NeqFour}^{\rm tree}(1,2,3,4)
\left( A^{(1)}_{{\cal N}=0}(1,2,3,4)+A^{(1)}_{{\cal N}=0}(1,3,4,2)+A^{(1)}_{{\cal N}=0}(1,4,2,3)
\right) \ .
\nonumber
\eea
where $A^{(1)}_{{\cal N}=0}$ are pure or matter-coupled YM theory color ordered amplitudes.
For the desired field configuration the relevant $A^{(1)}_{{\cal N}=0}$ is the all-plus four-point amplitude 
given by \cite{Bern:1995db}
\be
A^{(1)}_{{\cal N}=0} = \frac{2i}{(4\pi)^2}\frac{[12][34]}{\langle 12 \rangle\langle 34\rangle}
\left(-\frac{1}{6}+{\cal O}(\epsilon)\right) \ . 
\label{4ptallplus_app}
\ee
As in Appendix~\ref{4pt_oneminus} we shall use the anti-chiral superspace expression
of the $A^{\rm tree}_{{\cal N}=4}(1,2,3,4)$ factor
%
\iffalse
%\be
%A^{\rm tree}_{{\cal N}=4}(1,2,3,4) = i\frac{1}{\langle 12 \rangle\langle 23\rangle
%\langle 34\rangle\langle 41\rangle} \delta^{(8)}(\sum_{i=1}^4\eta_i^A\lambda_i)\ ,
%\label{4pttreesuper}
%\ee
\fi
%
\be
A^{\rm tree}_{{\cal N}=4}(1,2,3,4) = \frac{i}{[12][23][34][41]} \delta^{(8)}(\sum_{i=1}^4{\tilde \eta}_{i,A}{\tilde \lambda}_i)\ .
\label{4per_conj}
\ee
Alternatively, we can use the color/kinematics-satisfying representation of the one-loop ${\cal N}=0$ amplitude 
obtained by dimension shifting \cite{Bern:1996ja} from the ${\cal N}=4$ four-point MHV gluon amplitude (i.e. the coefficient 
of e.g. ${\tilde\eta}_1^4{\tilde\eta}_2^4$ in the first equation below):
\bea
&&{\cal A}^{(1);{\cal N}=4}(1,2,3,4)=i\frac{\langle 12\rangle\langle 34\rangle}{[12][34]}
\delta^{(8)}(\sum_{i=1}^4{\tilde \eta}_{i,A}{\tilde \lambda}_i)\left(I_{1234}[1]C_{1234}+I_{1342}[1]C_{1342}+I_{1423}[1]C_{1423}\right), \no 
\\
&&{\cal A}^{(1);{\cal N}=0}(1^+,2^+,3^+,4^+)=2i\frac{[12][34]}{\langle 12\rangle\langle 34\rangle}
\left(I_{1234}[\mu^4]C_{1234}+I_{1342}[\mu^4]C_{1342}+I_{1423}[\mu^4]C_{1423}\right) 
\label{amps_4pts}
\eea
Here $C_{abcd}$ are the color factors of a box integral with external legs ordered as $(a,b,c,d)$ and 
$I_{1234}[\mu^4]$ is
\be
I_{abcd}[\mu^4] = -\epsilon(1-\epsilon) I_{abcd}^{8-2\epsilon} = -\frac{1}{(4\pi)^2}\frac{1}{6}
\ee
with $I_{abcd}^{8-2\epsilon}$ the eight-dimensional box interval with external legs ordered as $(a,b,c,d)$. The 
argument $\mu^4$ of the integral on the left-hand side represents the insertion in the numerator of a four-dimensional 
box integral of the fourth power of the $(-2\epsilon)$-dimensional of the loop momentum. All external momenta are taken to be four dimensional.

Putting together the amplitudes \rf{amps_4pts} following the double-copy construction, we find
\bea
{\cal M}_{4}^{(1);{\cal N}=4\,\text{PSG}} (1,2, 3,4)&=&-i\left(\frac{\kappa}{2}\right)^4\left(-2\times 3\times \frac{1}{6}\right)
\frac{[12][34]}{\langle 12\rangle\langle 34\rangle}
\frac{\langle 12\rangle\langle 34\rangle}{[12][34]}
\delta^{(8)}(\sum_{i=1}^4{\tilde\eta}_{i,A}{\tilde\lambda}_i)
\cr
&=&i\left(\frac{\kappa}{2}\right)^4\delta^{(8)}(\sum_{i=1}^4{\tilde\eta}_{i,A}{\tilde\lambda}_i) \ ,
\label{finalappD}
\eea
which is the expression quoted in eq.~\rf{anomalous_4pt}. In the presence of $\nv$ real scalar fields coupled to the 
YM theory or, equivalently, in the presence of additional $\nv$ vector multiplets in $\NeqFour$ supergravity,  the amplitudes in 
eq.~\rf{4ptallplus_app}, the second eq.~\rf{amps_4pts} and eq.~\rf{finalappD} each acquire a factor of $(2+\nv)/2$.

\refstepcounter{section}
\def\theequation{E.\arabic{equation}}
\setcounter{equation}{0}

%%%%%%%%%%%%%%%%%%%%%%%%%%%%%%%%
\section*{Appendix E: \ \ Five-point superamplitudes \label{5pt_allplus}}

\addcontentsline{toc}{section}{Appendix E: \ \ Five-point superamplitudes }

In this appendix we include some of the details  of  the calculations leading to eqs.~\rf{examples} 
and \rf{final5ptsuperamlitude}. The main ingredients are the five-point superamplitude in a 
form obeying color/kinematic duality \cite{Carrasco:2011mn} and the corresponding all-plus 
amplitude in a similar color/kinematic-satisfying representation obtained though dimension-shifting.

\subsection*{E.1 \ \ Ingredients}

%%%%%%%%% FIGURE %%%%%%%%%%%%%%%%%%
\begin{figure}[t]
\begin{center}
\includegraphics[width=0.7\textwidth]{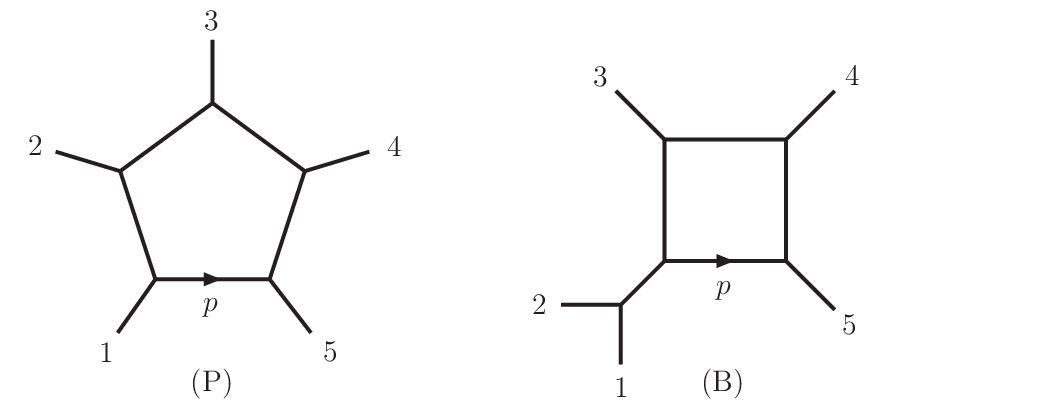}
\end{center}
\caption{\small The two integral topologies that appear in the five-point one-loop amplitudes.}
\label{1L}
\end{figure}
%%%%%%%%%%%%%%%%%%%%%%%%%%%%%%%%

The one-loop  five-point ${\cal N}=4~\text{sYM}$ 
MHV amplitude in a form obeying color/kinematics duality is \cite{Carrasco:2011mn}
\be
{\cal A}_5^{(1);\,  {\cal N}=4} =  i g^{5} \, \sum_{S_5} \, 
\Bigl( 
{\frac{1}{10}}\beta_{12345}C^{(\rm P)} I^{(\rm P)}+{\frac{1}{4}}\gamma_{12} C^{(\rm B)}I^{(\rm B)}_{12}
\Bigr) \,, 
\label{OneLoopSYMAmplitude}
\ee
where $g$ is the coupling constant, and the sum is over all 120 permutations, $S_5$, of the external leg labels; the symmetry factors 1/10 and 1/4 compensate for the overcount in this sum, $I^{(P)}$ and $I^{(B)}_{12}$ are the integrals shown in fig.~\ref{1L} and given by
\bea
I^{(\rm P)}&=&\int \frac{d^Dp}{(2\pi)^D} \frac{1}{p^2(p+k_1)^2(p+k_1+k_2)^2(p-k_4-k_5)^2(p-k_5)^2}\,, \no \\
I^{(\rm B)}&=&\frac{1}{s_{12}}\int \frac{d^Dp}{(2\pi)^D} \frac{1}{p^2(p+k_1+k_2)^2(p-k_4-k_5)^2(p-k_5)^2}\, .
\eea
The coefficients $\beta$ and $\gamma$ coefficients are: 
\bea
\beta_{12345}&=& i \delta^{(8)}(Q_5)\frac{\spb{1}.{2}\spb{2}.{3}\spb{3}.{4}\spb{4}.{5}\spb{5}.{1}}{\spa{1}.{2}\spb{2}.{3}\spa{3}.{5}\spb{5}.{1}-\spb{1}.{2}\spa{2}.{3}\spb{3}.{5}\spa{5}.{1}}=\delta^{(8)}(Q_5)\frac{\spb{1}.{2}\spb{2}.{3}\spb{3}.{4}\spb{4}.{5}\spb{5}.{1}}{4\, \varepsilon(1,2,3,4)}\,,
\nonumber
\\
\gamma_{12}&=&\beta_{12345}-\beta_{21345}=\delta^{(8)}(Q_5)\frac{\spb{1}.{2}^2\spb{3}.{4}\spb{4}.{5}\spb{3}.{5}}{4\, \varepsilon(1,2,3,4)}\,.
\label{gammaform}
\eea
The argument of the $\delta$-function is the usual supermomentum
\be
Q_5^{\alpha A}=\sum_{i=1}^{5}\lambda_i^{\alpha}\eta_i^{A} \ .
\label{gdelta}
\ee
%%%%%%%%%%%
%When the external states are scalars the integrand of the pentagon integral receives additional 
%$\mu$-integral contributions which integrate to zero because they are odd in $\mu$ and are not 
% relevant for our discussion.
%%%%%%%%%%%
The color factors can be read directly form the graphs in fig.~\ref{1L}:
\bea
C^{(\rm P)}&=& \f^{g a_1b}\f^{b a_2 c}\f^{c a_3 d}\f^{d a_4 e }\f^{e a_5 g} \,, \nonumber \\
C^{(\rm B)}&=& \f^{a_1a_2 b}\f^{b c g}\f^{c a_3 d}\f^{d a_4 e }\f^{e a_5 g} \,,
\eea
where $a_i$ are the external color labels.

\

We also need the all-plus five-point color-dressed gluon amplitude in pure Yang-Mills theory; to construct it
we may use the following color-ordered  all-plus five-point amplitude \cite{Bern:1996ja}, valid to all 
orders in~$\epsilon$:
\bea
&&
A_5^{(1)}(1^+,2^+,3^+,4^+,5^+)=
\frac{i}{\langle 12\rangle\langle 23\rangle\langle 34\rangle\langle 45\rangle\langle 51\rangle}
\frac{\epsilon(1-\epsilon)}{(4\pi)^{2-\epsilon}}
\cr
&&\quad\times\Big[s_{23}s_{34}I_{4;51}^{D=8-2\epsilon}+s_{34}s_{45}I_{4;12}^{D=8-2\epsilon}
+s_{45}s_{51}I_{4;23}^{D=8-2\epsilon} \cr
&&\quad +s_{51}s_{12}I_{4;34}^{D=8-2\epsilon}+s_{12}s_{23}I_{4;45}^{D=8-2\epsilon}
+(4-2\epsilon)\Tr[\gamma_5 k_1k_2k_3k_4]I^{(P),D=10-2\epsilon} \Big]
\label{allplus5_2}
\eea
To find the color-dressing we use \cite{DelDuca:1999rs}; the color factors are given 
by a color-space pentagon graph with one structure constant at each vertex:
\be
{\cal A}_5^{(1)}=g^2\sum_{\sigma\in S_{4}/{\cal R}} \Tr[F^{a_{\sigma_1}}F^{a_{\sigma_2}}
F^{a_{\sigma_3}}F^{a_{\sigma_4}}F^{a_{\sigma_5}}] A_5^{(1)}
({\sigma_1},{\sigma_2},{\sigma_3}, {\sigma_4},{\sigma_5}) \ .
\ee
Here the sum is over all non-cyclic permutations and ${\cal R}$ is the reflection, 
${\cal R}(1,2,3,4,5)=(5,4,3,2,1)$,  {etc.}  This color dressing applies to all one-loop amplitudes.

The integrals that appear in the all-plus amplitude (\ref{allplus5_2}) are \cite{Bern:1996ja}:
\be
\epsilon(1-\epsilon)I_{4;i,i+1}^{D=8-2\epsilon}=\frac{1}{6}\ , 
\qquad
\epsilon(1-\epsilon)I^{(P),D=10-2\epsilon} =\frac{1}{24} \ .
\label{valueintegrals}
\ee
Using them (\ref{allplus5_2}) becomes
\bea
&&
A_5^{(1)}(1^+,2^+,3^+,4^+,5^+)=
\frac{i}{\langle 12\rangle\langle 23\rangle\langle 34\rangle\langle 45\rangle\langle 51\rangle}
\frac{1}{6\,(4\pi)^{2}}
\cr
&&\qquad\qquad\qquad\times\Big[s_{23}s_{34}+s_{34}s_{45}
+s_{45}s_{51}+s_{51}s_{12}+s_{12}s_{23}+\Tr[\gamma_5 k_1k_2k_3k_4] \Big]
\label{allplus5v1}
\\
&&\qquad\qquad\qquad\qquad\quad
=-\frac{i}{48\pi^2}
\frac{1}{\langle 12\rangle\langle 23\rangle\langle 34\rangle\langle 45\rangle\langle 51\rangle}
\sum_{1\le i_1<i_2<i_3<i_4\le 5}\Tr_{+}[k_{i_1}k_{i_2}k_{i_3}k_{i_4}]~.
\nonumber
\eea

\

We can also construct the color-dressed all-plus one-loop five-point amplitude in a color/kinematics-satisfying form 
by using its relation to the dimensionally-shifted MHV amplitude amplitude in ${\cal N}=4$ sYM theory \cite{Bern:1996ja} 
and starting with \rf{OneLoopSYMAmplitude}. The result is
\be
{\cal A}_5^{(1);\,  {\cal N}=0} =  2 i g^{5} \, \sum_{S_5} \, 
\Bigl( 
{\frac{1}{10}}{\widehat\beta}_{12345}C^{(\rm P)} I^{(\rm P)}[\mu^4]+{\frac{1}{4}}{\widehat\gamma}_{12} 
C^{(\rm B)}I^{(\rm B)}_{12}[\mu^4] \Bigr) \,, 
\label{OneLoopSYMAmplitudeNeq0}
\ee
with
\bea
{\widehat\beta}_{12345}&=&
i \frac{\spb{1}.{2}\spb{2}.{3}\spb{3}.{4}\spb{4}.{5}\spb{5}.{1}}{\spa{1}.{2}\spb{2}.{3}\spa{3}.{5}\spb{5}.{1}-\spb{1}.{2}\spa{2}.{3}\spb{3}.{5}\spa{5}.{1}}=\frac{\spb{1}.{2}\spb{2}.{3}\spb{3}.{4}\spb{4}.{5}\spb{5}.{1}}{4\, \varepsilon(1,2,3,4)}\,,
\label{betaform0}
\\
{\widehat\gamma}_{12}&=&{\widehat\beta}_{12345}-{\widehat\beta}_{21345}=\frac{\spb{1}.{2}^2\spb{3}.{4}\spb{4}.{5}\spb{3}.{5}}{4\, \varepsilon(1,2,3,4)}\,,
\label{gammaform0}
\eea
These quantities are the same as $\beta$ and $\gamma$ but with the $\delta^{(8)}(Q)$ stripped off. 
This expression as well as \rf{OneLoopSYMAmplitudeNeq0} also match (after suitable manipulations) 
the one found in \cite{Bern:1995db}.

Evaluating the integrals using \rf{valueintegrals}
\be
I^{(\rm P)}[\mu^4]=0+{\cal O}(\epsilon)\ , 
\qquad
I^{(\rm B)}_{i, i+1}[\mu^4]=-\epsilon(1-\epsilon)\frac{I^{D=8-2\epsilon}_{4;i, i+1}}{s_{i, i+1}}
=-\frac{1}{(4\pi)^2}\frac{1}{6}\frac{1}{s_{12}}+{\cal O}(\epsilon)\ , 
\ee
implies that
\be
{\cal A}_5^{(1);\, {\cal N}=0}(1^+,2^+,3^+,4^+,5^+) =  -2\,\frac{i g^{5}}{6\,(4\pi)^2}\, \sum_{S_5} \, 
{\frac{1}{4}}\frac{{\widehat\gamma}_{12} }{s_{12}}
C^{(\rm B)} \, . 
\label{EvalOneLoopAllPlusAmplitude}
\ee
From here it is not difficult to extract the color ordered amplitude ${A}_5^{(1);\,  {\cal N}=0}(1^+,2^+,3^+,4^+,5^+)$ 
\be
{A}_5^{(1);\, {\cal N}=0}(1^+,2^+,3^+,4^+,5^+) 
=\frac{i g^{5}}{3\,(4\pi)^2}\Big[
\frac{\gamma_{12}}{s_{12}}+\frac{\gamma_{23}}{s_{23}}
 +\frac{\gamma_{34}}{s_{34}}+\frac{\gamma_{45}}{s_{45}}
 +\frac{\gamma_{51}}{s_{51}}\Big]\  
 \label{COallplus}
\ee
and check that reproduces \rf{allplus5v1} after suitable manipulations. In the presence of $\nv$ scalar fields both equations
pick up a factor of $(2+\nv)/2$.

%%%%%%%%%%%%%%%%%%%%%%%%%
\subsection*{E.2 \ \ The 5-point superamplitude containing ${\cal M}_5^{(1)}(h^{++},h^{++},\bartaua,\bartaua,\bartaua)$}

Using the ingredients above we can construct the superamplitude containing the two-graviton-three-scalar component 
amplitude${\cal M}_5^{(1)}(h^{++},h^{++},\bartaua,\bartaua,\bartaua)$. It is not difficult to see that, to have the 
desired field content, we in fact need the ${\overline {\rm MHV}}$ $\NeqFour$ superamplitude. As mentioned previously, 
it may be obtained from \rf{OneLoopSYMAmplitude} by simply $\lambda_{\alpha i}\leftrightarrow {\tilde\lambda}_{{\dot\alpha}, i} $ 
and $\eta\leftrightarrow {\tilde\eta}$. Using \cite{BCJLoop} or just eq.~\rf{DoubleCopy} we find
\bea
\label{5ptsuperamp_final1}
&&
{\cal M}_5^{(1);\,  {\cal N}=4\,\text{PSG}}(1,2, 3,4,5) 
=  i\left(\frac{\kappa}{2}\right)^5 \,\frac{\delta^{(8)}({\overline Q}_5)}{3\,(4\pi)^2}  \,
\sum_{S_5} \,  \frac{1}{4}\frac{{\widehat{\overline\gamma}}_{12}{\widehat\gamma}_{12}}{s_{12}} 
\no \\
&& 
= i\left(\frac{\kappa}{2}\right)^5 \,\frac{\delta^{(8)}({\overline Q}_5)}{(4\pi)^2}  \,
\Bigl[
\frac{{\widehat\gamma}_{12}  {\widehat{\overline\gamma}}_{12}}{s_{12}}
+\frac{{\widehat\gamma}_{13} {\widehat{\overline\gamma}}_{13}}{s_{13}}
+\frac{{\widehat\gamma}_{14} {\widehat{\overline\gamma}}_{14}}{s_{14}}
+\frac{{\widehat\gamma}_{15} {\widehat{\overline\gamma}}_{15}}{s_{15}}
+\frac{{\widehat\gamma}_{23} {\widehat{\overline\gamma}}_{23}}{s_{23}}
\cr
&&\qquad\qquad\quad\quad~
+\frac{{\widehat\gamma}_{24} {\widehat{\overline\gamma}}_{24}}{s_{24}}
+\frac{{\widehat\gamma}_{25} {\widehat{\overline\gamma}}_{25}}{s_{25}}
+\frac{{\widehat\gamma}_{34} {\widehat{\overline\gamma}}_{34}}{s_{34}}
+\frac{{\widehat\gamma}_{35} {\widehat{\overline\gamma}}_{35}}{s_{35}}
+\frac{{\widehat\gamma}_{45} {\widehat{\overline\gamma}}_{45}}{s_{45}}
\Bigr] \  , 
\label{final5ptsuperamlitude_app}
\eea
where ${\widehat\beta}_{12345}$ and ${\widehat\gamma}_{ij}$ are defined in eqs.~\rf{betaform0} and \rf{gammaform0}
and
\bea
{\overline Q}_5&=&\sum_{i=1}^5 {\tilde\eta}_{i,A}{\tilde\lambda}_i \ , \la{quqaagdelta} 
\\
{\widehat{\overline\gamma}}_{12}&=&{\widehat{\overline\beta}}_{12345}-{\widehat{\overline\beta}}_{21345}=\frac{\spa{1}.{2}^2\spa{3}.{4}\spa{4}.{5}\spa{3}.{5}}{4\, \varepsilon(1,2,3,4)}\, ,
\\
{\widehat{\overline\beta}}_{12345}&=&
i \frac{\spa{1}.{2}\spa{2}.{3}\spa{3}.{4}\spa{4}.{5}\spa{5}.{1}}{\spa{1}.{2}\spb{2}.{3}\spa{3}.{5}\spb{5}.{1}-\spb{1}.{2}\spa{2}.{3}\spb{3}.{5}\spa{5}.{1}}=\frac{\spa{1}.{2}\spa{2}.{3}\spa{3}.{4}\spa{4}.{5}\spa{5}.{1}}{4\, \varepsilon(1,2,3,4)} \ .
\eea
The sum in eq.~\rf{final5ptsuperamlitude_app} evaluates to
\bea
&&
\frac{{\widehat\gamma}_{12}  {\widehat{\overline\gamma}}_{12}}{s_{12}}
+\frac{{\widehat\gamma}_{13} {\widehat{\overline\gamma}}_{13}}{s_{13}}
+\frac{{\widehat\gamma}_{14} {\widehat{\overline\gamma}}_{14}}{s_{14}}
+\frac{{\widehat\gamma}_{15} {\widehat{\overline\gamma}}_{15}}{s_{15}}
+\frac{{\widehat\gamma}_{23} {\widehat{\overline\gamma}}_{23}}{s_{23}}
\cr
&&\qquad\qquad
+\frac{{\widehat\gamma}_{24} {\widehat{\overline\gamma}}_{24}}{s_{24}}
+\frac{{\widehat\gamma}_{25} {\widehat{\overline\gamma}}_{25}}{s_{25}}
+\frac{{\widehat\gamma}_{34} {\widehat{\overline\gamma}}_{34}}{s_{34}}
+\frac{{\widehat\gamma}_{35} {\widehat{\overline\gamma}}_{35}}{s_{35}}
+\frac{{\widehat\gamma}_{45} {\widehat{\overline\gamma}}_{45}}{s_{45}}
 =2
\eea
and thus the superamplitude containing the anomalous amplitude 
${\cal M}_5^{(1)}(h^{++},h^{++},\bartaua,\bartaua,\bartaua)$ is
\be
{\cal M}_5^{(1);\,  {\cal N}=4}(1,2, 3,4,5) 
= 2i\left(\frac{\kappa}{2}\right)^5 \,\frac{\delta^{(8)}({\overline Q}_5)}{(4\pi)^2} \ .
\label{super5ptconj}
\ee
This superamplitude is local and is the result quoted in \rf{examples}. 
%%%%%%%%%%%
%An additional factor of $(2+\nv)/2$ appears in the presence of $\nv$ vector multiplets.
%%%%%%%%%%%

\subsection*{E.3 \ \ Five-point superamplitude containing ${\cal M}_5^{(1)}(h^{++},h^{++}, \taua, A^+, A^+)$}

Using the ingredients described above  we can also construct the superamplitude containing 
${\cal M}(h^{++},h^{++}, \taua, A^+, A^+)$. This field content implies that it is natural to present this 
superamplitude in chiral superspace. Using \cite{BCJLoop} or just eq.~\rf{DoubleCopy} we find that 
it is given by:
\bea
\label{5ptsuperamp_final2}
&&
{\cal M}_5^{(1);\,  {\cal N}=4 \,\text{PSG}}(1,2, 3,4,5) 
= i \left(\frac{\kappa}{2}\right)^5 \,\frac{\delta^{(8)}(Q_5)}{3\,(4\pi)^2}  \,
\sum_{S_5} \,  \frac{1}{4}\frac{({\widehat\gamma}_{12})^2}{s_{12}} \no 
\\
&& 
= i\left(\frac{\kappa}{2}\right)^5 \,\frac{\delta^{(8)}(Q_5)}{(4\pi)^2}  \,
\Bigl[
\frac{{\widehat\gamma}_{12}^2}{s_{12}}+\frac{{\widehat\gamma}_{13}^2}{s_{13}}
+\frac{{\widehat\gamma}_{14}^2}{s_{14}}+\frac{{\widehat\gamma}_{15}^2}{s_{15}}
+\frac{{\widehat\gamma}_{23}^2}{s_{23}}+\frac{{\widehat\gamma}_{24}^2}{s_{24}}
+\frac{{\widehat\gamma}_{25}^2}{s_{25}}+\frac{{\widehat\gamma}_{34}^2}{s_{34}}
+\frac{{\widehat\gamma}_{35}^2}{s_{35}}+\frac{{\widehat\gamma}_{45}^2}{s_{45}}
\Bigr] \; 
\label{final5ptsuperamlitude_app1}
\eea
This is the expression quoted in eq.~\rf{final5ptsuperamlitude}. An additional factor of $(2+\nv)/2$ appears in 
the presence of $\nv$ vector multiplets.

It is not difficult to check (numerically) that this expression agrees with the result of the double-copy construction 
that uses the standard form of the all-plus five-point YM amplitude $A^{(1)}_{{\cal N}=0}(1^+,\dots, 5^+)$ given 
in \cite{Bern:1995db, Bern:1996ja} and in eq.~\rf{allplus5_2}.

To extract the amplitude ${\cal M}(h^{++},h^{++}, \bartaua, A^+, A^+)$ from (\ref{final5ptsuperamlitude}) one   
isolates the appropriate combination of $\eta$-variables, i.e. $\eta_3^4\eta_4^A\eta_4^B\eta_5^C\eta_5^D\epsilon_{ABCD}$; 
this leads to the simple replacement $\delta^{(8)}(Q)\mapsto\langle 35\rangle^2\langle 34\rangle^2$

\newpage

%%%%%%%%%%%%%%%%%%%%%%%%%%%%%%%%%%%%%%%%%

%%%%%%%%%%%%%%%%%%%%%%%%%%%%%%%

\end{document}